\documentclass[acmlarge]{acmart}

\usepackage{booktabs} 
\usepackage{subcaption}
\usepackage{color}
\usepackage{graphicx} 
\usepackage{amsmath}
\usepackage{bbold}
\usepackage{array}
\usepackage{arydshln}
\usepackage[ruled]{algorithm2e} 

\newcommand{\ra}[1]{\renewcommand{\arraystretch}{#1}}
\definecolor{blue}{rgb}{0,0,0}
\definecolor{green}{rgb}{0,0.5,0}
\SetAlFnt{\small}
\SetAlCapFnt{\small}
\SetAlCapNameFnt{\small}
\SetAlCapHSkip{0pt}
\IncMargin{-\parindent}
\usepackage{lineno}
\usepackage{wrapfig,lipsum,booktabs}

\begin{document}
\title[Translating Behavioral Theory into Technological Interventions]{Translating Behavioral Theory into Technological Interventions: Case Study of an mHealth App to Increase  Self-reporting of Substance-Use Related Data} 

\author{Mashfiqui Rabbi}
\authornote{Mashfiqui Rabbi is the corresponding author}
\affiliation{%
  \institution{Harvard University}
  \streetaddress{1 Oxford Street}
  \city{Cambridge}
  \state{MA}
  \postcode{02134}
  \country{USA}}
\email{mashfiqui.r.s@gmail.com}

\author{Meredith Philyaw-Kotov}
\affiliation{%
  \institution{University of Michigan}
  \city{Ann Arbor, MI}
  \country{USA}
}
\email{mphilyaw@med.umich.edu}

\author{Jinseok Li}
\affiliation{%
  \institution{University of Michigan}
  \city{Ann Arbor, MI}
  \country{USA}
}
\email{jinseok@umich.edu}

\author{Katherine Li}
\affiliation{%
  \institution{University of Michigan}
  \city{Ann Arbor, MI}
  \country{USA}
}
\email{likat@umich.edu}

\author{Bess Rothman}
\affiliation{%
  \institution{University of Michigan}
  \city{Ann Arbor, MI}
  \country{USA}
}
\email{bessar@umich.edu}

\author{Lexa Giragosian}
\affiliation{%
  \institution{Yale University}
  \city{Ann Arbor, MI}
  \country{USA}
}
\email{lgiragos@med.umich.edu}

\author{Maya Reyes}
\affiliation{%
  \institution{University of Michigan}
  \city{Ann Arbor, MI}
  \country{USA}
}
\email{mayarey@umich.edu}

\author{Hannah Gadway}
\affiliation{%
  \institution{University of Michigan}
  \city{Ann Arbor, MI}
  \country{USA}
}
\email{gadwayh@med.umich.edu}

\author{Rebecca Cunningham}
\affiliation{%
  \institution{University of Michigan}
  \city{Ann Arbor, MI}
  \country{USA}
}
\email{stroh@med.umich.edu}

\author{Erin Bonar}
\affiliation{%
  \institution{University of Michigan}
  \city{Ann Arbor, MI}
  \country{USA}
}
\email{erinbona@med.umich.edu}

\author{Inbal Nahum-Shani}
\affiliation{%
  \institution{University of Michigan}
  \city{Ann Arbor, MI}
  \country{USA}
}
\email{inbal@umich.edu}

\author{Maureen Walton}
\affiliation{%
  \institution{University of Michigan}
  \city{Ann Arbor, MI}
  \country{USA}
}
\email{waltonma@med.umich.edu}

\author{Susan Murphy}
\affiliation{%
  \institution{Harvard University}
  \city{Cambridge, MA}
  \country{USA}
}
\email{samurphy@fas.harvard.edu}

\author{Predrag Klasnja}
\affiliation{%
  \institution{University of Michigan}
  \city{Ann Arbor, MI}
  \country{USA}
}
\email{klasnja@umich.edu}

\renewcommand{\shortauthors}{Rabbi et al.}


\begin{abstract}
 Mobile health (mHealth) applications are a powerful medium for providing behavioral interventions, and systematic reviews  suggest that theory-based interventions are more effective. However, how exactly  theoretical concepts should be translated into features of technological interventions is often not clear. There is a gulf between the abstract nature of psychological theory and the concreteness of the designs needed to build health technologies. In this paper, we use SARA, a mobile app we developed to support substance-use research among adolescents and young adults, as a case study of a process of translating behavioral theory into mHealth intervention design. SARA was designed to increase adherence to daily self-report in longitudinal epidemiological studies. To achieve this goal, we implemented a number of constructs from the operant conditioning theory. We describe our design process and discuss how we operationalized theoretical constructs in the light of design constraints, user feedback, and empirical data from four formative studies.
\end{abstract}
\maketitle


%
%


\section{\textbf{Introduction}}
%
%
%

HCI research on health often focuses on the development of novel technologies for health behavior change~\cite{hekler2013mind,Rabbi:2015:MAP:2750858.2805840,consolvo2008activity,lane2011bewell,lee2014real,rabbi2018feasibility}. Such technologies use various \textit{intervention strategies}---goal-setting, feedback, rewards, etc.---to try to actuate \textit{mechanisms of change} (e.g., habit formation, operant learning) that can lead to desired changes in behavior. Commonly, both the intervention strategies that these technologies implement and the mechanisms they target are drawn from behavioral theories. These theories---social cognitive theory, theory of planned behavior, operant conditioning, etc.---are supported by extensive empirical evidence, and many have been successfully used by behavioral scientists to guide behavior-change interventions for decades~\cite{glanz2008health,redding2000health,rothman2000toward}. HCI researchers often follow a similar path, adopting widely used  theories to design their technological interventions. Since these theories are supported by extensive evidence base, theory-based interventions enable designers to maximize the likelihood of successfully influencing behavior change~\cite{Rabbi:2015:MAP:2750858.2805840,consolvo2008activity,consolvo2009theory}.\\

\noindent

\noindent
Despite the wide use of theory in intervention development, its translation into features of a concrete technological intervention is rarely straightforward~\cite{kok2004intervention}. One challenge is that theoretical constructs---i.e., the basic determinants postulated by a theory to influence behavior~\cite{hekler2013mind,michie2014abc}---are formulated at an abstract level that does not afford straightforward implementation. Take, for instance, a seemingly simple case of the construct ``goal.'' There is extensive evidence that concrete goals work better than abstract goals (e.g., ``do your best''), that challenging but doable goals work better than easy goals, and that goals work best when an individual feels ownership over them~\cite{locke2006new,gulotta2016fostering}. Yet, to use this evidence for intervention design---say, a mobile app to support physical activity---requires a great deal of specification. The designer has to decide how to make the goal concrete by specifying what units to use (steps, minutes of moderate-to-vigorous physical activity, number of exercise sessions?), what time frame (day, week, open-ended?), and what kinds of activities will count toward the goal (all physical activity, only sessions longer than 10 minutes, only activities above a heart-rate threshold?). Defining a ``challenging goal'' is still more complicated: Should the designer use national guidelines? Some percentage increase over baseline activity level (if so, what percentage)? Just let the user set what he/she thinks the right challenging goal is? And so on. Even for well specified constructs, implementation requires a myriad of decisions.\\

\noindent
Another challenge arises from design constraints. Coherence of the user experience, development resources, requirements of target population, cultural norms, intended duration of use, among other factors, all come into play when a new intervention is being designed. Such considerations can create design tensions and limit how a particular theory can be implemented. Finally, user reactions and feedback can override even the most careful theoretical and design thinking. If users are turned off by a feature, keeping it risks poor adoption or even abandonment.\\  

\noindent
Designing theoretically-based interventions thus has to be an intricate balancing act where theoretical constructs are iteratively concretized and operationalized in the light of constraints and user feedback. How that process proceeds is rarely made explicit. What is usually presented in papers is the final artifact, accompanied by a list of theories or constructs that the intervention embodies. What was involved in translating those theories into individual features and how those features ended up in the form they did typically remains unstated even in papers that report on the design process in some detail. Yet, theory translation is an essential aspect of health technology design and we need robust methods for doing it effectively.\\ 

%
%
%
\noindent
The primary methodological contribution of this paper is a process for theory translation during technology development. We describe this process through a case study---accompanied by empirical data from several formative studies---of the theory translation process during the design of a new technology. We describe the development of SARA (\textbf{\underline{S}}ubstance \textbf{\underline{A}}buse \textbf{\underline{R}}esearch \textbf{\underline{A}}ssistant), an app for collecting self-report data in epidemiological studies on substance use among adolescents (ages 14-17) and emerging adults (ages 18-24). We discuss how we used concepts from operant conditioning theory  ~\cite{skinner2011behaviorism,reynolds1975primer,staddon2003operant,ferster1957schedules}  to design a variety of incentives that aim to increase self-report adherence over time while minimizing the need for financial compensation. As we will see, even though operant conditioning has a robust evidence base and its constructs are well specified, the translation process was far from straightforward. We had to undertake an extensive user-centered design process to iteratively build numerous design elements that conformed to a large number of practical constraints. 
By presenting our experience with SARA, we hope we can make explicit the general issues that are involved in theory translation and begin formulating a more systematic method for translating theoretical constructs into high-fidelity technological interventions.\\

%
%
\noindent
 Our second contribution is the novel SARA app itself. SARA tackles the problem of low self-report adherence in mHealth~\cite{flurry2015,dorsey2017use} in the context of a uniquely new population for technical interventions---adolescents and young adults at high risk of substance abuse. The resulting self-report adherence rates from a 30-day SARA deployment are encouraging, and the rates are similar to prior substance use epidemiological studies that paid nearly seven times more money to participants to collect data (details in section 3.10)~\cite{bonar2018feasibility}.\\

\noindent

\vspace{-0.3cm}

%
%

\section{\textbf{Background and motivations}}
\subsection{\textbf{What we mean by theory translation}}
Every behavior change theory postulates a set of constructs and mechanisms by which the constructs interact to change behavior. For example, the goal-setting theory defines 14 constructs and  four mechanisms~\cite{locke1994goal,michie2014abc}. Two of these 14 constructs are `directive function' and `goal specificity', and these two constructs interact by the following mechanism: a specific goal acts as a directive function that takes attention away from irrelevant stimuli and behavior and by doing so increases the frequency of target behavior. This is just one example of the relationship of theory, constructs and mechanisms, and there are many more. For a fuller definition of theory, constructs, and mechanisms, see Hekler et al.~\cite{hekler2013mind} and Michie et al.~\cite{michie2014abc}. When designers use a theory like goal-setting as the basis for one or more components of a technological intervention, they implement constructs from that theory as concrete design elements---say, a graph that shows goal progress,  interface through which users set their weekly activity goals, or the algorithm the system uses to calculate the goal levels suggested to users~\cite{hekler2013mind}\footnote{Note that it is rare for an intervention to implement all constructs from a theory. Most interventions, both technological and non-technological, typically implement only a subset of constructs from a single theory and will often mix and match constructs from multiple theories~\cite{hekler2013mind,rabbi2018feasibility,adams2014towards}}. The issue is that while the constructs being implemented are general and abstract, the design elements that embody the interventions have to be made concrete. We define the process that designers go through to bridge this gap as the theory translation process.

\noindent
\subsection{\textbf{Theory translation  in HCI literature}}
\noindent
Theory translation happens every time designers incorporate an element in their technology that is based on a theoretical construct. {\color{blue} 
Theory  translation is a central step of several intervention development frameworks. For example, the Multiple Optimization Strategy (MOST) framework includes a ``preparation'' phase~\cite{collins2018optimization} during which researchers construct theory-based intervention components, and the Agile approach~\cite{hekler2016agile} includes a ``behavior change module development'' phase, where the goal is to build small units of intervention based on theory, expert knowledge, and secondary analysis of data.\footnote{
Note, other intervention optimization frameworks (e.g., micro-randomized trials~\cite{klasnja2015microrandomized},Just-in-time interventions~\cite{nahum2014just}) also provide stage-by-stage processes for intervention development. However, these models and MOST largely focus on optimization trials, where the goal is to gather data that can help the researcher find an optimal combination and sequence of interventions. These methods are quantitative and conducting the trials requires well-developed intervention components. Our paper deals with how to create these well-developed intervention components. Klasnja et al.~\cite{klasnja2017toward} give a process for defining a ``proximal outcome'' of an already developed intervention. However, the Klasnja et al. does not address how to build an intervention stage-by-stage.} Theory is a key source of intervention components in all these framework, and is one commonly used to develop HCI interventions as well.} One would expect, then, that best practices for theory translation process are well understood in HCI, and that designers have principled methods to engage in this process efficiently and effectively. This is not the case, however. {\color{blue}In our review of the HCI literature on behavior change, we found that intentional theory translation in HCI is not the norm, and that even when it does occur, it is often done without the benefit of a systematic and well-described process}.\\


\noindent
To perform the review, we searched for the term ``behavior change'' in the ACM Digital Library. The search returned 591 papers from CHI, Ubicomp, and CSCW conferences. We read the abstracts of these papers, and we excluded papers that did not contain interventions. We also excluded work-in-progress papers because the design process is likely incomplete for the interventions described in those papers.
This filtering reduced the number of papers to 71. The resulting set included papers that described behavior change interventions in several domains, including health, environmental sustainability, and internet use, among others (see Supplementary file 2 for a complete list). We then read each of the 71 papers and we coded them for how they used theory to support intervention design.\footnote{A limitation of our review is we only reviewed HCI literature. Theory translation has been done in behavioral science, but they often focus on organizational aspects of human-to-human interaction (e.g., doctors, stake-holders, policymakers). HCI on the other hand focus on technical interventions and interactions between humans and computers.}\\  

\noindent  
Five categories of theory use emerged from the review: \textbf{(\underline{i})} 43\% (31/71) of the papers did not explicitly mention theory or constructs at all. The interventions described in those papers either have no theoretical basis or they use theory implicitly, providing no information about what theories were used or how. \textbf{(\underline{ii})} 7\% (5/71) of the papers used theory to explain data. In other words, these papers used theory post facto to make sense of the findings, rather than to design the interventions.  \textbf{(\underline{iii})} 17\% (12/71) of the papers referred to theories (usually multiple) as being the basis for the intervention, but provided no description of how those theories were used for intervention design. Theories and the system are presented separately, and these papers provide no explicit description of which aspects of the system are based on which theoretical constructs and how. \textbf{(\underline{iv})} 20\% (14/71) of the papers used theory to inform design and are explicit about which technology features were derived from which construct. However, these papers provide no information about the translation process---i.e., how and why the features ended up in the form that they did. As such, this category also provides little help for understanding how the translation process takes place.  \textbf{(\underline{v})} Finally, a small fraction of papers (12\%, 9/71) described their theory translation processes in at least some detail. Examining these papers revealed no standardized ways of approaching theory translation, nor a presence of design techniques to facilitate this process. Furthermore, none of the papers in this category discuss if or how they ensured that the final design is a faithful representation of the underlying constructs. For example, some of the papers started with a theoretical construct and implemented something different than that construct, making it unclear what theory is tested in the end. {\color{blue}  In summary our review reveals that the HCI literature on behavior change has so far largely ignored theory translation as a methodological aspect of the design of behavior change technologies.}

\subsection{\textbf{Theory translation in behavioral science}}
\label{sec:ttp-general}
Of course, we are not the first to think about the intricacies of applying theory to intervention design. Literature in behavioral science deals with many of these issues, although usually in non-technological settings. Several models  have been proposed, including Intervention Mapping~\cite{bartholomew1998intervention,kok2004intervention}, Behavior Change Wheel~\cite{michie2011behaviour,michie2013behavior,michie2014behavior}, and the logic model~\cite{julian1997utilization}), that provide guidance on how to apply theory to develop interventions. {\color{blue} 
In intervention mapping (IM), design a theory-based intervention includes several phases. IM starts with a need assessment phase where focus groups, online surveys, expert consensus gathering, etc. are held to identify what, if anything, needs to be changed and for whom. This need assessment phase is intended to develop objectives that the intervention has to achieve. Needs assessment is followed by a literature search phase, which aims to identify theories that can help the intervention achieve these objectives. Then the process moves to a phase where the intervention is implemented. This is done by using chosen theories to specify ``strategies''---operations that the intervention will use to bring about change---and developing intervention materials that implement these strategies. However, IM provides little detail about how this move from strategies to intervention materials takes place, other than to note that the population and resource constraints need to be taken into account when intervention materials are developed.\\

\noindent
Logic model is typically used for the theory selection purpose~\cite{den2019electronic}. Logic model asks designers to clearly and thoroughly specify the theories/constructs and the mechanisms through which these theories/constructs will influence various intended outcomes. Logic model's highly detailed description brings clarity to communicating the intervention and creating an action plan to implement the intervention.
Logic model has been extensively used for non-technical intervention, and it is beginning to be used to design technical interventions as well~\cite{edwards2018creating}. The process to derive the logic model can greatly differ across studies. Some studies used focus groups and online surveys~\cite{geraghty2016developing}, while others used literature review and expert feedback to create the logic model~\cite{cole2019understanding,den2019electronic,theofanopoulou2019smart}. In addition, sometimes a logic model is used not to develop an intervention but to describe the final intervention~\cite{edwards2018creating}. While it is extremely useful for clarifying the causal pathway through which an intervention is supposed to operate, logic model has on occasion been criticized for being too stringent or rigid~\cite{greenhalgh2010evaluations}. For the purposes of this discussion, however, the main weakness of the logic model is that it too does not specify how exactly constructs in the model should be concretize into specific intervention features.\\ 

\noindent
The Behavior Change Wheel model~\cite{michie2011behaviour,michie2013behavior,michie2014behavior} is another model that provides steps to guide the creation of theory-based interventions.} 
At a high level, Behavior Change Wheel model proposes a three step process: \textbf{(\underline{i})} identify target behaviors in detail. This is done via the COM-B model~\cite{michie2011behaviour} that helps intervention designers work through the target population's capabilities, motivations, and opportunities for change; \textbf{(\underline{ii})} once target behaviors are specified, use the behavior change wheel framework to find appropriate intervention categories (e.g., education, persuasion, coercion) for impacting those target behaviors~\cite{michie2011behaviour}; and \textbf{(\underline{iii})} once intervention categories are chosen, use the taxonomy of behavior change techniques (BCT) to specify more granular intervention strategies (e.g., provide situation-specific rewards or punishment, facilitate goal-setting) that are appropriate for targeting chosen behaviors~\cite{michie2013behavior}. The least abstract entity in this model are the behavior change techniques. BCTs are theoretically-derived operations---such as ``facilitate planning,'' ``monitor behavior,'' ``monitor goal progress,'' etc.---that specify the functionality of intervention components that serve a theoretically-motivated purpose, such as fostering the reactivity of self-monitoring or increasing individuals' self-efficacy~\cite{michie2013behavior}. Yet, BCTs are still highly abstract; they are consensus-based distillations of the behavior change literature that attempt to systematize what exactly different interventions \emph{do}. BCTs are defined at a level of generality that allows for a single BCT to be implemented in many different ways.
As such, BCTs are types of strategies for facilitating change, but these strategies still need to be made concrete in any given intervention. Michie et al.'s model tries to account for this by asking intervention designers to specify the ``mode'' for each intervention component (delivered in person, via a text message, etc.). This level of specificity is still far lower than what is needed to effectively design concrete features of a technology for a mobile app or a wearable activity tracker. Beyond specifying the mode of delivery, the Behavior Change Wheel assumes that the specifics of implementation will be guided by clinical judgment, and thus it provides little additional guidance on how these specifics should be determined.

\subsection{\textbf{Why we need better theory translation methods}}
\label{sec:ttp-general}
Since theory translation is something designers routinely do, one may wonder if there is really a need for better methods for this process. Certainly, one way to read the results of our review (section 2.2) is that designers get along just fine without such methods. We believe there are several reasons why such methods are important for increasing the impact of HCI work on health behavior change. First, better methods for theory translation can improve how we both describe our interventions and report findings, which can strengthen the \textit{replicability} of our research~\cite{michie2013behavior}. Replicability is the cornerstone of the scientific process, and it is especially important for the health sciences~\cite{dallery2013single,shneidermana2016dangers}. Second, better theory translation methods can lead to interventions with higher \textit{theoretical fidelity}---i.e., how faithfully the intervention operationalizes theoretical constructs or mechanisms it is trying to implement~\cite{Rovniak,raedeke2017high}. Fidelity is important for both designers and behavioral scientists. For designers interested in building systems, a higher fidelity implementation has a higher likelihood to be effective at supporting behavior change~\cite{raedeke2017high}. For behavioral scientists interested in theory development, a higher fidelity implementation increases the confidence that results from an evaluation of a technological intervention can be interpreted as evidence for the operation of the hypothesized theoretical mechanisms. Thus, high-fidelity implementations of theoretical constructs greatly increase the utility of findings from evaluations of behavior change technologies. Finally, standardizing theory translation can \textit{speed up scientific progress}~\cite{michie2013behavior}. This is because careful theory translation compels researchers and designers to specify exactly what constructs they are implementing. Well-specified constructs and intervention components facilitate comparisons of results across studies, as well as synthetic analyses like systematic reviews and meta-analyses that can characterize the effectiveness of different types of interventions for different conditions and populations~\cite{michie2013behavior}. In summary, robust theory translation methods could greatly enhance the ability of HCI researchers and designers to develop technologies that not only work well but also contribute to the larger scientific evidence base about what works, for whom, and in what contexts.\\

\noindent
The current paper takes the first step toward the development of a methodological framework for translating constructs and mechanisms from behavioral theory into concrete features of technological interventions. We do this by providing a detailed account of how we attempted this process in SARA, our mHealth application for self-reporting substance use behaviors, in order to surface the complexities of theory translation and the kinds of issues that designers have to grapple with in order to design a usable technology which also aims to implement a set of theoretical constructs with fidelity. Our hope is that as more such accounts are published, we can begin to synthesize a method for efficiently engaging in this important aspect of the design process. 

\vspace{-0.0cm}
\section{\textbf{Iterative design of SARA: A detailed case study of translating theory into technological interventions}}
Following sections describe the iterative design of the SARA app. We demonstrate how we handled the intricacies of balancing theoretical considerations with project constraints and user feedback to create an app deployable in a 30-day clinical trial. To make our theory translation process more transparent, we first describe theoretical considerations and then the design choices we made based on these considerations. Before we go into the details of the design process, however, we describe the final design of the SARA app to provide context for the later discussion.

\begin{figure}[b] 
    \centering 
    \includegraphics[width=\textwidth]{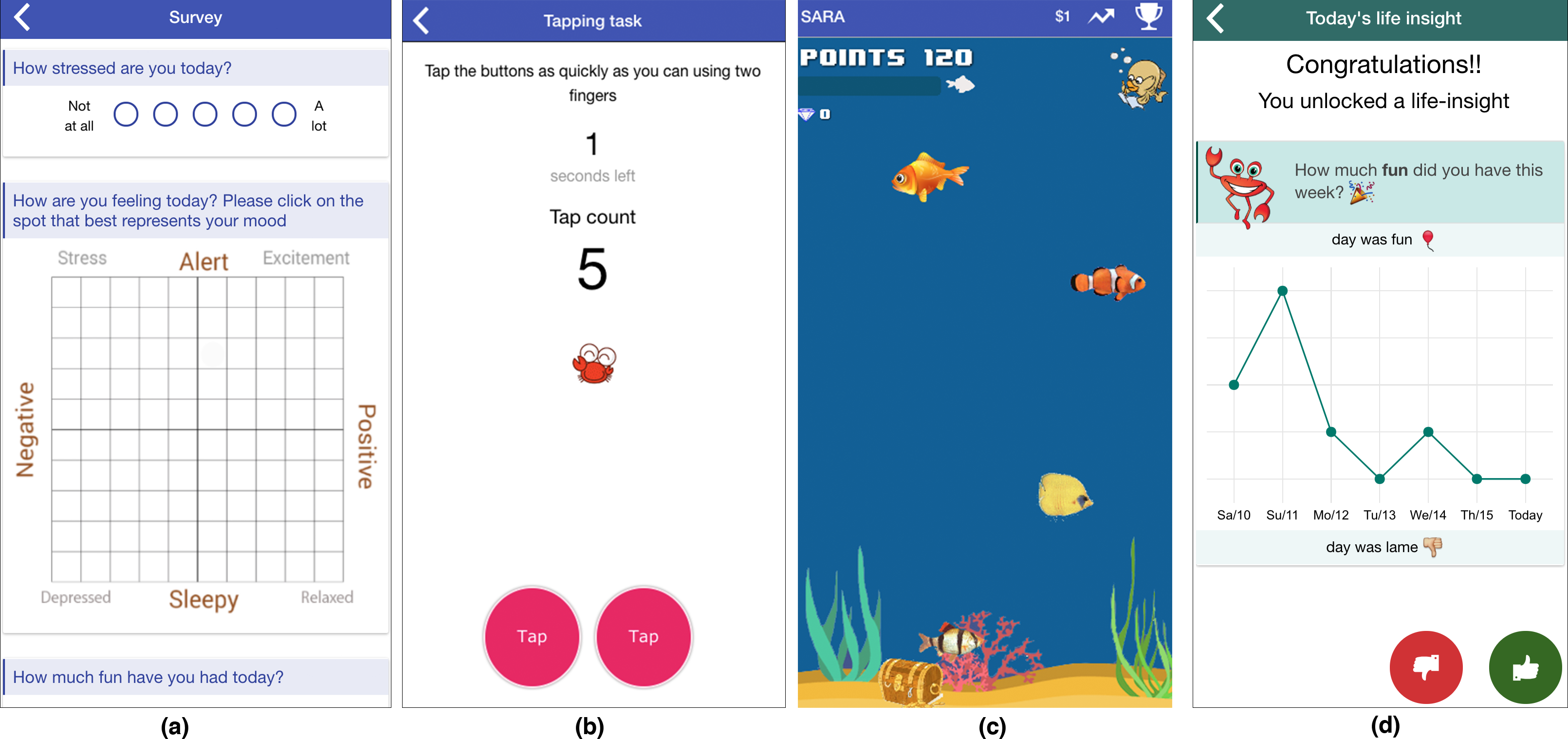}
    \caption{Screenshots of SARA application. (a) daily survey, (b) tapping task (c) virtual aquarium (d) visualization of past data}
    \vspace{-1em}
    \label{fig:overview}
\end{figure}

\vspace{-0.1cm}
\subsection{\textbf{An overview of the SARA application}}
\label{sec:sara-overview}
SARA (\textbf{\underline{S}}ubstance \textbf{\underline{A}}buse \textbf{\underline{R}}esearch \textbf{\underline{A}}ssistant) was created to support observational, epidemiological studies on \underline{\textbf{a}}dolescents and  \underline{\textbf{y}}oung \underline{\textbf{a}}dults (AYA) who are at high risk of substance abuse. We are interested in this problem because substance use is a public health issue and AYA are at high risk. A recent report shows that significant portion of AYA used substances in the last month~\cite{hedden2015behavioral}\footnote{In the past month, 6.1\%,7.4\%,7.4\% of adolescents and 
37.7\%,19.6\%,2.8\% of young adults respectively reported  binge drinking (5 or more drinks),   non-medical marijuana use, and misuse of prescription opioids~\cite{hedden2015behavioral}}. 
Consequences of substance use include hindered brain development and lifelong compromised decision making~\cite{gruber2012age,lopez2011altered,schepis2011gender}. However, there is a lack of fine-grained longitudinal data on how AYA use substances~\cite{shrier2013individual,shrier2014real,bonar2018feasibility}. While sensors can capture certain fine-grained behavior (e.g., alcohol use~\cite{bae2018mobile,marques2009field,scram2017}, ketamine~\cite{you2016kediary}), they cannot detect poly-substance use (e.g., marijuana or opioids) and important predictors of substance use (e.g., stress, mood, etc.)~\cite{choe2017semi,sinha2008chronic}. Thus, self-report remains a central way to capture AYA substance use data. However, long-term adherence to self-report is challenging for most mHealth apps (including substance use apps)~\cite{flurry2015,Helander2014,badawy2017texting,majeed2015apps,hoeppner2017there,cordeiro2015barriers}. Self-reporting can be increased by financial incentives~\cite{van2017gamification,lynn2001impact} and/or frequent human support by the research staff~\cite{mohr2011supportive}. But these methods are costly. Virtual rewards are cost-effective alternatives, but most published studies do not change virtual rewards over time~\cite{van2017gamification,werbach2012win,johnson2016gamification}, risking habituation and potentially failing to deal with participants' changing needs~\cite{eysenbach2005law,flurry2015,lazar2015we,epstein2015lived}. Furthermore, published studies do not provide any principled ways to combine virtual rewards with other incentives (e.g., money, data-visualizations). The goal of SARA was to develop a low-cost way to provide the right incentives at the right time to effectively support ongoing self-reporting.\\

\noindent
{\color{blue} The SARA version described in this paper was intended for use in a 30-day study, which  is a typical length for many observational substance use studies~\cite{suffoletto2012text,shrier2018pilot,clark2010project}. This period is also sufficient to examine engagement, as compliance with daily assessments declines over 30 days in samples of substance-using youth~\cite{bonar2018feasibility,buu2017assessment,comulada2015compliance,suffoletto2012text,wen2017compliance}. Thus, focusing on a 30-day study was an efficient way to iterate on our design decisions while keeping the study duration in line with other research in this area.}  Note also that SARA's goal was not to investigate whether self-reporting decreases substance use behaviors through self-regulation or self-reflection~\cite{you2015soberdiary}, which are sometimes referred to as assessment reactivity in the substance use literature. Although self-reflection is well-theorized in HCI, such as to increase exercise, the potential effects of self-reflection are less clear for AYA alcohol or marijuana use. Most substance-use focused EMA studies focus on tobacco cessation~\cite{serre2015ecological}, with few studies showing evidence of reactivity on smoking behavior (potentially due to habituation;~\cite{shiffman2009ecological}), and on samples with greater problem severity (e.g., injection drug users;~\cite{roth2017potential}). As reactivity may be lower among those with lower problem severity~\cite{wray2014using}, such as SARA's target population, we did not expect SARA to function as a therapeutic intervention. Rather, its goal was to support regular self-reporting so that the dynamics and determinants of AYA substance use can be better understood.\\

\noindent
Self-reporting in SARA involves completing one survey and two active tasks each day between 6 PM and midnight (see Fig~\ref{fig:overview}a-b).  The survey asks about emotions (e.g., stress, mood), hopefulness, and reflections about the day (e.g., amount of free time, level of excitement)~\cite{cranford2006procedure,ramirez1998sexual,hoyle2002reliability,lippman2014flourishing}. On Sundays, the survey asks an additional 14 questions about past week substance use (i.e., alcohol, cannabis, tobacco) frequency and motives, perceived risk of regular substance use, impulsivity, and behavioral intentions to avoid substance use in the following week~\cite{johnson2006monitoring,patton1995factor,stephens2002marijuana,simons1998validating,grant2007psychometric}. The two active tasks~\cite{appleactivetask,mariakakis2018drunk} in SARA are: a spatial memory task, in which a random sequence of five seashells light up in a 2-dimensional grid of nine seashells and participants are asked to repeat the sequence; and a reaction time task involving tapping two buttons alternately for 10 seconds. {\color{blue} 
The reaction task measures motor speed and a number of past studies have shown reaction time changes based on alcohol~\cite{celio2014we,nicholson1992predicting} or marijuana use~\cite{peeke1976effects}. Spatial memory task, on the other hand, measures executive function and visuo-spatial memory (i.e., remembering different locations of objects as well as spatial relations between objects), and past research has shown similar tasks can predict alcohol intoxication~\cite{celio2014we} and cannabis use~\cite{nestor2008deficits}.
}\\

\noindent
SARA uses the operant conditioning theory~\cite{skinner2011behaviorism} to provide a slew of virtual and monetary incentives to reinforce self-reporting  (see Fig~\ref{fig:overview}c-d). A virtual aquarium fills with  fish as daily self-reports are completed. Other incentives like visualizations of past data, funny or inspirational content, and small amounts of money are provided periodically to  further reinforce self-reporting. With these various combined incentives, SARA's self-report adherence rate  was similar to that of a prior AYA substance use study that used seven times more money (more details in section 3.10). This figure is especially significant because less than 1\% of AYA substance users are currently under treatment~\cite{abuse20162015} and a low-cost solution like SARA can democratize data collection at scale.\\

\noindent
This encouraging result was achieved largely due to our efforts to translate the operant conditioning theory~\cite{skinner2011behaviorism} in order to optimize the frequency and timing of incentives in SARA. This theory translation process, however, was not straightforward, and we undertook a lengthy iterative design process (five iteration and four studies) to pick and operationalize 16 constructs from the theory. We ended up working on 233 design elements while balancing 17 constraints during our process of theory translation (see Supplementary file 1 for a complete list of design choices, constructs, and constraints). It is to the details of this process that we now turn.

{\color{blue}
\subsection{\textbf{An outline of our agile approach to translate theory into SARA}}

\begin{figure*}[b] 
    \centering 
    \includegraphics[width=\textwidth]{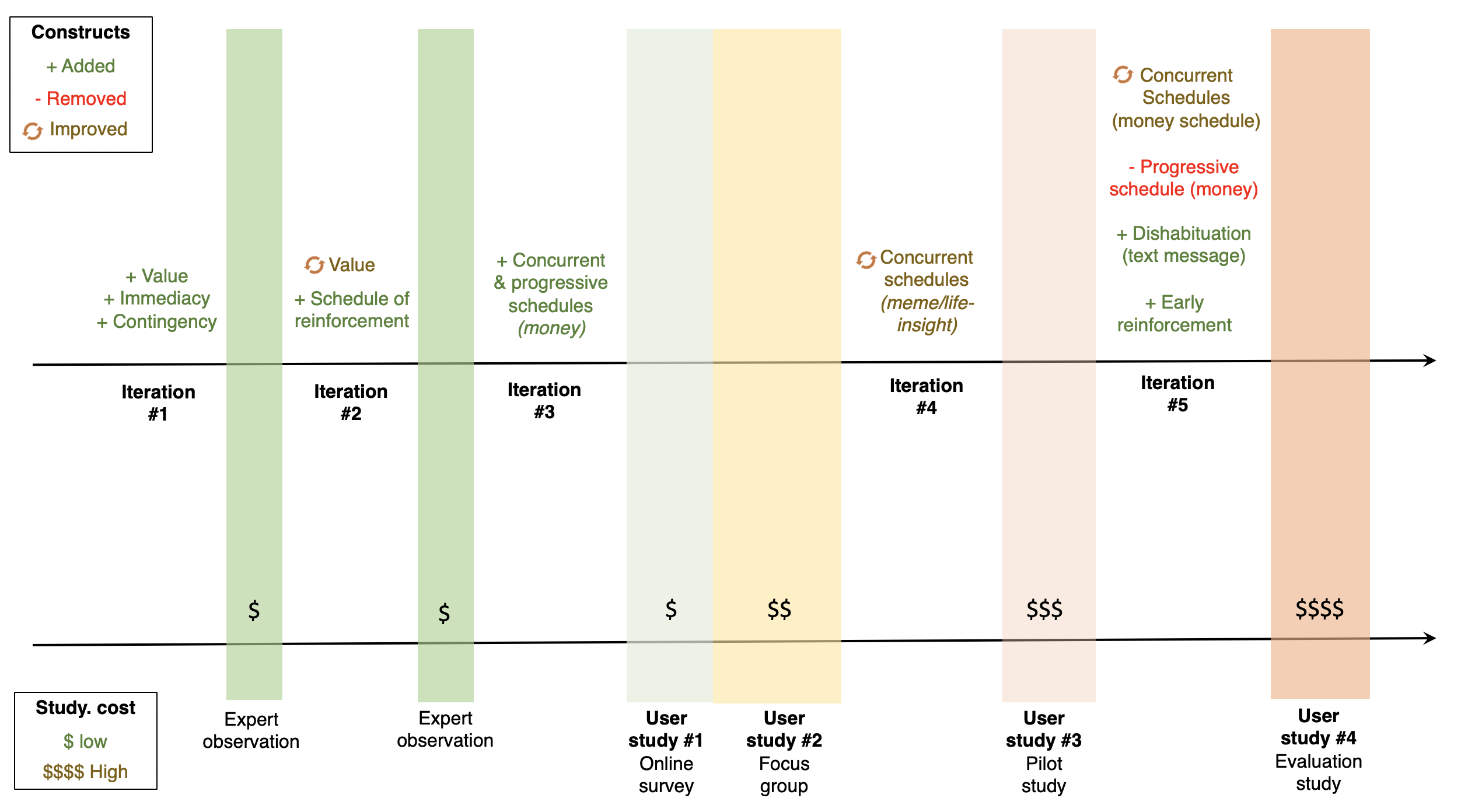}
    \caption{Iterative approach to translate theory in SARA. Left to right is the start and end of the design cycle. As we cycled through several iterations, we added, improved or removed constructs. Also, earlier evaluation strategies, shown in the bottom row, used low-cost methods initially and higher cost methods at later stages of the design process.}
    \vspace{-.2cm}
    \label{fig:iterativeprocess}
\end{figure*}

\noindent
A key question for theory translation is what kind of development process  the designer should use. While one can follow the classic waterfall model, we believe an agile approach is more appropriate for theory translation in HCI~\cite{hekler2016agile}. Below, we describe waterfall and agile models, and why agile models are more applicable for theory translation in the context of HCI.\\

\noindent
The waterfall model, as the name suggests, flows in one direction: after a need finding phase, a collection of features is identified and implemented. Waterfall models are classic software development models, where each development cycle is long and each release is feature rich. Waterfall's philosophy is ``just-in-case'' where a large number of features are implemented to minimize situations where the system cannot support a particular scenario. But, like any large systems, waterfall models are costly and they are less adaptable to change. Agile, on the other hand, starts with a minimal viable product with only a few features, and it then rapidly improves the product by  iterating, evaluating, and adapting to changes. So, agile is more adept at dealing with newly discovered needs. Agile approaches are also more cost-effective.\\

\noindent
Most health intervention development models (e.g., logic models, COM-B), including quite recent work~\cite{vilardaga2018user}, use a form of a waterfall model where theory selection and implementation of constructs is done once, at the start of the design process. While such approaches can be theoretically rigorous, they provide no built-in methods for early discovery of failures of construct operationalization or the need to incorporate additional constructs. Agile approaches are better suited for addressing such issues, especially in the context of HCI work, for several reasons: (i) development resources for novel HCI interventions are typically limited, and agile is more cost-effective, (ii) the need for changes are commonly discovered in HCI studies of behavior change applications: users get habituated or bored, rewards as designed are not found to be sufficiently rewarding by the target group, the application ends up creating more user burden than anticipated, etc. Rapid iterations can efficiently address such findings; and (iii) agile's iterative approach is similar to the iterative design approach in HCI, allowing HCI researchers and designers to align their design and theory-translation work.\\

\noindent
In SARA, we used agile principles in two important ways. First, we added, removed, and refined theories or theoretical constructs iteratively. We did so to address an important challenge of intervention design, choosing what constructs to implement from a large collection of theories/constructs from the behavioral change literature (nearly 83 theories and over 1600 constructs~\cite{michie2014abc}). A concrete implementation of any abstract theory/construct requires financial resources, requiring judicious selection of intervention components, especially in the context of resource-constrained HCI work. To deal with this issue, in SARA we started with a minimal set of constructs. We then made low-fidelity prototypes of these constructs, and used low-cost participant feedback methods to evaluate them. After the initial evaluations, we identified which constructs should be carried forward for higher fidelity implementations, how they should be revised, and which additional constructs should be added to create a better intervention. The top portion of Figure~\ref{fig:iterativeprocess} shows the theory/construct selection process used in SARA.\\

\noindent
The second challenge that an agile approach to theory translation helps address is how to assess theoretical fidelity: i.e., to what degree the implementation represents the underlying theory or construct. By focusing on fast, iterative evaluation, agile provides guidance on how to sequence different kinds of evaluations to check how effectively constructs have been implemented and if there is a need to implement additional constructs. Here, a tradeoff the designer has to make is between more informative user studies (e.g., a pilot study measuring actual use and health behaviors) that are also more costly and require more time, and faster, cheaper, but less rigorous evaluations. Since uncertainty is high in early stages of the design process, high cost methods can be wasteful. In SARA, we initially used low-cost methods like design reflection, online surveys, and focus groups that did not involves actual use of the app. For instance, in the SARA online survey, we asked users to rate perceived benefits of gamification, rewards, etc., without showing the app screenshots. In the focus group, we showed the app to users to get feedback about the current design of different theory-based features, but the participants did not use the app in their day-to-day lives. Only once several design iterations were done based on low-cost feedback, that we moved to more expensive means of validation, such as pilot and evaluation studies. These studies focused on the actual use of the app, and we also gathered qualitative feedback to gain insights for future improvements in our construct operationalization.\\ 

\noindent
A final consideration is sample size. The appropriate sample size depends on multiple factors, including the level of evidence required at each stage of the design process, the effort required by participants and study staff, and the constraints on time and other resources. Table~\ref{table:costuserstudy} shows the trade-offs of different user studies, and the bottom portion of Figure~\ref{fig:iterativeprocess} shows the different user studies we used at different phases of SARA's design.\\

\begin{table*}[h] \centering
\ra{1.2}
\begin{small}
\begin{tabular}{@{}lccccccc@{}}\toprule
\textbf{Evaluation method/} & \textbf{cost/} & \textbf{participant} & \textbf{cost/study} & \textbf{uncertainty in} & \textbf{Actual} & \textbf{Sample} & \textbf{when used in} \\ 
\textbf{user study} & \textbf{participant} & \textbf{burden} & \textbf{staff} & \textbf{design cycle} & \textbf{use} & \textbf{size} & \textbf{design cycle} \\\midrule

\textbf{Desiger's reflection} & none & none & none & high & no & -- & early on\\ \hdashline

\textbf{Online survey} & low & low & low & high & no & $N \ge 100$ & early on\\ \hdashline

\textbf{Focus group} & low & moderate & moderate & high & no & $N=20-30$ & early on\\ \hdashline

\textbf{Pilot study} & high & high & high & moderate & yes & $N=8-20$ & later on\\ \hdashline
\textbf{Evaluation study} & high & high & high & low & yes & $N \ge 30$ & later  on\\ 
\bottomrule
\end{tabular}
\end{small}
\caption{Types and relative trade-off of different user study types}
 \vspace{-2.0em}
\label{table:costuserstudy}
\end{table*}

}

\vspace{-0.1cm}
\subsection{\textbf{Iteration 1: A virtual aquarium}}
\textbf{Theoretical considerations}: The first challenge of developing SARA was to choose a theory that can be useful for supporting self-report adherence. {\color{blue} We looked at the substance use literature for theories to improve self-report adherence. However, as a recent meta-analysis in substance use research suggests, engagement theory is limited to the provision of monetary rewards~\cite{jones2019compliance}. While gamification~\cite{boyle2017pnf} and data-visualization~\cite{you2015soberdiary} have been used in substance use research, they were used to reduce alcohol use behavior and not to improve self-report adherence. Due to this lack of guidance on how to support self-report adherence in substance use research, we decided to look at the theories from psychology.}
While several theories are applicable~\cite{fogg2009behavior,petty1986elaboration}, we chose the \textbf{{O}}perant \textbf{{C}}onditioning \textbf{{T}}heory (OCT) ~\cite{skinner2011behaviorism,reynolds1975primer,staddon2003operant} for two reasons: (i) OCT provides a detailed account of how consequences of prior actions influence the frequency of future behavior. Most importantly for our purposes, OCT describes how positive reinforcement (valued consequences, rewards) can be used to increase the frequency of a target behavior. This directly matches SARA's aims of increasing the frequency of self-report completion; and (ii) OCT is a well-developed theory. First proposed in the 1930s, over the years, OCT has become one of the most precise and well supported theoretical accounts of learning. OCT has been successfully applied to a broad range of problems, from animal training~\cite{staddon2003operant,ferster1957schedules}, to education~\cite{weegar2012comparison,phillips2012behaviorism,shield2000critical,villamarin2010using}, and psychotherapy~\cite{wolpe1968psychotherapy,wolf1963application,burchard1964modification}.\\ 

\noindent
OCT describes a number of constructs that determine how reinforcement affects a target behavior. 
Three of these core constructs are: (i) \underline{value of the reinforcement}: extensive literature on Matching Law~\cite{hodos1961progressive,herrnstein1970law} demonstrates that organisms allocate their behavior in proportion to the perceived value of reinforcement associated with different behavioral choices. An effective way of increasing the frequency of a target behavior is to reinforce it with something the organism finds valuable; (ii) \underline{immediacy}: a reinforcement delivered immediately after a behavior is more effective than a reinforcement that is more temporally distant. The main reason for this is that organisms discount the value of reinforcements based on their temporal distance; i.e., the same reinforcement is perceived to be more valuable if it is received sooner~\cite{simpson2000temporal,ainslie2001breakdown}; and (iii) \underline{contingency}: reinforcement that an organism can clearly associate with a behavior influences that behavior more strongly than a reinforcement that could have resulted from multiple behaviors. A way to ensure contingency is to provide reinforcement only after the desired behavior~\cite{miltenberger2011behavior}.\\

\begin{figure}[htbp] 
    \centering 
    \includegraphics[width=0.88\textwidth]{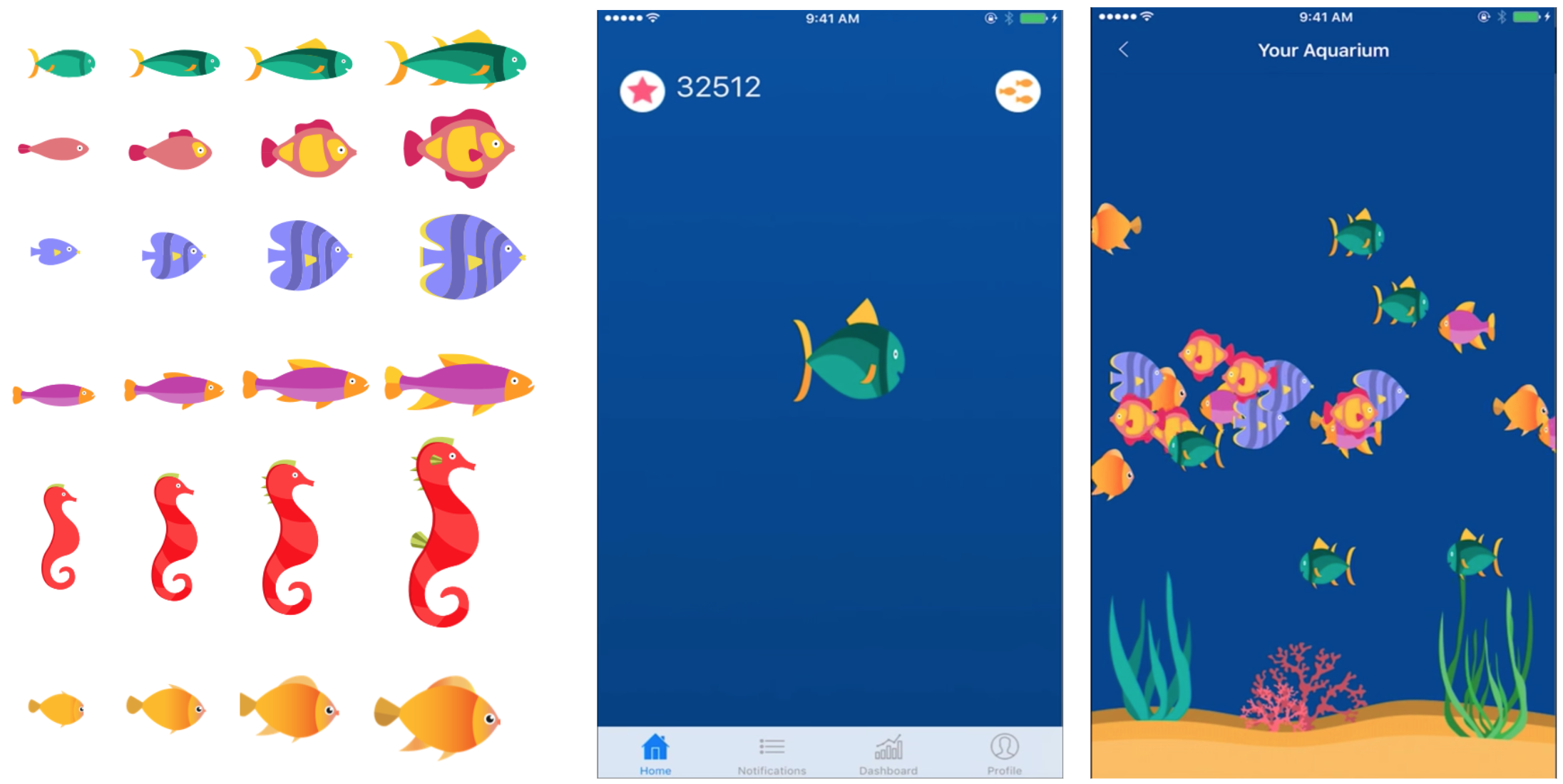}
    \caption{Iteration 1 of SARA. Left image shows the 6 fish and their stages of growth. Middle and right columns show the aquarium.}
    \label{fig:iteration1}
    \vspace{-1.0em}
\end{figure}

\noindent
\textbf{Design considerations}: In designing SARA, we first tried to translate the above-mentioned three OCT constructs: value of reinforcement, immediacy and contingency. Translating immediacy and contingency was straightforward: we could provide the reinforcement immediately and only after self-report completion. Translating the notion of a valuable reinforcement turned out to be more complex, however. While in animal research a strong reinforcement is routinely achieved by using food, the use of such strong reinforcement (food, shelter, personal safety) is ethically unacceptable in most human-subjects research.\footnote{Exceptions occur in special circumstances: e.g., in reinforcement-based therapy, abstinence in drug-dependent patients is reinforced by providing housing, job training, etc.~\cite{tuten2012abstinence}.} In research with people, the most common form of reinforcement is money~\cite{ariely2013upside}, and money has been used effectively to motivate self-report completion~\cite{van2017gamification,lynn2001impact}. However, since SARA's goal is to reduce money in order to make data collection more scalable, we initially tried to focus on non-financial reinforcements.\\

\noindent
A common non-financial approach to reinforcement is gamification~\cite{deterding2012gamification,deterding2011game}, where game-like elements such as points, badges, progression, levels, and leaderboards are provided~\cite{van2017gamification,werbach2012win,johnson2016gamification,ariely2013upside,deterding2012gamification,deterding2011game,zuckerman2014deconstructing}. Several gamification features satisfied our design goals and we included them in SARA (more details below). However, we chose not to  use social features such as leaderboards for two reasons. First, if study participants were recruited over time, as is common in clinical trials, different participants would be potentially exposed to a very different leaderboard based on when they were recruited (i.e., it's not clear that a leaderboard with 3 people and 150 people are the same intervention). Second, social features can have unintended adverse effects among young substance users (e.g., social undermining~\cite{fogg2009behavior} or negative contagion~\cite{joiner1999contagion,prinstein2007moderators}), and planning for and managing these was beyond our financial resources.\\

\noindent
Without a leaderboard, we suspected that points by themselves would not be perceived as particularly valuable. However, since  points were attractive for other reasons---they could scale over time, they were free, and could be adjusted to deferentially reinforce both individual acts of self-report completion and patterns of adherence over time (see below)---we attempted to increase the perceived value of points. To do so, we decided to create a way to convert points into something that participants would find interesting and attractive. After extensive design ideation on different types of representations (e.g., virtual pets, avatars, growing tree, space objects), we settled on the idea of creating a virtual aquarium that would be populated with fish as points accumulate.\\

\noindent
The aquarium representation had a number of attractive features that made it a good candidate for reinforcement. First, unlike many other representations, fish (and an aquarium) were positively received by both men and women, as well as by individuals of varying ages~\cite{lin2006fish,lane2011bewell}. Second, the representational language of the aquarium was quite rich, allowing us to provide a large number of interesting reinforcements and, thus, scale the representation over time. Finally, aquarium representations had already been used successfully in the mHealth setting: Fish'n step~\cite{lin2006fish} and BeWell~\cite{lane2011bewell} used aquariums to promote healthy activity, and Abyssrium, a mobile game that involves growing a fish population, has been downloaded over 30 million times and received a game of the year award in 2016~\cite{abyssriumpress}. As such, we had strong preliminary evidence for the feasibility and acceptability of an aquarium representation. Figure~\ref{fig:iteration1} shows the first prototype of the SARA application. We had six different fish and each fish had four stages of growth. Each time a participant completed a self-report, he or she would earn 200 points and a fish would go through one stage of growth. Once a fish went through four stages of growth, we considered the fish fully grown and added it to the aquarium.\\

\noindent
\textbf{Design reflections}: Our initial prototype translated the concept of reinforcement using points, aquarium, and fish. We also combined points with fish to create what we hoped would be a more valuable reinforcement. These reinforcements were to be provided immediately and only after self-report completion to maximize their effect. {\color{blue} Regarding evaluation of these ideas, we deferred evaluation using online surveys or focus groups at this stage because we were at an early stage of design and the number of features in SARA was small. We wanted to use more costly focus groups and online surveys when we had more features to investigate, so at this stage we decided to only use design reflection and feedback within the research group. During this evaluation, we quickly realized} that in this first iteration we did not consider how our reinforcements should be scheduled so they remained effective at promoting self-report completion over time~\cite{ferster1957schedules}. The next iteration tried to address this issue. 

\vspace{-0.1cm}
\subsection{\textbf{Iteration 2: Reinforcement schedules}}
\textbf{Theoretical considerations}: Extensive research in OCT shows that different schedules, or timing, of reinforcement can produce different behavioral effects~\cite{ferster1957schedules}. Consider the simple case of \textit{satiation}, where a reinforcement is received too often in a short period of time and it temporarily loses efficacy. For instance, satiation can happen when an animal has eaten enough and is no longer hungry, and the availability of additional food temporarily loses the ability to influence the animal. The same phenomenon is found in humans: after binge-watching our favorite TV show, an opportunity to watch more of it may not be perceived as particularly desirable, no matter how much we love the show~\cite{staddon2003operant}. In cases of satiation, temporarily reducing the frequency of reinforcement can give an organism time to re-sensitize to it, and reinforcement then regains its ability to influence behavior.\\

\noindent
Another key idea is that reinforcing behavior intermittently can lead to high levels of responding.
Intermittent reinforcement can follow a fixed or variable-rate schedule. A fixed-rate schedule means that reinforcement is provided each time a fixed number of instances of the target behavior is completed. A variable-rate schedule means reinforcement is given after a variable number of target behaviors, but with the mean number of target behaviors before reinforcement held constant. Reinforcing less often using a variable schedule can generate similar frequencies of target behavior as a fixed schedule that reinforces more often. The variable-rate schedules have this effect due to uncertainty and anticipation~\cite{ferster1957schedules}. However, when a new behavior is learned for the first time, reinforcing more often using a fixed schedule can result in faster initial learning ~\cite{ferster1957schedules}. Furthermore, the perceived value of reinforcement matters. A less valuable reinforcement needs to be used more often---i.e., after fewer occurrences of target behavior---compared to a more valuable reinforcement~\cite{Trosclair-Lasserre}.\\

\noindent
\textbf{Design considerations}: We translated the above mentioned theoretical insights by adopting a fast, fixed-rate schedule, where we provided (almost) one fish for each day of self-reporting. We chose this schedule because (i) self-reporting is a new behavior and reinforcing more often  induces faster learning; and (ii) we suspected fish may not be perceived to be a very valuable reinforcement, so we needed to use them more often to increase self-report.  Lacking evidence for how valuable fish would be perceived to be, we opted to be conservative and assume they would be of low relative value and would require a fast schedule. Now, a potential side-effect of frequent reinforcement is that it can induce satiation. To prevent satiation, we included a gap day after every 4th to 5th fish, much as games use satiation and deprivation sequences to increase engagement~\cite{zichermann2011gamification}. \\

\begin{figure*}[t] 
    \centering 
    \includegraphics[width=\textwidth]{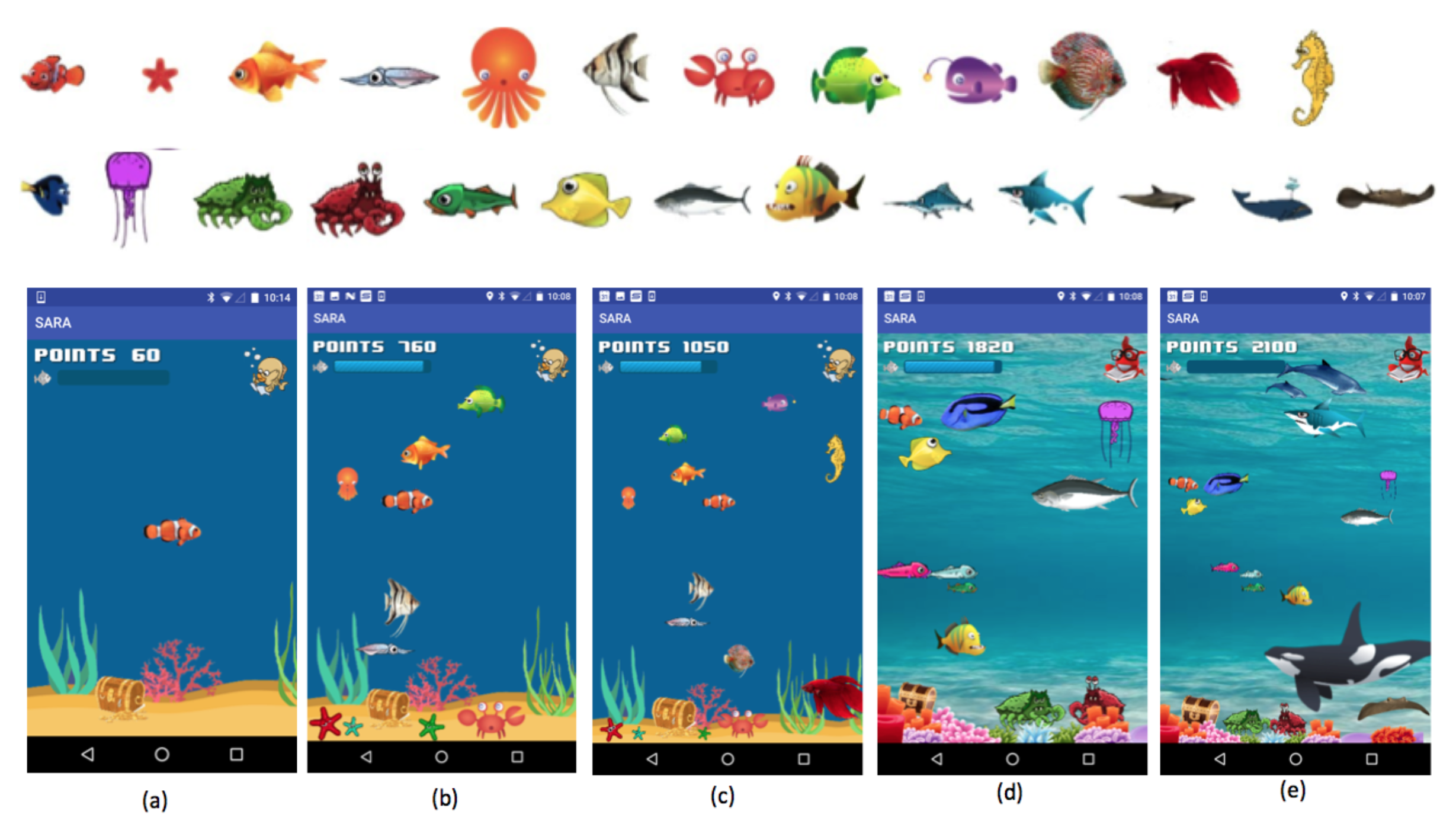}
    \caption{Top two rows show 25 different fish in Iteration 2. Bottom row shows different stages for the growing aquarium. a,b,c,d,e respectively shows the state of aquarium if a participant logs for 1,7,14, 24, 30 days.}
    \vspace{-.2cm}
    \label{fig:iteration2}
\end{figure*}

\noindent
We had to significantly redesign SARA in order to incorporate this fast reinforcement schedule for fish.
First, we retired the six fish from iteration  1 (Figure~\ref{fig:iteration1}) because in order to follow a fast schedule of one-fish-a-day, we would have needed to start reusing previously given fish after the first six days. Recycling old fish would mean that the  reinforcement would become less novel and hence potentially less valuable. We replaced the six fish from iteration  1 with 25 unique fish (see Figure~\ref{fig:iteration2}) so that a different fish could be awarded for almost each day of self-report completion in a 30-day study. \\

\noindent
In addition to reinforcement schedules, in this iteration we also attempted to increase the value of fish in several ways: (i) we used animated and better looking fish, because better aesthetics can improve reward value~\cite{perski2016conceptualising}; (ii) we displayed a fun fact about each fish when it was unlocked.  For example, when the goldfish was unlocked, we showed a fun fact ``Do you know goldfish can recognize faces?" The aim of the fun facts was to generate curiosity and thus increase the reward value of the fish; and (iii) we provided an outline for the next fish to be unlocked to increase anticipation and feeling of progression. Finally, a potential challenge for the one-fish-a-day schedule was that the aquarium could get over-populated. Therefore, we introduced levels as is commonly done in games~\cite{zichermann2011gamification}. For the 30-day planned study for SARA, we created two levels: participants began with a fish bowl environment level and unlocked fish, such as goldfish, which are commonly seen in household aquariums. After 15 days of self-reporting, participants graduated to a sea environment and unlocked sea creatures, such as sharks and dolphins. Figure~\ref{fig:iteration2} shows iteration 2 of SARA's design and a progression of the aquarium over a 30-day study.\\

\noindent
\textbf{Design reflections}: The major improvement in this iteration, from  a theoretical perspective, was the explicit inclusion of a reinforcement schedule. However, due to our lack of knowledge about how valuable the fish would be perceived to be, we selected a fast, fixed-rate reinforcement schedule. But the use of a fast schedule also created problems. Even for a relatively short 30-day study, this schedule necessitated the inclusion of a lot more fish, as well as a way to deal with aquarium overcrowding. The later issue, in particular, is a good example of how theoretical concepts must be considered in light of design constraints, in this case limitations related to the number of fish that could be comfortably shown on a mobile phone screen.\\

\noindent
{\color{blue} For evaluation of the design ideas in Iteration 2, we brainstormed whether we were missing any obvious features because adding those features would give us an opportunity to ask about them in an online surveys and focus groups. So, here again design evaluation was done via  design reflection and feedback within the research group. During the design reflection we realized that} even with our best efforts to increase the reward value of fish and include reinforcement schedules, we would need additional forms of reinforcements due to the burden of ongoing self-reporting. In mHealth, the use of multiple reinforcement strategies is quite common. Nearly all commercial mHealth apps and health-gamification research projects use two or more kinds of reinforcements (e.g., stars,  badges, etc.) to promote health behaviors~\cite{fitbit,johnson2016gamification}. In addition, inclusion of additional reinforcements would also allow us to introduce other types of reinforcement schedules, which could more effectively target sustainability of self-reporting behavior. Thus, for the next design iteration, we focused on introducing additional forms of reinforcement.

\noindent 
\vspace{-0.15cm}
\subsection{\textbf{Iteration 3: Additional reinforcements and conjoint schedules}}
\textbf{Theoretical considerations}:  For guidelines on how to schedule multiple types of reinforcement, we again turned to OCT. We found that OCT supports the idea that multiple types of reinforcement can be more valuable together than a single reinforcement. One piece of evidence for this came from Hursh~\cite{hursh1980economic}, who proposed a microeconomic conceptualization of OCT. Hursh posits that demand for reinforcement is created after a target behavior happens; the obtainable reinforcement is the supply for meeting that demand. Hursh argues that multiple reinforcements are additive and increase the overall supply. 
Another, less obvious part of Hursh's argument is that multiple types of reinforcement can create a \textit{substitution effect}: if one type of reinforcement is ineffective but other types of reinforcement are available, they can substitute for it and still maintain the effectiveness of the overall reinforcement supply. For instance, at the start, the fish may not be seen as particularly rewarding since the aquarium is mostly empty and participants are learning its value to them. Giving some money, which most people care about, may reduce the demand for the fish to be immediately highly reinforcing. Usefulness of multiple types of reinforcement was further supported by the OCT literature on conjoint reinforcement schedules.This literature suggests that conjoint schedules generally increase target behavior more than single schedules, unless the  rate of reinforcement is so slow that the amount of reinforcement from single and conjoint schedules cannot be distinguished~\cite{weatherly1996within}. \\

\noindent
\textbf{Design considerations}: As with the initial design iteration, a key design question was what to use for this new form of reinforcement and how to schedule it. Given that providing some form of financial incentive is common in most studies, we decided to use small financial reinforcements that would not compromise scaling up SARA for larger studies. However, we had to make the financial reinforcements coherent with the overall gamified theme of the SARA app. We decided to tie monetary reinforcements to badges: if participants completed several back-to-back days of self-report, SARA would award them badges that came with small monetary incentives. Specifically, if participants completed daily surveys three days in a row, or they completed active tasks three days in a row, they received a badge and 25 cents. For longer streaks, participants could earn different badges and higher monetary incentives. SARA could reward 3, 6, 12, 18, and 30-day streaks; for these streaks, participants could earn 0.25, 0.50, 1, 2, and 3 dollars, respectively. Under this schedule, if we assume 90\% adherence, then there was less than 5\% chance that participants would earn more than 10 dollars in a 30-day study. This amount is lower than most substance use studies that generally pay \$1-4 per day of study participation~\cite{bonar2018feasibility,buu2017assessment}. We hoped that this conjoint schedule of being able to receive money every 3 days and fish nearly every day (from iteration 2) would be fast enough so that its effect would be greater than a single schedule of either money or fish~\cite{weatherly1996within}. \\

\noindent
\textbf{Design reflections}: The primary challenge of this design iteration was how to structure the schedule of financial reinforcements. We decided to provide more money for longer streaks; this progressive monetary schedule was intended to encourage longer stretches of daily self-reporting. At the same time, we wanted to keep the application scalable, so we opted for the smallest amounts of money that we thought might still be effective in reinforcing self-reporting in the AYA population.\\

\noindent 
Up to this point, the app design was based on our best attempt to translate several principles from OCT in a way that took into account various constraints: target population, the app's scalability etc. Along the way, we made a number of assumptions and design choices that we thought would help us to effectively translate OCT. When we got to this stage of design, {\color{blue}we felt that, given the considerable number of theory-driven features we had developed, we needed to bring in user feedback to check our assumptions and design choices. To do this, we conducted our first two formative studies--an online survey and a focus group study.}


\noindent
\subsection*{\textbf{Formative user feedback}}
We conducted two formative studies to gather early developmental input on SARA. The first study involved an online survey and the second study was a focus group. Both studies targeted undergraduate students of University of Anonymous who were within our target age group. {\color{blue} Since we were still early in the design cycle, we did not opt to use higher-cost methods, such as a study focusing on actual use. Furthermore, since online surveys are lower-participant-burden and less expensive than focus groups, we decided to do an online survey first. Given the low cost of conducting a survey, we aimed to recruit at least 100 participants, to increase the confidence in any potential findings. We further reduced evaluation costs by using the same online survey to recruit focus group participants as we describe below.}

\vspace{-0.15cm}
\subsection{\textbf{User study 1: Online survey (N=124)}}
The online survey intended to gather information on perceptions of different reinforcement types from a sample of our target population. After asking a set of questions about demographics and prior experience with mHealth apps, the survey asked the  participants to rate the likelihood of how money, progression, and unlockable features would motivate them to self-report regularly. We initially also intended to ask questions about risky substance use. However, since many responders could be under the age of 21, the IRB needed special approval to include substance-use-related questions. Since binge drinking and marijuana use is common in college students,\footnote{Among undergraduates at University of A, 53.6\% reported binge drinking and 34.4\% reported marijuana use in past 3 months [--]} we decided to drop these questions.\\

\noindent  
\textbf{\textit{Results}}: We used the university registrar to send the survey to a random sample of 2000 undergraduate students. Of the 280 students who started the survey, 124 (38\% male) answered the questions regarding different types of reinforcement. Mean age of these participants was 19.9 (SD=2.59). There were no significant differences in demographics between those starting the survey and those completing the survey. A 4-point Likert scale was used to rate the anticipated impact of reinforcements on self-report in mHealth apps. The scale ranged from 1=not likely to 4=very likely. As we can see in Figure~\ref{fig:ratings}a-\ref{fig:ratings}c, money was rated as the most likely reinforcement to influence self-reporting $(\mu=3.72,\sigma=0.53)$, followed by gamification features such as points $(\mu=2.57,\sigma=0.98)$ and unlockable features $(\mu=2.73,\sigma=0.93)$. The high ratings of points, unlockable features, and money suggested that our target population might perceive these reinforcements as valuable, providing preliminary evidence for our efforts to operationalize reinforcement.

\subsection{\textbf{User study 2: Focus groups (N=21)}}
The second formative study focused on getting in-depth qualitative feedback on SARA's design. We invited 21 participants (47\% male) from the 124 respondents of the online survey for focus groups. {\color{blue} During the recruitment, we balanced gender, ethnicity, and age. We held three one-hour-long, semi-structured, mixed-gender focus group sessions $(N=7,5,9)$. All sessions were audio recorded. Our initial goal was to recruit 30 participants as is often done in focus group studies~\cite{schueller2018discovery,metting2018assessing}. Given the early stage of the design process, however, we decided to prioritize resource efficiency. So, we kept the number of sessions to three and recruited 21 participants}. During each focus group session, after a few ice-breaking questions, participants received phones with the SARA app installed. Moderators then demonstrated how to self-report in SARA and showed the different reinforcements of SARA (aquarium, levels, money, etc.) using a storyboard. We created temporary buttons which could be pressed to simulate daily self-reports and give a feel for how the aquarium evolved and money was rewarded. After participants played with the SARA app for some time, they answered questions about which features they liked or disliked and what additional features would make SARA feel more rewarding. After the focus group, participants completed a debriefing survey on how likely they thought money, aquarium, progression, and fish would be to affect self-reporting. Since the debriefing survey was anonymous, participants also reported past 3-month alcohol and marijuana use.\footnote{61.9\% and 28.5\% focus group participants  reported they binge drank and used marijuana, respectively, in the past 3 months.} Participants received 20 dollars for participation.\\

\noindent 
\textbf{\textit{Results}}: We analyzed the focus group discussions using thematic analysis~\cite{braun2006using}. In the following, we present a summary of the results that are relevant to translating OCT.\\

\begin{figure}[b] 
    \centering 
    \includegraphics[width=0.78\textwidth]{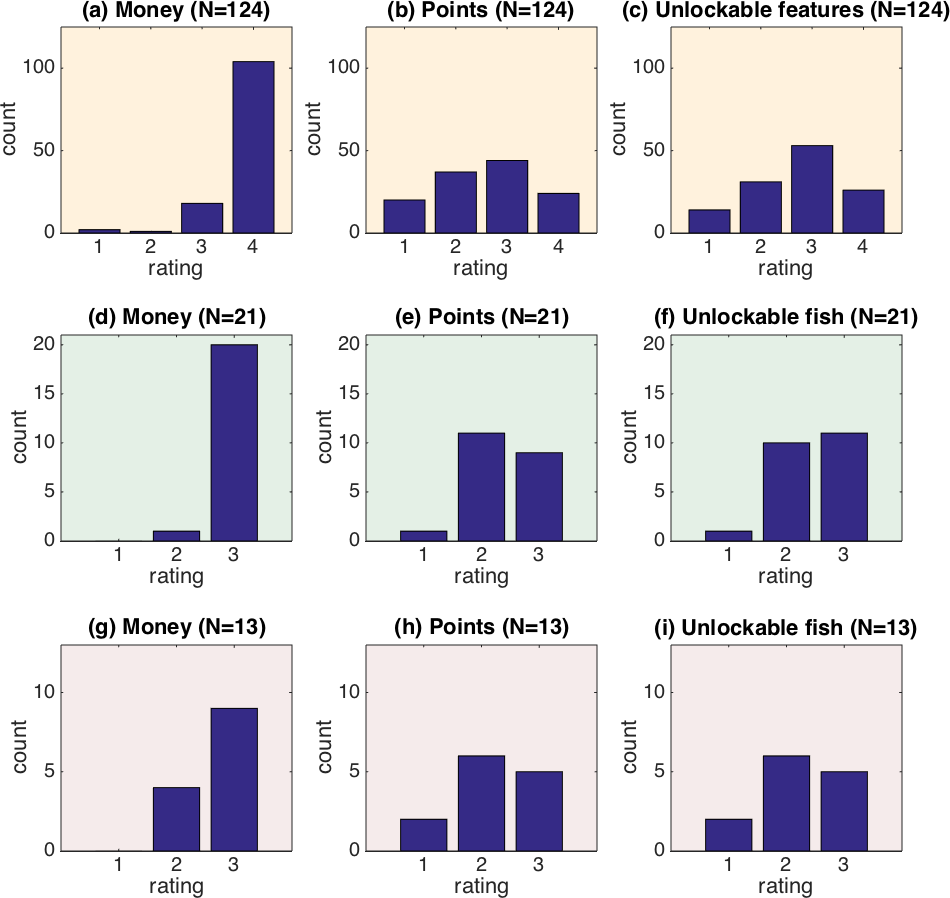}
    \caption{Distribution of ratings for various reinforcement types. Top row is for online survey (User study 1) with a 4-point Likert scale (1=not at all, 4=very much). Middle and bottom rows are for debriefing survey of focus group (User study 2) and follow-up survey after a 30-day pilot trial (User study 3), which use 3-point Likert scales (1=not at all, 3=very much)}. 
    \vspace{-.2cm}
    \label{fig:ratings}
\end{figure}

\noindent  
\underline{\textit{Fish and aquarium}}: When asked about the fish and aquarium, participants appreciated the aquarium theme and characterized the aquarium as relaxing and calming.  Participants gave mixed responses when asked how they would like the aquarium to be improved. Some participants mentioned that the app's appearance was not as polished as other commercial apps; some participants wanted more consistent animations of fish. As one participant stated, \textit{``I guess it's mildly off-putting to me that the fish are all kind of different kinds of animation, some look more like clip art, some look more like actual fish kind of a deal.''} Other participants were more accepting of the fish animations; one participant stated, \textit{``you could make them goldfish, like literally just pure gold, or diamond fish, or platinum fish. [chuckle] Because people aren't gonna care. It's an app.''}. Participants also liked when the aquarium changed levels to the sea environment. For example, as one participant stated,\textit{``I think it's a good concept. I think it'd be really cool... you graduate from this aquarium and go to the next.''} We also provided an outline of the next unlockable fish and several participants mentioned the outline created anticipation. Like the survey, these observations indicated that participants thought they would find the aquarium and the fish to be rewarding, providing support for  our efforts to design non-financial reinforcements AYA found valuable.\\

\noindent 
Participants also appreciated how the aquarium progressed over time. One participant mentioned that \textit{``it's really good that there's an objective you can work towards and you can see what's happening each day.''} However, participants found the aquarium more appealing when it was  full. As one participant stated, \textit{``If you don't have that many [fish] then it's kind of a bland thing to look at. What I'm looking at right now, I only have one fish. If I have something like that [fuller aquarium], it's more pleasing to the eye.''}. These comments provided initial evidence for our decision to use a fast, fixed-ratio schedule, but indicated that the fish schedule may not be sufficiently fast.\\

\noindent  
\underline{\textit{Money and streaks}}: When we asked about money, participants unanimously agreed that money was a huge pull, particularly in their age group. As one participant noted,  \textit{``I think the monetary reward is one of the highest things I think for our age group, that should definitely be pretty clear and specified at the beginning. That is probably one of the main things that'll keep people coming back.''} As expected, the focus groups supported the idea to use money for reinforcement. Furthermore, participants provided no indication they thought that money would conflict with the aquarium, providing preliminary support for our decision to pair these forms of reinforcement (Iteration 3). However, one participant pointed out that the streaks may be too hard to complete and that, once interrupted, there was no way to recover a lost streak. Given how monetary incentives were structured, he said, losing a streak, especially a longer more valuable one, would feel like punishment. Although only one participant commented on this issue, it raised the question of whether the schedule of monetary reinforcements would behave as we intended it to, encouraging regular, uninterrupted self-reporting.\\

\noindent  
\underline{\textit{Additional novelty}}: A majority of participants wanted more novelty. Participants wanted additional themes such as gardens, jungles or car collections. One participants said \textit{``It would be cool to be able to unlock other little worlds... Maybe a little forest or something? Or a garden, you could do butterflies or something. Everybody likes butterflies... Maybe for people who like bugs or something, you can do a bug garden if that's even a thing...''}  When the moderators mentioned the possibility of having visualizations of personal data, participants were enthusiastic about the idea: \textit{``It'd be cool if there were like a reinforcements page and then also data but kind of... Like how the Apple health app gives you graphs and stuff like that. I think that would be kinda cool... Put data like mood, maybe as a graph for energy level.''} We interpreted these observations as indicating that we had not over-saturated SARA with reinforcements and that AYAs thought there was still a need for additional forms of reinforcement.\\

\noindent
\underline{\textit{Choice and interactivity}}: When asked about what features they would like to change, participants wanted more control over the aquarium and to be able to interact with the fish. One participant wanted to change the background color of the aquarium. Several participants wanted to feed the fish and for the fish to do something interesting (e.g., tell a fun fact) when they touched the fish. Participants also wanted to have control over their points and be able to use their points to get the fish they preferred or restore a broken streak of money. They also wanted to choose at what level they would start. These findings raised two interesting possibilities for operationalizing reinforcement: first, giving choice meant people could choose the reinforcements they found valuable, providing a way to deal with the heterogeneity in people's preference. And second, participants' comments indicated that interactivity might increase their ownership over the aquarium, which, in turn, would and make it feel more valuable~\cite{sundar2007social}. We considered these lessons in future design iterations.\\

\noindent
\textit{\textbf{Debriefing survey}}: Following the focus group, participants answered a debriefing survey, where they reported how they thought different reinforcements would increase a participant's use of the app. All the ratings were recorded on a three point Likert scale: 1=not at all, 2=somewhat, 3=very much. Figures~\ref{fig:ratings}d-\ref{fig:ratings}f show the results. Money $(\mu=2.95,\sigma=0.22)$ was again rated the highest, followed by unlockable fish $(\mu=2.45,\sigma=0.59)$, and points $(\mu=2.38,\sigma=0.58)$.\\

\noindent
\textbf{Formative study summary and design reflections}: The two formative studies provided initial positive evidence for how we translated OCT constructs. Participants generally liked the representations we developed (aquarium, fish, etc.), as well as the financial incentives, which indicated that these reinforcements held promise for reinforcing self-reporting. We also learned that providing choice and interactivity were additional ways of increasing reinforcement value. Furthermore, participants liked having several reinforcement types, which provided support for Hursh's reinforcement supply idea~\cite{hursh1980economic} and indicated that we might be able to effectively combine different forms of reinforcement using a conjoint schedule. Participants appreciated the fast schedule for fish, providing support for the use of such a schedule at the start of the study, in line with theory suggestions. However, we got an indication that the schedule of financial incentives might be problematic since larger monetary reinforcements were tied to long uninterrupted streaks which participants felt would be difficult to achieve. Given, though, that focus participants were just trying to imagine how they would experience the monetary schedule, it was difficult to tell how it would perform in a deployment.

\vspace{-0.11cm}
\subsection{\textbf{Iteration 4: Memes and life insights}}
\textbf{Design constraints and theoretical considerations}: While our formative studies provided us with a number of insights about how to improve the perceived value of reinforcement in SARA, we could not implement every request due to resource constraints. In particular, both substantial improvements in the app's aesthetics and implementation of reinforcement choice and interactivity required additional designers and developers, which were beyond our level of development resources. Adding novelty, however, was easier to implement because we could simply add more reinforcement types without extensive development.\\

\noindent 
\textbf{Design considerations}: In this design iteration, we introduced two new types of reinforcements to create additional novelty. The first kind was what we called \textit{life insights}; life insights are visualizations of past tracked data. We decided to include life insights because focus-group participants wanted to see their past data and prior mHealth work found that seeing patterns in one's own data was intrinsically motivating and could encourage regular self-tracking~\cite{vorauer2006information,van2009making,nahum2014supervisor,leonardelli2010new,weary1997causal}. We created seven different life insights that visualized past week's (i) stress, (ii) loneliness, (iii) level of fun, (iv) how new and exciting their days were, (v) free hours each day, (vi) tap count, and (vii) the number of seconds required to complete the spatial task. The data for life insights were pulled from the daily surveys (i-v) and active tasks (vi-vii). \\

\noindent
The second kind of reinforcement we introduced were memes. We included memes because they are widely available on the internet and are popular among SARA's target age group. Memes can also make people laugh and evoke positive emotions, which are powerful intrinsic rewards~\cite{perski2016conceptualising,miller2009toward}. To enhance the novelty of this reinforcer, we incorporated two types of memes: funny and inspirational. 120 memes were compiled and filtered by Amazon MTurk workers and undergraduate research assistants to increase their relevance to AYA.\\

\noindent
A question that remained was how to schedule memes and life insights. We decided to use a variable schedule, where (i) a meme would be provided with 0.5 probability if a daily survey was completed and (ii) a life insight would be provided with 0.5 probability if the two active tasks were completed. We decided on a variable schedule because we did not want to satiate participants with too frequent reinforcement. In addition, using variable schedules for these reinforcements let us micro-randomize their delivery to empirically decide on the best schedule in a post-study analysis~\cite{klasnja2015microrandomized,boruvka2017assessing}.

\begin{figure}[h] 
    \centering 
    \includegraphics[width=0.98\textwidth]{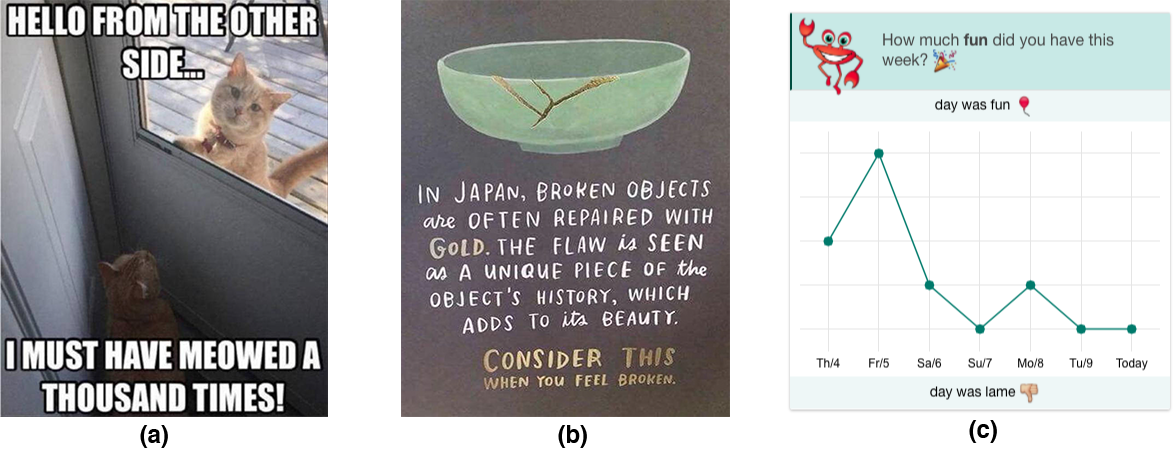}
    \caption{Examples of funny memes (left), inspirational memes (middle) and a life-insight (right).}
    \vspace{-.2cm}
    \label{fig:memes}
\end{figure}

\vspace{-0.1cm}
\subsection{\textbf{User study 3: Pilot trial (N=13)}}
At this point, we pushed SARA's design as far as we could without empirical data on how participants would experience and respond to the various components of SARA. To make further changes, we needed to understand how the reinforcements we designed would affect self-reporting and how their effects would change over time. {\color{blue} Furthermore, while we could have run focus groups to test the features we added in iteration 3, we thought that the number of new features from iteration 3 was too small to warrant investing in another set of focus groups. Instead, we opted to move directly to} a 30-day pilot study. {\color{blue} Since further changes may have been  needed based on how people respond to SARA, we first did a pilot study before conducting a larger scale evaluation study. The sample size of such pilot studies  are typically 8-20 people~\cite{consolvo2008activity,lane2011bewell,burns2011harnessing}. We recruited 16 participants, but three initial participants dropped out due to a software bug that was unrelated to the theory-based incentives in SARA. So, the final sample size was 13.}
\\

\noindent
Pilot study participants were recruited from the University of Anonymous Hospital Pediatric and Adult Emergency Department. Patients were eligible (i) if they were between the ages of 14-24,  understood English, medically stable, able to provide informed consent/assent (e.g., not cognitively  impaired/intoxicated), and accompanied by a parent/guardian (for participants between the ages of 14-17), (ii) screened positive for binge drinking (>4 drinks female, >5 drinks male;~\cite{bush1998audit}) in the past month, or any past-month cannabis use without a medical marijuana card. Then a research assistant installed SARA  on the participant's phone and demonstrated how to use the app. After the 30-day study, participants completed a 45-60 minute telephone interview, where we asked a sequence of close and open-ended questions about their experience with SARA~\cite{hoeppner2010good,stoyanov2015mobile}.\\

\noindent     
\textbf{\textit{Results}}: Recruiters approached 241 individuals out of which 26 participants met study inclusion criteria and 17 were enrolled in the study. The average age of participants was 21.2 years $(\sigma=1.9$, range=18-24, 60\% male). 45\% screened in for past-month binge drinking only, 35\% for past-month marijuana use only, and the rest for using both substances. Four participants dropped out of the study due to software bugs or malfunctioning phones. We  excluded these participants from the following analyses since their adherence was not related to the reinforcements in SARA.\\

\noindent
\underline{\textit{Adherence}}: Figure~\ref{fig:adherencepilot}a shows participants' adherence over the 30-day study. Participants counted as being adherent if they answered the daily survey and active tasks for the day. We found adherence to decrease with time: adherence was 63.8\% for days 1 to 10, 49.2\% for days 11 to 20, and 34.6\% for days 21 to 30. We also identified 3 clusters of participants: \textbf{(\underline{i})} 5 out of 13 participants, represented as green in Figure~\ref{fig:adherencepilot}a, were engaged for the entirety of the study and self-reported for more than 20 days (mean=25.4 days); \textbf{(\underline{ii})} 4 participants, represented as blue in Figure~\ref{fig:adherencepilot}a, self-reported on more than 10 days (mean=12.5 days). Two of these participants provided no data past the 15th day of study participation, while the other two provied no data after the 23rd day in the study; and \textbf{(\underline{iii})} 4 participants, represented as red in Figure~\ref{fig:adherencepilot}a, self-reported on fewer than 10 days (mean=3.75 days), and only one of them provided any data past the 13th day in the study. While SARA clearly did not work for this last group of participants, the presence of both the second and the third group indicated the need for further design refinements to reduce these types of non-adherence.\\

\noindent
\underline{\textit{Monetary incentives}}: On average, participants earned \$4.60 $(\sigma=5.5, q_{50}=2.2)$. Most participants failed to complete longer streaks and earned less money as a result: the total number of 3, 6, 12, 18, and 30 daily streaks completed was 22, 10, 6, 3, and 1, respectively.\footnote{Note that when a longer streak was completed, we did not count its shorter constituent streaks. For instance, for a 12 day streak, we did not count the 3 or 6-day streaks that were completed on the way to the 12-day streak.} The low number of long self-reporting streaks suggested, in line with the focus-group participant's intuitions, that our initial schedule of financial reinforcements was ineffective for encouraging long, uninterrupted periods of self-reporting.\\

\noindent
\underline{\textit{Between self-report distance}}: We also measured the gap between two successive self-reports. If our implementation of schedules of reinforcement in Iteration 2 worked as intended, then we should see regular self-reporting and the intra-day gap between self-reports would be small. Figure~\ref{fig:adherencepilot}b shows the distribution of day gaps between two successive self-reports; these numbers are for participants prior to finishing the 30-day study or prior to their complete disengagement (i.e., before they stopped using the app completely). The gap was one day in 78.2\% of cases,  two days in 13.9\% of cases and three days in 3.91\% of cases.  Of the remaining 3.91\% of the cases where the gaps were 4 days or longer, 71.4\% of the gaps were for people who replied fewer than 10 times in total during the study (i.e., the red participants in Figure~\ref{fig:adherencepilot} top). For participants in blue and green clusters, the gaps between responses were low, which indicates that SARA's reinforcement schedules likely influenced regular self-reporting as OCT suggests, at least to some extent.\\

\begin{figure}[h] 
    \centering 
    \includegraphics[width=1.00\textwidth]{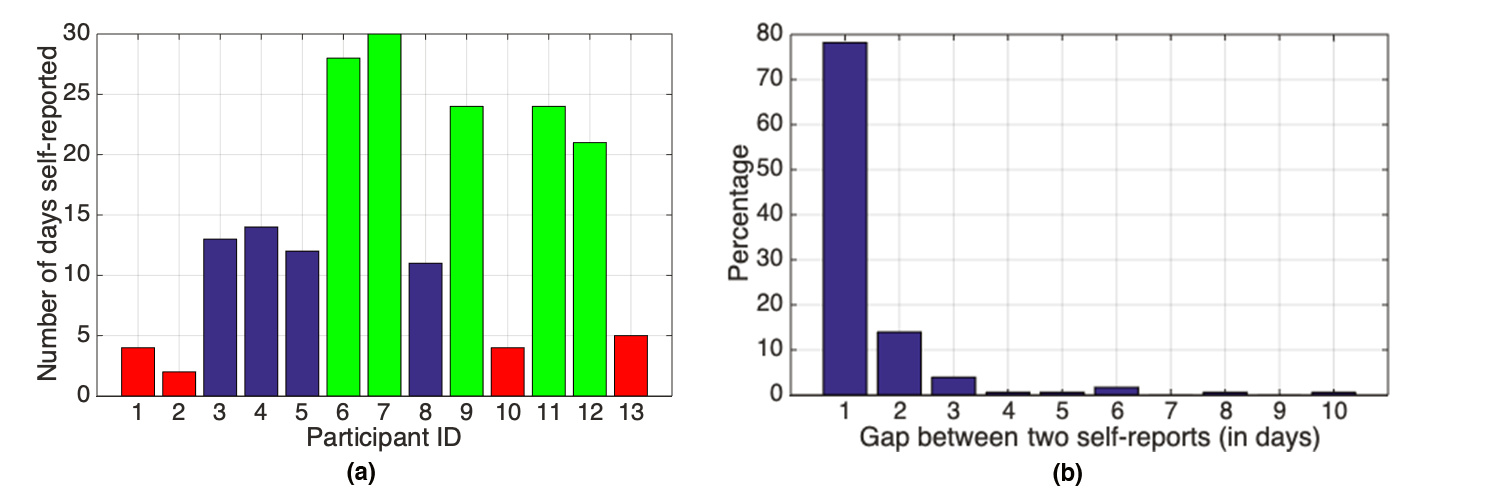}
    \caption{(a) Number of days participants self-reported in a 30-day study. The colors represent three clusters. The green, blue and red color participants self-reported for >20, 11-19, and <10 days, respectively. (b) Distribution of the gap between two consecutive self-report days.}
    \label{fig:adherencepilot}
\end{figure}

\noindent
\underline{\textit{Follow-up survey}}: After the study, we asked participants to rate how they thought various reinforcements influenced their app use over the 30 days. For the new reinforcements from Iteration 4, memes and life insights, participants rated their perceived influence on a five point scale, from 1=not at all to 5=very much. Life insights  $(\mu=3.92,\sigma=0.8)$ and  memes $(\mu=3.8,\sigma=0.5)$ were rated similarly. These results indicated that that the delivery of memes and life insights was perceived as being rewarding after they were experienced in use, whether or not they actually changed participants' behavior. For other reinforcements, we used the same three point Likert scale we used in the debriefing survey after the focus groups, where 1=not at all, 2=somewhat, and 3=very much. Figure~\ref{fig:ratings}g-\ref{fig:ratings}i show the results. Participants rated money most highly $(\mu=2.7,\sigma=0.49)$, followed by points $(\mu=2.4,\sigma=0.9)$, and the fish $(\mu=2.2,\sigma=0.7)$. Note that these ratings are on average lower than those obtained from the focus group debriefing survey, where participants did not use SARA in their daily lives (see Figure~\ref{fig:ratings} for a visual comparison). There are several possible reasons for these lower ratings after participants used SARA for 30 days: (i) they may indicate that the reinforcements lost their value over time (e.g., due to habituation or wearing off of novelty); or (ii) focus group participants could not accurately predict how they would experience the reinforcements and the burden of self-reporting over the long-term, and thus they misjudged how valuable they would find SARA's reinforcements in the future.\\

\noindent 
\underline{\textit{The Follow-up interview}}: After the 30-day study, we  conducted semi-structured telephone interviews with participants. Below, we discuss the interview themes related to reinforcement design.\\

\noindent 
\underline{Aquarium and fish}: Participants' reactions to the aquarium  mirrored those of the focus group participants. One participant noted: \textit{``I liked the fish a lot. I thought they were very cute, and I liked when the app moved from the fish bowl to the ocean.''}  However, similarly to focus group participants, they also indicated a need for better aesthetics and more interactivity.\\

\noindent
\underline{Financial reinforcements}: While participants liked receiving money for self-reporting, they did not like financial reinforcements in small fractions: \textit{``Bigger rewards would be more exciting... don't have \$0.25 rewards and instead have \$1.00 rewards every once in a while.''} One participant also suggested that monetary reinforcements be ramped up as participants reported more data in the study. \textit{``I would prefer that the app gradually gave me more money as I took more surveys and active tasks. So at the beginning, I would only receive a small amount and gradually receive more.''} Note that the schedule of financial reinforcements actually increased incentives for longer streaks, but most participants did not see this since streaks were broken even if they missed a single day.\\

\noindent   
\underline{Memes and life insights}: Some participants liked  the memes, indicating that they were funny, but others wanted memes to be personalized. One participant said \textit{``Maybe you could let people choose what type of memes they want to see... or add a like/don't like button for the memes (like you have for the inspirational messages) and then if someone doesn't like one type of memes, you could push a different type of meme.''} Life insights were generally liked. One participant said: \textit{``I like the tracking and life insights on a daily basis. For people my age, [it is] totally helpful. [It is] awesome.''} However, participants also wanted life insights that combined different types of data. Overall, the memes and life insights were generally well liked, suggesting they may be effective as reinforcers for AYA.\\

\noindent
\underline{Habituation}: Participants mentioned that fish became ``trivial'' over time. Some also mentioned that self-reporting became repetitive and they wanted more variety in active tasks or more information about why certain information was being collected. These observations indicate that participants might have habituated to the reinforcements, and that self-reporting was increasingly seen as boring. To improve the experience, participants suggested to add a variety of new active tasks and add educational content on why the data were being collected.

\subsection{\textbf{Iteration 5: Final improvements before the trial}}
\textbf{Design constraints and theoretical considerations}: While the overall results were positive, our preliminary studies pointed to several shortcomings of our operationalizations of concepts from OCT. Iteration 5 tried to address some of these shortcomings. The first shortcoming was related to the schedule for financial reinforcements. Our initial schedule was progressive, where more money was paid for longer streaks. Progressive schedules work well for highly valued reinforcements~\cite{hodos1961progressive}, but our findings indicated they were not well suited for the small amounts we were paying. Second, for participants who disengaged early in the study, the available reinforcement might have been insufficient to overcome self-reporting burden. OCT suggests fast reinforcement schedules at the start because more reinforcement early on can speed up learning~\cite{ferster1957schedules}. The amount of reinforcement we provided early on might not have been high enough for these participants. Finally, the interview data suggested that participants became habituated to the fish and other reinforcements (habituation is the decrease in efficacy of a reinforcement after repeated exposure~\cite{thompson2015habituation,Zang2007,rankin2009habituation}). OCT suggests two ways to deal with habituation: (i) allow for ``break time'' to resensitize to the existing reinforcements; and (ii) increase reinforcement  variety. Given that we were under time pressure to run a trial, introducing ``break time'' was the more feasible way to proceed.

\noindent
\begin{figure}[htbp] 
    \centering 
    \includegraphics[width=\textwidth]{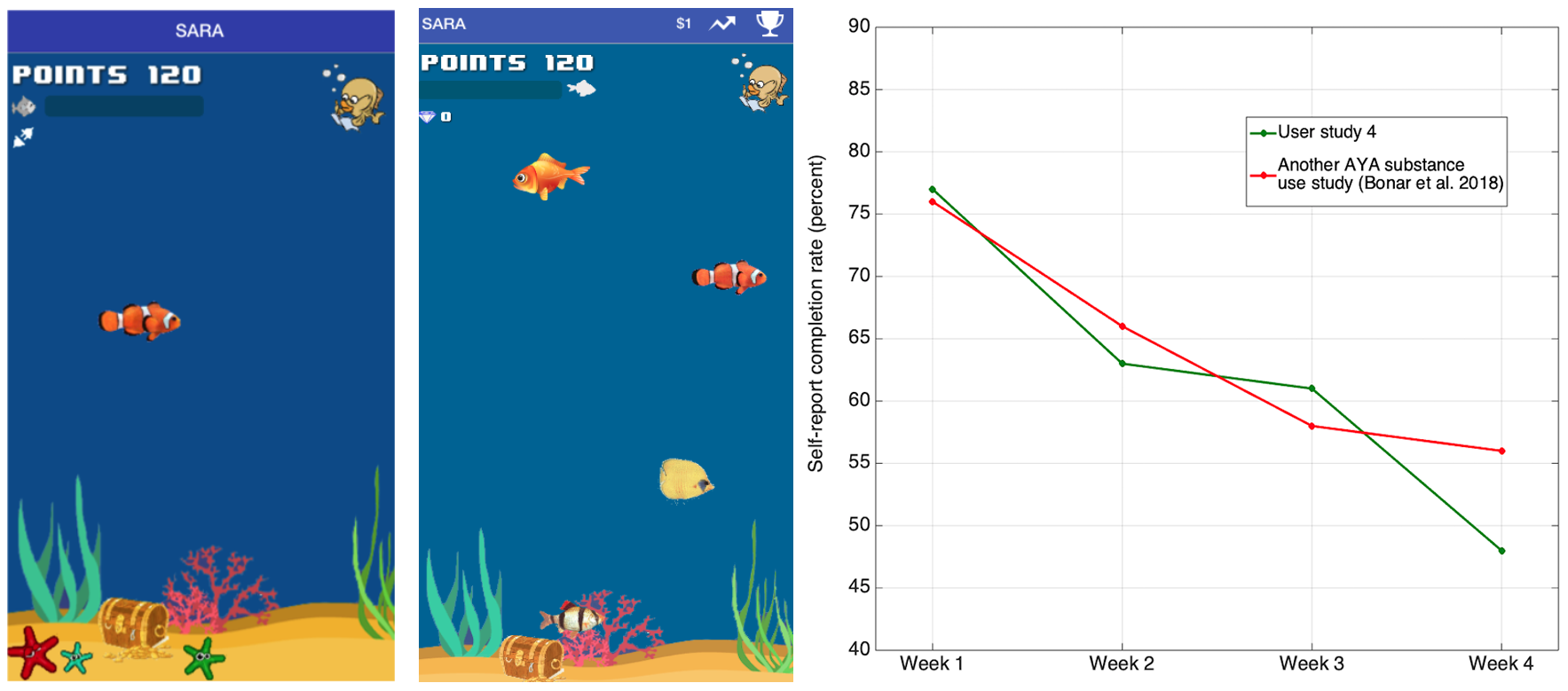}
    \caption{Left and middle are aquarium screenshots after 2-days of data collection for iteration 2 and iteration 5 respectively. Right screenshot shows adherence rates of User study 4 in comparison with Bonar et al.~\cite{bonar2018feasibility}}
    \vspace{-.2em}
    \label{fig:iteration5}
\end{figure}

\noindent
\textbf{Design considerations}: In iteration 5, we made three modifications to our reinforcement design. The first modification was to use a simpler schedule for monetary reinforcements where participants earned \$1 whenever they completed a 3-day streak of self-reporting. This change addressed the difficulties of achieving longer streaks and participants' preference for reinforcements in whole dollar amounts. Note that the new schedule did not substantially increase projected participant earnings: 90\% adherence would lead, on average, to the earning of only \$12 over a 30-day study. The second modification attempted to increase reinforcement early in the study. To do this, we front-loaded reinforcements at the start of the study: (i) we awarded participants \$1 after they self-reported the first day; and (ii) in the first two days, we  awarded two fish for each self-report, enabling participants to earn four fish in two days (see Figure~\ref{fig:iteration5}). The third modification meant to address habituation. 
We implemented a simple protocol to try to bring participants back to the study after they have had time to re-sensitize to the reinforcements in SARA: we decided to send text messages to participants after a few days of non-response. The first text message was to be sent after 2 days of no self-reporting. If non-adherence continued, an additional text message would be sent every three days. Text messaging stopped if participants did not self-report for 3 weeks.


\subsection{\textbf{User study 4: An evaluation study of SARA (N=37)}}
We evaluated the final design of SARA's adherence interventions in a 30-day field study with 37 high-risk AYA substance users (49\% male; 73.5\% Caucasian; age: $\mu$=20.4, SD=2.1; 53\% binge drinking-only, 47\% any marijuana use in last month). {\color{blue} We did not test the new features from Iteration 5 in a focus group or another pilot, because we had a deadline on our funding and the number of new features was small to necessitate additional pilot evaluations. For the field evaluation, we used }the same 30-day study protocol as we used in the user study 3. We recruited high-risk AYA who were admitted to {\color{blue} the University of Anonymous Hospital Pediatric and Adult Emergency Department}  for risky substance use. Potential participants were approached as they were being discharged and were screened for eligibility and interest in the study. Interested individuals were consented and SARA was installed on their personal phones, after which they began 30 days of data collection. As we noted above, the 30-day study duration is common for observational studies on substance use~\cite{suffoletto2012text,shrier2018pilot,clark2010project,bonar2018feasibility,buu2017assessment,comulada2015compliance,suffoletto2012text,wen2017compliance}, so this duration allowed us to compare our adherence rates to those in the literature.  {\color{blue} Furthermore, the sample of $N=37$ is similar to or higher than other mHealth deployment studies that focused on novel intervention technologies ~\cite{consolvo2008flowers,lane2014bewell,bardram2013designing}. The sample size is also sufficient to draw design insights using both qualitative and quantitative measures.}\\

{\color{blue}

\noindent
The findings from Study 4 fall into several categories: \\ 

\noindent
\underline{Adherence}: We compared the adherence patterns from User Sstudy 4 with the adherence rates in User Study 3 as well as with a similar study by Bonar et al.~\cite{bonar2018feasibility}. User Study 3 used the same protocol as User Study 4, and the Bonar et al.~\cite{bonar2018feasibility} study involved answering a daily survey for 30 days and the population is similar to SARA's population of AYA at high risk of substance use. Bonar et al.'s asked 27 questions in the daily survey, but the average number of questions answered was 18. For SARA, average number of questions are close to seven but SARA had two additional assessments in terms of active tasks. Thus the overall self-report burden in SARA and Bonar et al.~\cite{bonar2018feasibility} are comparable.\\

\noindent
SARA's adherence rate  in User Study 4 was 71.7\% for the first 10 days, 64.4\% for days 11-20, and 50.8\% for days 21-30. The adherence rate of 50.8\% after 20 days is higher than for most self-report studies. These results suggest that some of the changes in Iteration 5 had a positive effect: beyond the first ten days, adherence rates increased substantially from User Study 3 ( 63.8\% for days 1 to 10, 49.2\% for days 11 to 20, and 34.6\% for days 21 to 30).  Figure~\ref{fig:iteration5} shows the adherence rate of SARA and Bonar et al.~\cite{bonar2018feasibility} on week 1,2,3,4 of the study. While the adherence rates are similar, Bonar et al.~\cite{bonar2018feasibility} paid significantly more money.\\

\begin{figure}[t] 
    \centering 
    \includegraphics[width=1.00\textwidth]{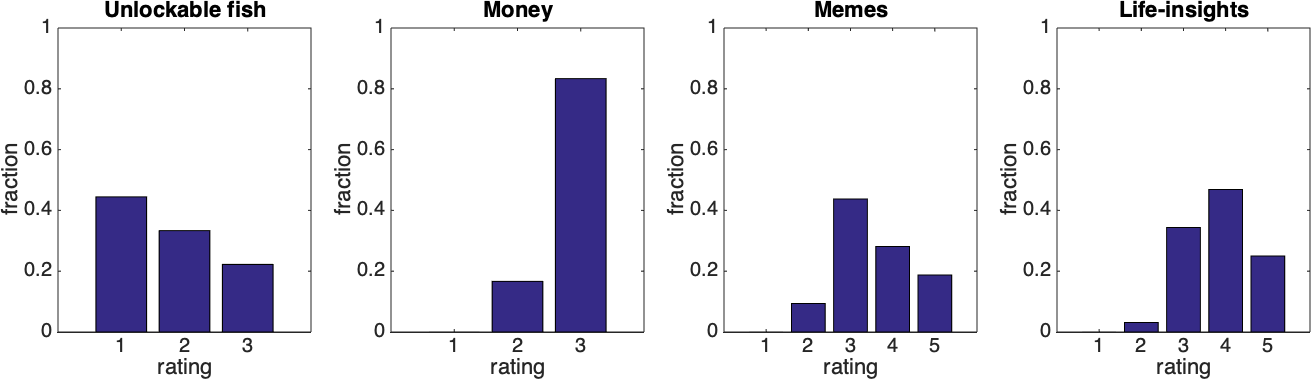}
    \caption{Distribution of ratings for various reinforcement types in user study 4. Left two subfigure uses a three point likert scale (1=not all, 2=somewhat,3=very much) to rate how these reinforcement influenced self-report. Righ two subfigure uses a five point likert scale (1=not all, 5=very much) to rate how these reinforcement influenced self-report.}
    \label{fig:adherencepilot}
\end{figure}

\noindent
\underline{Monetary incentives}: On average, participants earned \$6.53 each ($\sigma=3.8, q_{50}=6.5$) in User Study 4 for completion of the assessments over 30 days. In the exit survey, when we asked participants how money affected their use of the SARA app (``How much did earning money bonuses increase your use of the app?'' on a 3-point Likert scale (1=not at all, 2=somewhat, and 3=very much), 83.3\% reported very much and the rest reported somewhat. No participant answered `not at all.' In the exit interviews, both the amount of money and the schedule were perceived positively: one participant said \textit{``I also liked the money bonuses and thought the amount of money offered was just right.''} and another participant said  \textit{``I liked earning money because it kept me coming back for 3 days in a row.''} These results suggest that the changes in monetary schedule in Iteration 5 (i.e., \$1 for each three-day streak) had a positive impact.\\

\noindent
Furthermore, the average amount of money earned, \$6.53, is important. Bonar et al., described earlier, used higher financial incentives to achieve similar daily adherence rates to Iteration 5 of SARA. Bonar et al. paid \$2 for each daily survey completed and an additional \$5 bonus when participants completed at least 6 of 7 daily surveys each week (up to \$20 total bonus). Participants in Bonar et al.'s study earned on average \$46.20 $(SD=\$24.62)$ in 28 days (including bonuses), which is seven times more than what  SARA participants were paid.\\



\noindent
\underline{Aquarium and fish}: Participants on average spent 14.5 seconds $(SD=17.6 seconds)$ looking at the aquarium. However, proportion of participants looked at the aquarium everyday other than self-reporting decreased by 1\% per day over the course of study which may mean participants were likely losing interest. When we asked participants how unlocking fish affected their use of the SARA app (``How much did unlockable fish increase your use of the app?'') on a 3-point Likert scale (1=not at all, 2=somewhat, and 3=very much), the average rating was 1.7 $(SD=2.7)$. Note that these ratings are lower than those of the focus group (see section 3.7) whose participants did not use the app for 30-days. The qualitative feedback for the aquarium is similar to the feedback from the pilot study. Some participants liked the appearance of the aquarium: \textit{``I also liked the appearance of it, it was engaging. The contrasting colors were captivating and I also liked the idea of an aquarium.''}  and some liked the reinforcement schedule of regularly unlocking fish:
\textit{``The app made me want to use it every day because I wanted to unlock new fish.''} However, as in earlier user studies, some participants did not find the aquarium aesthetically pleasing: one participant said, \textit{``The color composition is not aesthetically appealing.''}. Another participant said,  \textit{`The images look like they're from the 1990s.''}. \\

\noindent
\underline{Memes and life insight}: Participants on average spent 5.6 seconds $(SD=7.7)$ looking at the memes and 7.1 seconds $(SD=8.9)$ looking at the life insights. In the exit survey, participants rated the perceived influence of memes and life-insight on a five-point scale, from 1=not at all to 5=very much. Life insights  $(\mu=4.2,\sigma=2.5)$ were rated higher than memes $(\mu=3.6,\sigma=2.3)$. Participants generally liked the life insights and they reported using life-insights to reflect on their week and to track their progress in tapping or spatial tasks. One participant said, \textit{``I liked having a way to record all of my days. I thought it was cool to look back on each day at the graphs and see how my weeks have been.''} Another participant said \textit{``[I] enjoyed seeing how performance on tasks reflected in insights.''} Adding more life insights was one area of improvement identified by participants.  Regarding memes, some participants liked their variety while others had recommendations for updating and personalizing them. For example, one participant said \textit{``I liked how there were different memes and quotes every day.''} and another said \textit{``the memes weren't actually memes. They were either 2009-style memes, which just look kinda bad, or they just looked like pictures used to make memes.''}\\

\noindent
These results suggest that life insights are generally viewed positively, which is consistent with past findings that younger adults like to explore their data~\cite{meng2018exploring}. However, the memes we selected using pilot procedures were not as appealing; user feedback indicates that memes have a temporal nature and people like memes on specific topics they care about. It is harder to design memes as reinforcement than we had anticipated. In part, because some memes were meant to appeal to user's sense of humor, anticipating and personalizing memes to such a highly personal trait such as sense of humor is challenging. It may be that the inspirational memes used were more appealing; however, this supposition requires future study.\\

\noindent
\underline{Text messages and phone call}: On average, participants received 5.1 $(SD=1.8)$ texts and phone calls. On 47\% of days, participants self-reported on the same day after they received a text message. Recall the text messages or phone calls were sent after participants did not 2 to 3 days. So, the contacts from study staff likely increased adherence as intended.\\

\noindent
\underline{Future app improvement opportunities}: In the exit interviews, participants provided additional suggestions for improving the future version of the app. The first type of feedback asked for options to choose the time of survey completion. One participant said, \textit{``The survey starting at 6pm was a barrier to adherence... it would've been better if it was even an hour earlier or earlier in the day so there's more time and more flexibility.''}  Another participant said \textit{``I would've preferred to answer the daily survey first thing in the morning about the previous day.''} The second type of feedback suggested categorizing the memes and letting participants choose which kind they see. One participant said \textit{``I think if there was an option to choose a specific category of memes/quotes to receive, then I would have been more captivated.''} Finally, participants asked for better instructions on using the app and asked for more features to give reasons to go the app. We believe the issue with the instructions can be resolved with additional user-centered design. Creating more value can be achieved in different ways. We can use data to provide therapeutic strategies to reduce substance use (e.g., coping skills). Another way to create value could be a ``viable research alliance'' where participants are informed of the scientific value of the data they are giving us~\cite{csikszentmihalyi2014validity}. Another idea to improve design could be to work with developmental psychologists to add more age appropriate content. 
Another thing to note is that this user feedback deals with issues of choice, interruptibility, and empowering users with knowledge. These issues, however, are not well-covered by OCT, so we need to move beyond OCT and use other theories to make SARA more engaging.\\

\noindent
\noindent
Nonetheless, the above results suggest that our theory translation of OCT and design iterations led to improved effectiveness of the reinforcement used in SARA. These results do not reach the level of rigor that a randomized controlled trial would yield. But given their purpose---to provide empirical data that can be triangulated with information from previous user studies to assess the quality of our theory translation efforts--- User Study 4's results indicate that constructs from operant conditioning may have been implemented successfully.\\

}

\vspace{-.1cm}
\section{\textbf{Theory translation in a nutshell}}
In the previous section, we gave a detailed description of how we translated a set of constructs from OCT into a number of features of the SARA application. In this section, we distill our experience and provide some general methodological guidelines for the process of translating theory into technical interventions. We hope these guidelines can act as a starting point for a more structured process that can help designers do theory translation in a more efficient and less error-prone way.\\ 

\noindent
Theory translation is an iterative design process. Although iterative deign is common in HCI, iterative design for theory translation is different in that its aim is to improve the \emph{theoretical fidelity} of the intervention---i.e., how faithfully the intervention features embody the constructs they are intended to implement. For example, the animated fish from the iteration 2 of SARA (section 3.3) are a higher-theoretical-fidelity representation of `valued reinforcement' than the static fish from Iteration 1.   Improving theoretical fidelity is important for reasons discussed in Section 2.4. However, improving theoretical fidelity is a complex iterative process of balancing user feedback, project constraints, theoretical insights, etc. Based on our experience with designing SARA, we have identified eight key steps that designers need to address during theory translation:


\begin{enumerate}
    \item \underline{Selecting a theory}: First, select an appropriate theory that will provide an account of how the technology under development is intended to influence the target behavior in the target population. Intervention development frameworks from behavioral science, such as the Behavior Change Wheel~\cite{michie2011behaviour,michie2013behavior,michie2014behavior} and Intervention Mapping~\cite{bartholomew1998intervention,kok2004intervention}, can be used to facilitate this step. A key thing to note at this step is that choosing a theory means that the theory will act, at least in part, as a blueprint of the causal influence that the technology is intended to have on the target behavior. It should describe how exactly the technology is intended to change people's behavior. It thus follows that how well a theory is specified greatly influences both how useful it is as the basis for an intervention and how easily it can be implemented. \\ 
    
     \item \underline{Selecting an initial set of constructs}: Once a theory is selected, select the constructs that will form the foundation of the intervention's functionality. In SARA, our goal was to increase the frequency of responding to the daily self-report and to maintain this responding over time. Thus, we started by trying to implement positive reinforcement and schedules of reinforcement, which are the core OCT constructs that describe the process that increases behavior frequency and affects its dynamics over time (Section 3.2-3.3). Additional constructs can be added later as needed (e.g., based on user feedback or to account for a specific aspect of the behavior change process), but we suggest trying to get the core functionality---and the corresponding intended causal process---clearly specified first.\\
     
     \item \underline{Developing initial designs}: For each initially selected construct, explore the design space for how that construct can be operationalized in the particular system that is being developed. The key goals of the ideation stage is to develop designs that (a) embody key properties of the construct, and (b) match the known constraints of the population, context of use, and system coherence. A useful structure for exploring the design space is to develop both alternative ways of implementing a construct and multiple variations of each of those designs. Generated designs can then be evaluated for feasibility, user acceptance, and preliminary theoretical fidelity.\\
     
     \item \underline{Specifying theoretical fidelity criteria}: After the basic set of designs have been developed, for each implemented construct articulate how you would be able to tell if your implementation of the construct got it ``right''---namely, whether the technology feature has captured the essence of the construct. This is arguably the hardest step in the whole theory translation process, as the criteria for establishing theoretical fidelity may not be at all obvious. For some constructs, theoretical fidelity may need to be established based on the qualitative user feedback; for instance, to operationalize ``reinforcement value'' we had to rely on participants' comments about whether they found proposed design features (e.g., fish) appealing and whether they thought those features would motivate them to self-report. For other constructs, it may be possible to articulate exactly what kinds of patterns of behavior one would expect if the feature was operating as described by the construct (e.g., this is possible for different types of reinforcement schedules). Despite the difficulties inherent in this step, it is a crucial one, as it allows designers to test whether their technology features, as designed, are in fact embodying the intended construct.\\  
     
    \item \underline{Preliminary testing of theoretical fidelity}: After the basic set of constructs have been selected and the criteria for theoretical fidelity articulated, implement low-fi prototypes of the features that embody these constructs and test their theoretical fidelity. This can be done in many ways, from informal internal testing to small user studies depending on the nature of the fidelity criteria. Whatever the method, the central task in this step is to get early information about whether the technology feature, as envisioned, may be able to act as described by the construct. As is typical for low-fi prototyping, ineffective prototypes can be discarded and the rest of the prototypes quickly improved to achieve higher-theoretical-fidelity implementation of the underlying constructs. The purpose of our user studies 1 and 2 was precisely to do this type of preliminary testing of theoretical fidelity, although, in retrospect, these studies could have likely been done in a cleaner way to achieve this goal.\\
    
    \item \underline{Adding constructs}: As the intervention develops, it will often be necessary to add additional constructs either from the same theory or from other theories. New constructs should be chosen either because they may amplify the functioning of the already implemented constructs or because designers believe the intervention requires additional forms of behavioral support. In either case, as the constructs are being chosen, it is important to articulate how they would interact with the already implemented features, so designers can assess, before development resources are expended, if the features based on these constructs would strengthen or negatively impact the existing functionality. Formal representations, such as causal diagrams, can often be helpful for this, but less formal approaches---like our considerations of using multiple reinforcers---may be sufficient.\\ 
    
    \item \underline{Selecting study methods for formative evaluations}: Early tests of theoretical fidelity, as well as formative evaluations of overall functionality will often require some form of data collection. What types of user studies will be useful will depend, of course on the specific questions they are intended to address. However, during the formative stage, online surveys or focus groups can often be cost-effective ways to assess user perceptions in order to improve the theoretical fidelity of the intervention. Later on, field trials will usually be needed to capture behavioral responses to confirm intended functioning. The key thing to note is that the studies need to match the level of evidence that is required to move to the next stage of the development process. Early on, a designer might just need a sanity check on an idea, as we did with the aquarium representation. The level of evidence needed at this point is much lower---and can be achieved with a much simpler study---than the evidence needed to establish the effect size for an intervention. Being clear about what exactly a study is intended to achieve, and then selecting the study design and sample size appropriately, is thus paramount both for doing theory translation in a resource-efficient way and, down the line, for producing reliable, trustworthy evidence on the completed intervention. Similar considerations apply to selecting the study duration as well. Behavioral interventions can have effects that change over time due to learning, habituation, habit formation and so on. Real-world use can uncover problems which may need to be addressed with additional constructs or fidelity improvements. For example, the length of the User Study 3 allowed us to learn about habituation and the problems with the initially implemented monetary schedule. As with the study type, the study length for field studies thus needs to be chosen thoughtfully, with the eye toward maximizing the designers' ability to learn about problems that need to be addressed while minimizing the need for resources and participant burden. \\
    
    \item \underline{Providing in-depth descriptions of intervention design}: One goal of theory-based interventions is to allow testing of theories in order to advance science. To further this goal, it is essential for designers to report what theoretical constructs were implemented in an intervention and how in detail. Such descriptions can help both with the interpretation of findings from any experimental trials of the intervention, as well as their comparison with results from other studies of other interventions that embody the same constructs. This level of description is currently unusual both in HCI and in behavioral science, but we firmly believe that it is essential for advancing our understanding of the behavior change process and the factors that influence intervention response.  
    
\end{enumerate}





\vspace{-.1cm}
\section{\textbf{Discussion and conclusions}}
In this paper, we provided a detailed account of our process of trying to translate a set of concepts from OCT into features of SARA, a mobile application for conducting substance use research with adolescents and young adults. As a case study of theory translation, OCT is in many ways the best case scenario. The constructs and processes postulated by the theory are well specified and are supported by---and have been refined through---decades of careful empirical work with both animals and humans. If there is a theory that should allow for straightforward implementation, OCT is it. Indeed, some of the concepts we could implement in a very straightforward way. We could implement the notions of immediacy and contingency just by providing reinforcement right after, and only after, a user provided self-report (Section 3.2). Similarly, the extensive empirical data on schedules of reinforcement allowed us to make informed decisions to use a fast fixed-ratio schedule for fish and to include variable-rate schedules to make the effects of reinforcement more sustainable (Section 3.3, 3.4, 3.7). OCT also provided clear guidance on the question of whether multiple forms of reinforcement could work together effectively, leading us to greatly expand the range of rewards that SARA could provide to reinforce self-reporting (Section 3.4, 3.7). Much more directly than many theories, then, OCT could tell us \textit{what} we needed to do, at least in broad terms.\\


\noindent
Yet, even with such a highly specified theory, \textit{how} exactly the various forms of reinforcement in SARA needed to work was left unclear. What forms of reinforcement would be found to be valuable, how fast the fish schedule needed to be, how exactly the schedule of financial reinforcements needed to be structured were among the many decisions we had to make based on design intuitions, user feedback, and resource constraints. The theory could tell us what should work in principle, but the many design details we had to decide on to operationalize its concepts for this  application and this particular population were left to us to work through. Yet, those design details mattered greatly. The aesthetics of the fish or whether the financial rewards were given in fractions or whole dollar amounts influenced how valuable---and thus reinforcing---our participants perceived them to be. To get the details right, we had to keep the theory in constant conversation with user feedback and other constraints and to iteratively make gradual design changes until we started to see reactions and behaviors that began to approach what OCT told us should happen for the concepts we were trying to implement. In other words, we could only tell that we implemented a construct from OCT adequately when we could see the behavior postulated by the theory to result from the construct.\\

\noindent 
Which brings us to a key challenge of theory translation: determining if a theoretical construct has been implemented with fidelity. Theory-based interventions play a dual role. On the one hand, they are designed to address a particular problem in a particular population. Insofar as an intervention achieves this goal, it can be considered to be a success. On the other hand, interventions act as tests of the theories they embody. For science to progress, theories have to be tested and refined, and there is no way to do that in the abstract. A behavioral theory can only be tested by studying the behavior of real people and their responses to concrete interventions. But for a study to provide evidence about a theory, the intervention used in the study needs to have implemented that theory with fidelity.\\

\noindent 
How to ensure that a construct has, in fact, been implemented with high fidelity is not trivial, however. As we noted, our test for theoretical fidelity in SARA was to look for behaviors that the theory postulated. For a precise theory like OCT that provides highly specific accounts of what should happen under different circumstances, such a test makes sense. But many of the theories commonly used to guide the development of technological interventions are not nearly so specific in their accounts of what to expect beyond assertions that a set of determinants influence behavior. For such theories, a designer is left without clear criteria for evaluating theoretical fidelity. Consulting with domain experts with experience in that theory can help with articulating theoretical fidelity criteria, but even this strategy can sometimes leave the designer with having to come up with something herself. As we noted in Section 4, making an effort to articulate fidelity criteria is paramount, however, and with the rapid move toward technological interventions in behavior change research the issue of theoretical fidelity will only grow in importance. As HCI researchers are increasingly participating in interdisciplinary collaborations with behavioral scientists, and insofar as we wish to more deeply understand why our technologies do or do not work, our discipline will need to deal with this issue head-on.\\

\noindent
All theories are by their nature abstract. Theories are formulated to account for concrete behaviors and events, but a theoretical explanation is only achieved by focusing on certain narrow aspects of a phenomenon, abstracting out its general features, and stripping out everything else.  The problem, from the standpoint of technology design, is going in the opposite direction---filling in all the concreteness that was left out in order to achieve theoretical generality. How best to approach this process of operationalizing theories in technological interventions is something that, we believe, our community needs to think about carefully. In presenting this case study, our goal was to surface the complexity and nuances involved in theory translation. We are not arguing that the process we followed is the right one, or that others should apply it to their work in the same form. Rather, we hope that the process we went through can serve as an example that can be critiqued and improved upon. Over time, we hope, we will arrive at a more robust and efficient process for theory translation and will more fully understand its challenges and pitfalls. There is much left to do to achieve this goal.

\bibliography{mash_bib}{}


\begin{thebibliography}{150}


\ifx \showCODEN    \undefined \def \showCODEN     #1{\unskip}     \fi
\ifx \showDOI      \undefined \def \showDOI       #1{#1}\fi
\ifx \showISBNx    \undefined \def \showISBNx     #1{\unskip}     \fi
\ifx \showISBNxiii \undefined \def \showISBNxiii  #1{\unskip}     \fi
\ifx \showISSN     \undefined \def \showISSN      #1{\unskip}     \fi
\ifx \showLCCN     \undefined \def \showLCCN      #1{\unskip}     \fi
\ifx \shownote     \undefined \def \shownote      #1{#1}          \fi
\ifx \showarticletitle \undefined \def \showarticletitle #1{#1}   \fi
\ifx \showURL      \undefined \def \showURL       {\relax}        \fi
\providecommand\bibfield[2]{#2}
\providecommand\bibinfo[2]{#2}
\providecommand\natexlab[1]{#1}
\providecommand\showeprint[2][]{arXiv:#2}

\bibitem[\protect\citeauthoryear{Abuse, Administration, et~al\mbox{.}}{Abuse
  et~al\mbox{.}}{2016}]%
        {abuse20162015}
\bibfield{author}{\bibinfo{person}{Substance Abuse}, \bibinfo{person}{Mental
  Health~Services Administration}, {et~al\mbox{.}}}
  \bibinfo{year}{2016}\natexlab{}.
\newblock \showarticletitle{2015 National Survey on Drug Use and Health}.
\newblock  (\bibinfo{year}{2016}).
\newblock


\bibitem[\protect\citeauthoryear{{Abyssrium}}{{Abyssrium}}{2018}]%
        {abyssriumpress}
\bibfield{author}{\bibinfo{person}{{Abyssrium}}.}
  \bibinfo{year}{2018}\natexlab{}.
\newblock \bibinfo{howpublished}{https://www.abyssrium.com/}.
\newblock
\newblock
\shownote{[Online; accessed 2 July 2018].}


\bibitem[\protect\citeauthoryear{Adams, Rabbi, Rahman, Matthews, Voida, Gay,
  Choudhury, and Voida}{Adams et~al\mbox{.}}{2014}]%
        {adams2014towards}
\bibfield{author}{\bibinfo{person}{Phil Adams}, \bibinfo{person}{Mashfiqui
  Rabbi}, \bibinfo{person}{Tauhidur Rahman}, \bibinfo{person}{Mark Matthews},
  \bibinfo{person}{Amy Voida}, \bibinfo{person}{Geri Gay},
  \bibinfo{person}{Tanzeem Choudhury}, {and} \bibinfo{person}{Stephen Voida}.}
  \bibinfo{year}{2014}\natexlab{}.
\newblock \showarticletitle{Towards personal stress informatics: Comparing
  minimally invasive techniques for measuring daily stress in the wild}. In
  \bibinfo{booktitle}{\emph{Proceedings of the 8th International Conference on
  Pervasive Computing Technologies for Healthcare}}. ICST (Institute for
  Computer Sciences, Social-Informatics and Telecommunications Engineering),
  \bibinfo{pages}{72--79}.
\newblock


\bibitem[\protect\citeauthoryear{Ainslie}{Ainslie}{2001}]%
        {ainslie2001breakdown}
\bibfield{author}{\bibinfo{person}{George Ainslie}.}
  \bibinfo{year}{2001}\natexlab{}.
\newblock \bibinfo{booktitle}{\emph{Breakdown of will}}.
\newblock \bibinfo{publisher}{Cambridge University Press}.
\newblock


\bibitem[\protect\citeauthoryear{Ariely and Jones}{Ariely and Jones}{2013}]%
        {ariely2013upside}
\bibfield{author}{\bibinfo{person}{Dan Ariely} {and} \bibinfo{person}{Simon
  Jones}.} \bibinfo{year}{2013}\natexlab{}.
\newblock \bibinfo{booktitle}{\emph{The upside of irrationality}}.
\newblock \bibinfo{publisher}{CNIB}.
\newblock


\bibitem[\protect\citeauthoryear{Badawy and Kuhns}{Badawy and Kuhns}{2017}]%
        {badawy2017texting}
\bibfield{author}{\bibinfo{person}{Sherif~M Badawy} {and}
  \bibinfo{person}{Lisa~M Kuhns}.} \bibinfo{year}{2017}\natexlab{}.
\newblock \showarticletitle{Texting and mobile phone app interventions for
  improving adherence to preventive behavior in adolescents: a systematic
  review}.
\newblock \bibinfo{journal}{\emph{JMIR mHealth and uHealth}}
  \bibinfo{volume}{5}, \bibinfo{number}{4} (\bibinfo{year}{2017}).
\newblock


\bibitem[\protect\citeauthoryear{Bae, Chung, Ferreira, Dey, and Suffoletto}{Bae
  et~al\mbox{.}}{2018}]%
        {bae2018mobile}
\bibfield{author}{\bibinfo{person}{Sangwon Bae}, \bibinfo{person}{Tammy Chung},
  \bibinfo{person}{Denzil Ferreira}, \bibinfo{person}{Anind~K Dey}, {and}
  \bibinfo{person}{Brian Suffoletto}.} \bibinfo{year}{2018}\natexlab{}.
\newblock \showarticletitle{Mobile phone sensors and supervised machine
  learning to identify alcohol use events in young adults: Implications for
  just-in-time adaptive interventions}.
\newblock \bibinfo{journal}{\emph{Addictive behaviors}}  \bibinfo{volume}{83}
  (\bibinfo{year}{2018}), \bibinfo{pages}{42--47}.
\newblock


\bibitem[\protect\citeauthoryear{Bardram, Frost, Sz{\'a}nt{\'o},
  Faurholt-Jepsen, Vinberg, and Kessing}{Bardram et~al\mbox{.}}{2013}]%
        {bardram2013designing}
\bibfield{author}{\bibinfo{person}{Jakob~E Bardram}, \bibinfo{person}{Mads
  Frost}, \bibinfo{person}{K{\'a}roly Sz{\'a}nt{\'o}}, \bibinfo{person}{Maria
  Faurholt-Jepsen}, \bibinfo{person}{Maj Vinberg}, {and}
  \bibinfo{person}{Lars~Vedel Kessing}.} \bibinfo{year}{2013}\natexlab{}.
\newblock \showarticletitle{Designing mobile health technology for bipolar
  disorder: a field trial of the monarca system}. In
  \bibinfo{booktitle}{\emph{Proceedings of the SIGCHI conference on human
  factors in computing systems}}. ACM, \bibinfo{pages}{2627--2636}.
\newblock


\bibitem[\protect\citeauthoryear{Bartholomew, Parcel, and Kok}{Bartholomew
  et~al\mbox{.}}{1998}]%
        {bartholomew1998intervention}
\bibfield{author}{\bibinfo{person}{L~Kay Bartholomew}, \bibinfo{person}{Guy~S
  Parcel}, {and} \bibinfo{person}{Gerjo Kok}.} \bibinfo{year}{1998}\natexlab{}.
\newblock \showarticletitle{Intervention mapping: a process for developing
  theory and evidence-based health education programs}.
\newblock \bibinfo{journal}{\emph{Health Education \& Behavior}}
  \bibinfo{volume}{25}, \bibinfo{number}{5} (\bibinfo{year}{1998}),
  \bibinfo{pages}{545--563}.
\newblock


\bibitem[\protect\citeauthoryear{Bonar, Cunningham, Collins, Cranford,
  Chermack, Zimmerman, Blow, and Walton}{Bonar et~al\mbox{.}}{2018}]%
        {bonar2018feasibility}
\bibfield{author}{\bibinfo{person}{Erin~E Bonar}, \bibinfo{person}{Rebecca~M
  Cunningham}, \bibinfo{person}{R~Lorraine Collins}, \bibinfo{person}{James~A
  Cranford}, \bibinfo{person}{Stephen~T Chermack}, \bibinfo{person}{Marc~A
  Zimmerman}, \bibinfo{person}{Frederic~C Blow}, {and}
  \bibinfo{person}{Maureen~A Walton}.} \bibinfo{year}{2018}\natexlab{}.
\newblock \showarticletitle{Feasibility and acceptability of text messaging to
  assess daily substance use and sexual behaviors among urban emerging adults}.
\newblock \bibinfo{journal}{\emph{Addiction research \& theory}}
  \bibinfo{volume}{26}, \bibinfo{number}{2} (\bibinfo{year}{2018}),
  \bibinfo{pages}{103--113}.
\newblock


\bibitem[\protect\citeauthoryear{Boruvka, Almirall, Witkiewitz, and
  Murphy}{Boruvka et~al\mbox{.}}{2017}]%
        {boruvka2017assessing}
\bibfield{author}{\bibinfo{person}{Audrey Boruvka}, \bibinfo{person}{Daniel
  Almirall}, \bibinfo{person}{Katie Witkiewitz}, {and} \bibinfo{person}{Susan~A
  Murphy}.} \bibinfo{year}{2017}\natexlab{}.
\newblock \showarticletitle{Assessing time-varying causal effect moderation in
  mobile health}.
\newblock \bibinfo{journal}{\emph{J. Amer. Statist. Assoc.}}
  \bibinfo{number}{just-accepted} (\bibinfo{year}{2017}).
\newblock


\bibitem[\protect\citeauthoryear{Boyle, Earle, LaBrie, and Smith}{Boyle
  et~al\mbox{.}}{2017}]%
        {boyle2017pnf}
\bibfield{author}{\bibinfo{person}{Sarah~C Boyle}, \bibinfo{person}{Andrew~M
  Earle}, \bibinfo{person}{Joseph~W LaBrie}, {and} \bibinfo{person}{Daniel~J
  Smith}.} \bibinfo{year}{2017}\natexlab{}.
\newblock \showarticletitle{PNF 2.0? Initial evidence that gamification can
  increase the efficacy of brief, web-based personalized normative feedback
  alcohol interventions}.
\newblock \bibinfo{journal}{\emph{Addictive behaviors}}  \bibinfo{volume}{67}
  (\bibinfo{year}{2017}), \bibinfo{pages}{8--17}.
\newblock


\bibitem[\protect\citeauthoryear{Braun and Clarke}{Braun and Clarke}{2006}]%
        {braun2006using}
\bibfield{author}{\bibinfo{person}{Virginia Braun} {and}
  \bibinfo{person}{Victoria Clarke}.} \bibinfo{year}{2006}\natexlab{}.
\newblock \showarticletitle{Using thematic analysis in psychology}.
\newblock \bibinfo{journal}{\emph{Qualitative research in psychology}}
  \bibinfo{volume}{3}, \bibinfo{number}{2} (\bibinfo{year}{2006}),
  \bibinfo{pages}{77--101}.
\newblock


\bibitem[\protect\citeauthoryear{Burchard and Tyler~Jr}{Burchard and
  Tyler~Jr}{1964}]%
        {burchard1964modification}
\bibfield{author}{\bibinfo{person}{John Burchard} {and} \bibinfo{person}{Vernon
  Tyler~Jr}.} \bibinfo{year}{1964}\natexlab{}.
\newblock \showarticletitle{The modification of delinquent behaviour through
  operant conditioning}.
\newblock \bibinfo{journal}{\emph{Behaviour Research and Therapy}}
  \bibinfo{volume}{2}, \bibinfo{number}{2-4} (\bibinfo{year}{1964}),
  \bibinfo{pages}{245--250}.
\newblock


\bibitem[\protect\citeauthoryear{Burns, Begale, Duffecy, Gergle, Karr,
  Giangrande, and Mohr}{Burns et~al\mbox{.}}{2011}]%
        {burns2011harnessing}
\bibfield{author}{\bibinfo{person}{Michelle~Nicole Burns},
  \bibinfo{person}{Mark Begale}, \bibinfo{person}{Jennifer Duffecy},
  \bibinfo{person}{Darren Gergle}, \bibinfo{person}{Chris~J Karr},
  \bibinfo{person}{Emily Giangrande}, {and} \bibinfo{person}{David~C Mohr}.}
  \bibinfo{year}{2011}\natexlab{}.
\newblock \showarticletitle{Harnessing context sensing to develop a mobile
  intervention for depression}.
\newblock \bibinfo{journal}{\emph{Journal of medical Internet research}}
  \bibinfo{volume}{13}, \bibinfo{number}{3} (\bibinfo{year}{2011}).
\newblock


\bibitem[\protect\citeauthoryear{Bush, Kivlahan, McDonell, Fihn, and
  Bradley}{Bush et~al\mbox{.}}{1998}]%
        {bush1998audit}
\bibfield{author}{\bibinfo{person}{Kristen Bush}, \bibinfo{person}{Daniel~R
  Kivlahan}, \bibinfo{person}{Mary~B McDonell}, \bibinfo{person}{Stephan~D
  Fihn}, {and} \bibinfo{person}{Katharine~A Bradley}.}
  \bibinfo{year}{1998}\natexlab{}.
\newblock \showarticletitle{The AUDIT alcohol consumption questions (AUDIT-C):
  an effective brief screening test for problem drinking}.
\newblock \bibinfo{journal}{\emph{Archives of internal medicine}}
  \bibinfo{volume}{158}, \bibinfo{number}{16} (\bibinfo{year}{1998}),
  \bibinfo{pages}{1789--1795}.
\newblock


\bibitem[\protect\citeauthoryear{Buu, Massey, Walton, Cranford, Zimmerman, and
  Cunningham}{Buu et~al\mbox{.}}{2017}]%
        {buu2017assessment}
\bibfield{author}{\bibinfo{person}{Anne Buu}, \bibinfo{person}{Lynn~S Massey},
  \bibinfo{person}{Maureen~A Walton}, \bibinfo{person}{James~A Cranford},
  \bibinfo{person}{Marc~A Zimmerman}, {and} \bibinfo{person}{Rebecca~M
  Cunningham}.} \bibinfo{year}{2017}\natexlab{}.
\newblock \showarticletitle{Assessment methods and schedules for collecting
  daily process data on substance use related health behaviors: A randomized
  control study}.
\newblock \bibinfo{journal}{\emph{Drug and alcohol dependence}}
  \bibinfo{volume}{178} (\bibinfo{year}{2017}), \bibinfo{pages}{159--164}.
\newblock


\bibitem[\protect\citeauthoryear{Celio, Usala, Lisman, Johansen,
  Vetter-O'Hagen, and Spear}{Celio et~al\mbox{.}}{2014}]%
        {celio2014we}
\bibfield{author}{\bibinfo{person}{Mark~A Celio}, \bibinfo{person}{Julie~M
  Usala}, \bibinfo{person}{Stephen~A Lisman}, \bibinfo{person}{Gerard~E
  Johansen}, \bibinfo{person}{Courtney~S Vetter-O'Hagen}, {and}
  \bibinfo{person}{Linda~P Spear}.} \bibinfo{year}{2014}\natexlab{}.
\newblock \showarticletitle{Are we drunk yet? Motor versus cognitive cues of
  subjective intoxication}.
\newblock \bibinfo{journal}{\emph{Alcoholism: clinical and experimental
  research}} \bibinfo{volume}{38}, \bibinfo{number}{2} (\bibinfo{year}{2014}),
  \bibinfo{pages}{538--544}.
\newblock


\bibitem[\protect\citeauthoryear{Choe, Fogarty, Lee, Matthews, Kientz,
  Abdullah, Rabbi, Thomaz, Epstein, Cordeiro, et~al\mbox{.}}{Choe
  et~al\mbox{.}}{2017}]%
        {choe2017semi}
\bibfield{author}{\bibinfo{person}{Eun~Kyoung Choe}, \bibinfo{person}{James
  Fogarty}, \bibinfo{person}{Bongshin Lee}, \bibinfo{person}{Mark Matthews},
  \bibinfo{person}{Julie~A Kientz}, \bibinfo{person}{Saeed Abdullah},
  \bibinfo{person}{Mashfiqui Rabbi}, \bibinfo{person}{Edison Thomaz},
  \bibinfo{person}{Daniel~A Epstein}, \bibinfo{person}{Felicia Cordeiro},
  {et~al\mbox{.}}} \bibinfo{year}{2017}\natexlab{}.
\newblock \showarticletitle{Semi-automated tracking: A balanced approach for
  self-monitoring applications}.
\newblock \bibinfo{journal}{\emph{IEEE Pervasive Computing}}
  \bibinfo{number}{1} (\bibinfo{year}{2017}), \bibinfo{pages}{74--84}.
\newblock


\bibitem[\protect\citeauthoryear{Clark, Ringwalt, Hanley, Shamblen, Flewelling,
  and Hano}{Clark et~al\mbox{.}}{2010}]%
        {clark2010project}
\bibfield{author}{\bibinfo{person}{Heddy~Kovach Clark},
  \bibinfo{person}{Chris~L Ringwalt}, \bibinfo{person}{Sean Hanley},
  \bibinfo{person}{Stephen~R Shamblen}, \bibinfo{person}{Robert~L Flewelling},
  {and} \bibinfo{person}{Mary~C Hano}.} \bibinfo{year}{2010}\natexlab{}.
\newblock \showarticletitle{Project SUCCESS'effects on the substance use of
  alternative high school students}.
\newblock \bibinfo{journal}{\emph{Addictive behaviors}} \bibinfo{volume}{35},
  \bibinfo{number}{3} (\bibinfo{year}{2010}), \bibinfo{pages}{209--217}.
\newblock


\bibitem[\protect\citeauthoryear{Cole-Lewis, Ezeanochie, and
  Turgiss}{Cole-Lewis et~al\mbox{.}}{2019}]%
        {cole2019understanding}
\bibfield{author}{\bibinfo{person}{Heather Cole-Lewis}, \bibinfo{person}{Nnamdi
  Ezeanochie}, {and} \bibinfo{person}{Jennifer Turgiss}.}
  \bibinfo{year}{2019}\natexlab{}.
\newblock \showarticletitle{Understanding Health Behavior Technology
  Engagement: Pathway to Measuring Digital Behavior Change Interventions}.
\newblock \bibinfo{journal}{\emph{JMIR formative research}}
  \bibinfo{volume}{3}, \bibinfo{number}{4} (\bibinfo{year}{2019}),
  \bibinfo{pages}{e14052}.
\newblock


\bibitem[\protect\citeauthoryear{Collins}{Collins}{2018}]%
        {collins2018optimization}
\bibfield{author}{\bibinfo{person}{Linda~M Collins}.}
  \bibinfo{year}{2018}\natexlab{}.
\newblock \bibinfo{booktitle}{\emph{Optimization of Behavioral, Biobehavioral,
  and Biomedical Interventions: The Multiphase Optimization Strategy (MOST)}}.
\newblock \bibinfo{publisher}{Springer}.
\newblock


\bibitem[\protect\citeauthoryear{Comulada, Lightfoot, Swendeman, Grella, and
  Wu}{Comulada et~al\mbox{.}}{2015}]%
        {comulada2015compliance}
\bibfield{author}{\bibinfo{person}{W~Scott Comulada},
  \bibinfo{person}{Marguerita Lightfoot}, \bibinfo{person}{Dallas Swendeman},
  \bibinfo{person}{Christine Grella}, {and} \bibinfo{person}{Nancy Wu}.}
  \bibinfo{year}{2015}\natexlab{}.
\newblock \showarticletitle{Compliance to cell phone-based EMA among Latino
  youth in outpatient treatment}.
\newblock \bibinfo{journal}{\emph{Journal of ethnicity in substance abuse}}
  \bibinfo{volume}{14}, \bibinfo{number}{3} (\bibinfo{year}{2015}),
  \bibinfo{pages}{232--250}.
\newblock


\bibitem[\protect\citeauthoryear{Consolvo, Klasnja, McDonald, Avrahami,
  Froehlich, LeGrand, Libby, Mosher, and Landay}{Consolvo
  et~al\mbox{.}}{2008a}]%
        {consolvo2008flowers}
\bibfield{author}{\bibinfo{person}{Sunny Consolvo}, \bibinfo{person}{Predrag
  Klasnja}, \bibinfo{person}{David~W McDonald}, \bibinfo{person}{Daniel
  Avrahami}, \bibinfo{person}{Jon Froehlich}, \bibinfo{person}{Louis LeGrand},
  \bibinfo{person}{Ryan Libby}, \bibinfo{person}{Keith Mosher}, {and}
  \bibinfo{person}{James~A Landay}.} \bibinfo{year}{2008}\natexlab{a}.
\newblock \showarticletitle{Flowers or a robot army?: encouraging awareness \&
  activity with personal, mobile displays}. In
  \bibinfo{booktitle}{\emph{Proceedings of the 10th international conference on
  Ubiquitous computing}}. ACM, \bibinfo{pages}{54--63}.
\newblock


\bibitem[\protect\citeauthoryear{Consolvo, McDonald, and Landay}{Consolvo
  et~al\mbox{.}}{2009}]%
        {consolvo2009theory}
\bibfield{author}{\bibinfo{person}{Sunny Consolvo}, \bibinfo{person}{David~W
  McDonald}, {and} \bibinfo{person}{James~A Landay}.}
  \bibinfo{year}{2009}\natexlab{}.
\newblock \showarticletitle{Theory-driven design strategies for technologies
  that support behavior change in everyday life}. In
  \bibinfo{booktitle}{\emph{Proceedings of the SIGCHI conference on human
  factors in computing systems}}. ACM, \bibinfo{pages}{405--414}.
\newblock


\bibitem[\protect\citeauthoryear{Consolvo, McDonald, Toscos, Chen, Froehlich,
  Harrison, Klasnja, LaMarca, LeGrand, Libby, et~al\mbox{.}}{Consolvo
  et~al\mbox{.}}{2008b}]%
        {consolvo2008activity}
\bibfield{author}{\bibinfo{person}{Sunny Consolvo}, \bibinfo{person}{David~W
  McDonald}, \bibinfo{person}{Tammy Toscos}, \bibinfo{person}{Mike~Y Chen},
  \bibinfo{person}{Jon Froehlich}, \bibinfo{person}{Beverly Harrison},
  \bibinfo{person}{Predrag Klasnja}, \bibinfo{person}{Anthony LaMarca},
  \bibinfo{person}{Louis LeGrand}, \bibinfo{person}{Ryan Libby},
  {et~al\mbox{.}}} \bibinfo{year}{2008}\natexlab{b}.
\newblock \showarticletitle{Activity sensing in the wild: a field trial of
  ubifit garden}. In \bibinfo{booktitle}{\emph{Proceedings of the twenty-sixth
  annual SIGCHI conference on Human factors in computing systems}}. ACM,
  \bibinfo{pages}{1797--1806}.
\newblock


\bibitem[\protect\citeauthoryear{Cordeiro, Epstein, Thomaz, Bales, Jagannathan,
  Abowd, and Fogarty}{Cordeiro et~al\mbox{.}}{2015}]%
        {cordeiro2015barriers}
\bibfield{author}{\bibinfo{person}{Felicia Cordeiro}, \bibinfo{person}{Daniel~A
  Epstein}, \bibinfo{person}{Edison Thomaz}, \bibinfo{person}{Elizabeth Bales},
  \bibinfo{person}{Arvind~K Jagannathan}, \bibinfo{person}{Gregory~D Abowd},
  {and} \bibinfo{person}{James Fogarty}.} \bibinfo{year}{2015}\natexlab{}.
\newblock \showarticletitle{Barriers and negative nudges: Exploring challenges
  in food journaling}. In \bibinfo{booktitle}{\emph{Proceedings of the 33rd
  Annual ACM Conference on Human Factors in Computing Systems}}. ACM,
  \bibinfo{pages}{1159--1162}.
\newblock


\bibitem[\protect\citeauthoryear{Cranford, Shrout, Iida, Rafaeli, Yip, and
  Bolger}{Cranford et~al\mbox{.}}{2006}]%
        {cranford2006procedure}
\bibfield{author}{\bibinfo{person}{James~A Cranford},
  \bibinfo{person}{Patrick~E Shrout}, \bibinfo{person}{Masumi Iida},
  \bibinfo{person}{Eshkol Rafaeli}, \bibinfo{person}{Tiffany Yip}, {and}
  \bibinfo{person}{Niall Bolger}.} \bibinfo{year}{2006}\natexlab{}.
\newblock \showarticletitle{A procedure for evaluating sensitivity to
  within-person change: Can mood measures in diary studies detect change
  reliably?}
\newblock \bibinfo{journal}{\emph{Personality and Social Psychology Bulletin}}
  \bibinfo{volume}{32}, \bibinfo{number}{7} (\bibinfo{year}{2006}),
  \bibinfo{pages}{917--929}.
\newblock


\bibitem[\protect\citeauthoryear{Csikszentmihalyi and Larson}{Csikszentmihalyi
  and Larson}{2014}]%
        {csikszentmihalyi2014validity}
\bibfield{author}{\bibinfo{person}{Mihaly Csikszentmihalyi} {and}
  \bibinfo{person}{Reed Larson}.} \bibinfo{year}{2014}\natexlab{}.
\newblock \showarticletitle{Validity and reliability of the experience-sampling
  method}.
\newblock In \bibinfo{booktitle}{\emph{Flow and the foundations of positive
  psychology}}. \bibinfo{publisher}{Springer}, \bibinfo{pages}{35--54}.
\newblock


\bibitem[\protect\citeauthoryear{Dallery, Cassidy, and Raiff}{Dallery
  et~al\mbox{.}}{2013}]%
        {dallery2013single}
\bibfield{author}{\bibinfo{person}{Jesse Dallery}, \bibinfo{person}{Rachel~N
  Cassidy}, {and} \bibinfo{person}{Bethany~R Raiff}.}
  \bibinfo{year}{2013}\natexlab{}.
\newblock \showarticletitle{Single-case experimental designs to evaluate novel
  technology-based health interventions}.
\newblock \bibinfo{journal}{\emph{Journal of medical Internet research}}
  \bibinfo{volume}{15}, \bibinfo{number}{2} (\bibinfo{year}{2013}).
\newblock


\bibitem[\protect\citeauthoryear{den Bakker, Schaafsma, van~der Meij,
  Meijerink, van~den Heuvel, Baan, Davids, Scholten, van~der Meij, van Baal,
  et~al\mbox{.}}{den Bakker et~al\mbox{.}}{2019}]%
        {den2019electronic}
\bibfield{author}{\bibinfo{person}{Chantal~M den Bakker},
  \bibinfo{person}{Frederieke~G Schaafsma}, \bibinfo{person}{Eva van~der Meij},
  \bibinfo{person}{Wilhelmus~JHJ Meijerink}, \bibinfo{person}{Baukje van~den
  Heuvel}, \bibinfo{person}{Astrid~H Baan}, \bibinfo{person}{Paul~HP Davids},
  \bibinfo{person}{Petrus~C Scholten}, \bibinfo{person}{Suzan van~der Meij},
  \bibinfo{person}{W~Marchien van Baal}, {et~al\mbox{.}}}
  \bibinfo{year}{2019}\natexlab{}.
\newblock \showarticletitle{Electronic Health Program to Empower Patients in
  Returning to Normal Activities After General Surgical and Gynecological
  Procedures: Intervention Mapping as a Useful Method for Further Development}.
\newblock \bibinfo{journal}{\emph{Journal of medical Internet research}}
  \bibinfo{volume}{21}, \bibinfo{number}{2} (\bibinfo{year}{2019}),
  \bibinfo{pages}{e9938}.
\newblock


\bibitem[\protect\citeauthoryear{Deterding}{Deterding}{2012}]%
        {deterding2012gamification}
\bibfield{author}{\bibinfo{person}{Sebastian Deterding}.}
  \bibinfo{year}{2012}\natexlab{}.
\newblock \showarticletitle{Gamification: designing for motivation}.
\newblock \bibinfo{journal}{\emph{interactions}} \bibinfo{volume}{19},
  \bibinfo{number}{4} (\bibinfo{year}{2012}), \bibinfo{pages}{14--17}.
\newblock


\bibitem[\protect\citeauthoryear{Deterding, Dixon, Khaled, and Nacke}{Deterding
  et~al\mbox{.}}{2011}]%
        {deterding2011game}
\bibfield{author}{\bibinfo{person}{Sebastian Deterding}, \bibinfo{person}{Dan
  Dixon}, \bibinfo{person}{Rilla Khaled}, {and} \bibinfo{person}{Lennart
  Nacke}.} \bibinfo{year}{2011}\natexlab{}.
\newblock \showarticletitle{From game design elements to gamefulness: defining
  gamification}. In \bibinfo{booktitle}{\emph{Proceedings of the 15th
  international academic MindTrek conference: Envisioning future media
  environments}}. ACM, \bibinfo{pages}{9--15}.
\newblock


\bibitem[\protect\citeauthoryear{Dorsey, McConnell, Shaw, Trister, Friend,
  et~al\mbox{.}}{Dorsey et~al\mbox{.}}{2017}]%
        {dorsey2017use}
\bibfield{author}{\bibinfo{person}{E~Ray Dorsey}, \bibinfo{person}{Michael~V
  McConnell}, \bibinfo{person}{Stanley~Y Shaw}, \bibinfo{person}{Andrew~D
  Trister}, \bibinfo{person}{Stephen~H Friend}, {et~al\mbox{.}}}
  \bibinfo{year}{2017}\natexlab{}.
\newblock \showarticletitle{The use of smartphones for health research}.
\newblock \bibinfo{journal}{\emph{Academic Medicine}} \bibinfo{volume}{92},
  \bibinfo{number}{2} (\bibinfo{year}{2017}), \bibinfo{pages}{157--160}.
\newblock


\bibitem[\protect\citeauthoryear{Edwards, Caton, Lumsden, Rivas, Steed,
  Pirunsarn, Jumbe, Newby, Shenvi, Mazumdar, et~al\mbox{.}}{Edwards
  et~al\mbox{.}}{2018}]%
        {edwards2018creating}
\bibfield{author}{\bibinfo{person}{Elizabeth~A Edwards}, \bibinfo{person}{Hope
  Caton}, \bibinfo{person}{Jim Lumsden}, \bibinfo{person}{Carol Rivas},
  \bibinfo{person}{Liz Steed}, \bibinfo{person}{Yutthana Pirunsarn},
  \bibinfo{person}{Sandra Jumbe}, \bibinfo{person}{Chris Newby},
  \bibinfo{person}{Aditi Shenvi}, \bibinfo{person}{Samaresh Mazumdar},
  {et~al\mbox{.}}} \bibinfo{year}{2018}\natexlab{}.
\newblock \showarticletitle{Creating a theoretically grounded, gamified health
  app: lessons from developing the Cigbreak smoking cessation mobile phone
  game}.
\newblock \bibinfo{journal}{\emph{JMIR serious games}} \bibinfo{volume}{6},
  \bibinfo{number}{4} (\bibinfo{year}{2018}), \bibinfo{pages}{e10252}.
\newblock


\bibitem[\protect\citeauthoryear{Epstein, Ping, Fogarty, and Munson}{Epstein
  et~al\mbox{.}}{2015}]%
        {epstein2015lived}
\bibfield{author}{\bibinfo{person}{Daniel~A Epstein}, \bibinfo{person}{An
  Ping}, \bibinfo{person}{James Fogarty}, {and} \bibinfo{person}{Sean~A
  Munson}.} \bibinfo{year}{2015}\natexlab{}.
\newblock \showarticletitle{A lived informatics model of personal informatics}.
  In \bibinfo{booktitle}{\emph{Proceedings of the 2015 ACM International Joint
  Conference on Pervasive and Ubiquitous Computing}}. ACM,
  \bibinfo{pages}{731--742}.
\newblock


\bibitem[\protect\citeauthoryear{Eysenbach}{Eysenbach}{2005}]%
        {eysenbach2005law}
\bibfield{author}{\bibinfo{person}{Gunther Eysenbach}.}
  \bibinfo{year}{2005}\natexlab{}.
\newblock \showarticletitle{The law of attrition}.
\newblock \bibinfo{journal}{\emph{Journal of medical Internet research}}
  \bibinfo{volume}{7}, \bibinfo{number}{1} (\bibinfo{year}{2005}).
\newblock


\bibitem[\protect\citeauthoryear{Ferster and Skinner}{Ferster and
  Skinner}{1957}]%
        {ferster1957schedules}
\bibfield{author}{\bibinfo{person}{Charles~B Ferster} {and}
  \bibinfo{person}{Burrhus~Frederic Skinner}.} \bibinfo{year}{1957}\natexlab{}.
\newblock \showarticletitle{Schedules of reinforcement.}
\newblock  (\bibinfo{year}{1957}).
\newblock


\bibitem[\protect\citeauthoryear{{Fitbit, Inc.}}{{Fitbit, Inc.}}{2013}]%
        {fitbit}
\bibfield{author}{\bibinfo{person}{{Fitbit, Inc.}}}
  \bibinfo{year}{2013}\natexlab{}.
\newblock \bibinfo{howpublished}{http://www.fitbit.com/}.
\newblock
\newblock
\shownote{[Online; accessed 19 March 2013].}


\bibitem[\protect\citeauthoryear{Fogg}{Fogg}{2009}]%
        {fogg2009behavior}
\bibfield{author}{\bibinfo{person}{BJ Fogg}.} \bibinfo{year}{2009}\natexlab{}.
\newblock \showarticletitle{A behavior model for persuasive design}. In
  \bibinfo{booktitle}{\emph{Proceedings of the 4th international Conference on
  Persuasive Technology}}. ACM, \bibinfo{pages}{40}.
\newblock


\bibitem[\protect\citeauthoryear{Geraghty, Mu{\~n}oz, Yardley, Mc~Sharry,
  Little, and Moore}{Geraghty et~al\mbox{.}}{2016}]%
        {geraghty2016developing}
\bibfield{author}{\bibinfo{person}{Adam~WA Geraghty},
  \bibinfo{person}{Ricardo~F Mu{\~n}oz}, \bibinfo{person}{Lucy Yardley},
  \bibinfo{person}{Jennifer Mc~Sharry}, \bibinfo{person}{Paul Little}, {and}
  \bibinfo{person}{Michael Moore}.} \bibinfo{year}{2016}\natexlab{}.
\newblock \showarticletitle{Developing an unguided Internet-delivered
  intervention for emotional distress in primary care patients: Applying common
  factor and person-based approaches}.
\newblock \bibinfo{journal}{\emph{JMIR mental health}} \bibinfo{volume}{3},
  \bibinfo{number}{4} (\bibinfo{year}{2016}), \bibinfo{pages}{e53}.
\newblock


\bibitem[\protect\citeauthoryear{Glanz, Rimer, and Viswanath}{Glanz
  et~al\mbox{.}}{2008}]%
        {glanz2008health}
\bibfield{author}{\bibinfo{person}{Karen Glanz}, \bibinfo{person}{Barbara~K
  Rimer}, {and} \bibinfo{person}{Kasisomayajula Viswanath}.}
  \bibinfo{year}{2008}\natexlab{}.
\newblock \bibinfo{booktitle}{\emph{Health behavior and health education:
  theory, research, and practice}}.
\newblock \bibinfo{publisher}{John Wiley \& Sons}.
\newblock


\bibitem[\protect\citeauthoryear{Grant, Stewart, O'Connor, Blackwell, and
  Conrod}{Grant et~al\mbox{.}}{2007}]%
        {grant2007psychometric}
\bibfield{author}{\bibinfo{person}{Valerie~V Grant}, \bibinfo{person}{Sherry~H
  Stewart}, \bibinfo{person}{Roisin~M O'Connor}, \bibinfo{person}{Ekin
  Blackwell}, {and} \bibinfo{person}{Patricia~J Conrod}.}
  \bibinfo{year}{2007}\natexlab{}.
\newblock \showarticletitle{Psychometric evaluation of the five-factor Modified
  Drinking Motives Questionnaire-Revised in undergraduates}.
\newblock \bibinfo{journal}{\emph{Addictive behaviors}} \bibinfo{volume}{32},
  \bibinfo{number}{11} (\bibinfo{year}{2007}), \bibinfo{pages}{2611--2632}.
\newblock


\bibitem[\protect\citeauthoryear{Greenhalgh and Russell}{Greenhalgh and
  Russell}{2010}]%
        {greenhalgh2010evaluations}
\bibfield{author}{\bibinfo{person}{Trisha Greenhalgh} {and}
  \bibinfo{person}{Jill Russell}.} \bibinfo{year}{2010}\natexlab{}.
\newblock \showarticletitle{Why do evaluations of eHealth programs fail? An
  alternative set of guiding principles}.
\newblock \bibinfo{journal}{\emph{PLoS medicine}} \bibinfo{volume}{7},
  \bibinfo{number}{11} (\bibinfo{year}{2010}), \bibinfo{pages}{e1000360}.
\newblock


\bibitem[\protect\citeauthoryear{Gruber, Dahlgren, Sagar, G{\"o}nenc, and
  Killgore}{Gruber et~al\mbox{.}}{2012}]%
        {gruber2012age}
\bibfield{author}{\bibinfo{person}{Staci~A Gruber},
  \bibinfo{person}{Mary~Kathryn Dahlgren}, \bibinfo{person}{Kelly~A Sagar},
  \bibinfo{person}{Atilla G{\"o}nenc}, {and} \bibinfo{person}{William~DS
  Killgore}.} \bibinfo{year}{2012}\natexlab{}.
\newblock \showarticletitle{Age of onset of marijuana use impacts inhibitory
  processing}.
\newblock \bibinfo{journal}{\emph{Neuroscience letters}} \bibinfo{volume}{511},
  \bibinfo{number}{2} (\bibinfo{year}{2012}), \bibinfo{pages}{89--94}.
\newblock


\bibitem[\protect\citeauthoryear{Gulotta, Forlizzi, Yang, and Newman}{Gulotta
  et~al\mbox{.}}{2016}]%
        {gulotta2016fostering}
\bibfield{author}{\bibinfo{person}{Rebecca Gulotta}, \bibinfo{person}{Jodi
  Forlizzi}, \bibinfo{person}{Rayoung Yang}, {and} \bibinfo{person}{Mark~Wah
  Newman}.} \bibinfo{year}{2016}\natexlab{}.
\newblock \showarticletitle{Fostering engagement with personal informatics
  systems}. In \bibinfo{booktitle}{\emph{Proceedings of the 2016 ACM Conference
  on Designing Interactive Systems}}. ACM, \bibinfo{pages}{286--300}.
\newblock


\bibitem[\protect\citeauthoryear{Hedden}{Hedden}{2015}]%
        {hedden2015behavioral}
\bibfield{author}{\bibinfo{person}{Sarra~L Hedden}.}
  \bibinfo{year}{2015}\natexlab{}.
\newblock \bibinfo{booktitle}{\emph{Behavioral health trends in the United
  States: results from the 2014 National Survey on Drug Use and Health}}.
\newblock \bibinfo{publisher}{Substance Abuse and Mental Health Services
  Administration, Department of Heath \& Human Services}.
\newblock


\bibitem[\protect\citeauthoryear{Hekler, Klasnja, Froehlich, and Buman}{Hekler
  et~al\mbox{.}}{2013}]%
        {hekler2013mind}
\bibfield{author}{\bibinfo{person}{Eric~B Hekler}, \bibinfo{person}{Predrag
  Klasnja}, \bibinfo{person}{Jon~E Froehlich}, {and} \bibinfo{person}{Matthew~P
  Buman}.} \bibinfo{year}{2013}\natexlab{}.
\newblock \showarticletitle{Mind the theoretical gap: interpreting, using, and
  developing behavioral theory in HCI research}. In
  \bibinfo{booktitle}{\emph{Proceedings of the SIGCHI Conference on Human
  Factors in Computing Systems}}. ACM, \bibinfo{pages}{3307--3316}.
\newblock


\bibitem[\protect\citeauthoryear{Hekler, Klasnja, Riley, Buman, Huberty,
  Rivera, and Martin}{Hekler et~al\mbox{.}}{2016}]%
        {hekler2016agile}
\bibfield{author}{\bibinfo{person}{Eric~B Hekler}, \bibinfo{person}{Predrag
  Klasnja}, \bibinfo{person}{William~T Riley}, \bibinfo{person}{Matthew~P
  Buman}, \bibinfo{person}{Jennifer Huberty}, \bibinfo{person}{Daniel~E
  Rivera}, {and} \bibinfo{person}{Cesar~A Martin}.}
  \bibinfo{year}{2016}\natexlab{}.
\newblock \showarticletitle{Agile science: creating useful products for
  behavior change in the real world}.
\newblock \bibinfo{journal}{\emph{Translational behavioral medicine}}
  \bibinfo{volume}{6}, \bibinfo{number}{2} (\bibinfo{year}{2016}),
  \bibinfo{pages}{317--328}.
\newblock


\bibitem[\protect\citeauthoryear{Helander, Kaipainen, Korhonen, and
  Wansink}{Helander et~al\mbox{.}}{2014}]%
        {Helander2014}
\bibfield{author}{\bibinfo{person}{Elina Helander}, \bibinfo{person}{Kirsikka
  Kaipainen}, \bibinfo{person}{Ilkka Korhonen}, {and} \bibinfo{person}{Brian
  Wansink}.} \bibinfo{year}{2014}\natexlab{}.
\newblock \showarticletitle{{Factors related to sustained use of a free mobile
  app for dietary self-monitoring with photography and peer feedback:
  retrospective cohort study.}}
\newblock \bibinfo{journal}{\emph{Journal of medical Internet research}}
  \bibinfo{volume}{16}, \bibinfo{number}{4} (\bibinfo{date}{apr}
  \bibinfo{year}{2014}), \bibinfo{pages}{e109}.
\newblock
\showISSN{1438-8871}
\urldef\tempurl%
\url{https://doi.org/10.2196/jmir.3084}
\showDOI{\tempurl}


\bibitem[\protect\citeauthoryear{Herrnstein}{Herrnstein}{1970}]%
        {herrnstein1970law}
\bibfield{author}{\bibinfo{person}{Richard~J Herrnstein}.}
  \bibinfo{year}{1970}\natexlab{}.
\newblock \showarticletitle{On the law of effect 1}.
\newblock \bibinfo{journal}{\emph{Journal of the experimental analysis of
  behavior}} \bibinfo{volume}{13}, \bibinfo{number}{2} (\bibinfo{year}{1970}),
  \bibinfo{pages}{243--266}.
\newblock


\bibitem[\protect\citeauthoryear{Hodos}{Hodos}{1961}]%
        {hodos1961progressive}
\bibfield{author}{\bibinfo{person}{William Hodos}.}
  \bibinfo{year}{1961}\natexlab{}.
\newblock \showarticletitle{Progressive ratio as a measure of reward strength}.
\newblock \bibinfo{journal}{\emph{Science}} \bibinfo{volume}{134},
  \bibinfo{number}{3483} (\bibinfo{year}{1961}), \bibinfo{pages}{943--944}.
\newblock


\bibitem[\protect\citeauthoryear{Hoeppner, Schick, Kelly, Hoeppner, Bergman,
  and Kelly}{Hoeppner et~al\mbox{.}}{2017}]%
        {hoeppner2017there}
\bibfield{author}{\bibinfo{person}{Bettina~B Hoeppner},
  \bibinfo{person}{Melissa~R Schick}, \bibinfo{person}{Lourah~M Kelly},
  \bibinfo{person}{Susanne~S Hoeppner}, \bibinfo{person}{Brandon Bergman},
  {and} \bibinfo{person}{John~F Kelly}.} \bibinfo{year}{2017}\natexlab{}.
\newblock \showarticletitle{There is an app for that--Or is there? A content
  analysis of publicly available smartphone apps for managing alcohol use}.
\newblock \bibinfo{journal}{\emph{Journal of substance abuse treatment}}
  \bibinfo{volume}{82} (\bibinfo{year}{2017}), \bibinfo{pages}{67--73}.
\newblock


\bibitem[\protect\citeauthoryear{Hoeppner, Stout, Jackson, and
  Barnett}{Hoeppner et~al\mbox{.}}{2010}]%
        {hoeppner2010good}
\bibfield{author}{\bibinfo{person}{Bettina~B Hoeppner},
  \bibinfo{person}{Robert~L Stout}, \bibinfo{person}{Kristina~M Jackson}, {and}
  \bibinfo{person}{Nancy~P Barnett}.} \bibinfo{year}{2010}\natexlab{}.
\newblock \showarticletitle{How good is fine-grained Timeline Follow-back data?
  Comparing 30-day TLFB and repeated 7-day TLFB alcohol consumption reports on
  the person and daily level}.
\newblock \bibinfo{journal}{\emph{Addictive Behaviors}} \bibinfo{volume}{35},
  \bibinfo{number}{12} (\bibinfo{year}{2010}), \bibinfo{pages}{1138--1143}.
\newblock


\bibitem[\protect\citeauthoryear{Hoyle, Stephenson, Palmgreen, Lorch, and
  Donohew}{Hoyle et~al\mbox{.}}{2002}]%
        {hoyle2002reliability}
\bibfield{author}{\bibinfo{person}{Rick~H Hoyle}, \bibinfo{person}{Michael~T
  Stephenson}, \bibinfo{person}{Philip Palmgreen},
  \bibinfo{person}{Elizabeth~Pugzles Lorch}, {and} \bibinfo{person}{R~Lewis
  Donohew}.} \bibinfo{year}{2002}\natexlab{}.
\newblock \showarticletitle{Reliability and validity of a brief measure of
  sensation seeking}.
\newblock \bibinfo{journal}{\emph{Personality and individual differences}}
  \bibinfo{volume}{32}, \bibinfo{number}{3} (\bibinfo{year}{2002}),
  \bibinfo{pages}{401--414}.
\newblock


\bibitem[\protect\citeauthoryear{Hursh}{Hursh}{1980}]%
        {hursh1980economic}
\bibfield{author}{\bibinfo{person}{Steven~R Hursh}.}
  \bibinfo{year}{1980}\natexlab{}.
\newblock \showarticletitle{Economic concepts for the analysis of behavior}.
\newblock \bibinfo{journal}{\emph{Journal of the experimental analysis of
  behavior}} \bibinfo{volume}{34}, \bibinfo{number}{2} (\bibinfo{year}{1980}),
  \bibinfo{pages}{219--238}.
\newblock


\bibitem[\protect\citeauthoryear{Inc.}{Inc.}{2018}]%
        {appleactivetask}
\bibfield{author}{\bibinfo{person}{Apple Inc.}}
  \bibinfo{year}{2018}\natexlab{}.
\newblock \bibinfo{title}{Active tasks in Apple Research Kit}.
\newblock
\newblock
\urldef\tempurl%
\url{http://researchkit.org/docs/docs/ActiveTasks/ActiveTasks.html}
\showURL{%
\tempurl}


\bibitem[\protect\citeauthoryear{Johnson, Deterding, Kuhn, Staneva, Stoyanov,
  and Hides}{Johnson et~al\mbox{.}}{2016}]%
        {johnson2016gamification}
\bibfield{author}{\bibinfo{person}{Daniel Johnson}, \bibinfo{person}{Sebastian
  Deterding}, \bibinfo{person}{Kerri-Ann Kuhn}, \bibinfo{person}{Aleksandra
  Staneva}, \bibinfo{person}{Stoyan Stoyanov}, {and} \bibinfo{person}{Leanne
  Hides}.} \bibinfo{year}{2016}\natexlab{}.
\newblock \showarticletitle{Gamification for health and wellbeing: A systematic
  review of the literature}.
\newblock \bibinfo{journal}{\emph{Internet Interventions}}  \bibinfo{volume}{6}
  (\bibinfo{year}{2016}), \bibinfo{pages}{89--106}.
\newblock


\bibitem[\protect\citeauthoryear{Johnson, O'Malley, Bachman, and
  Schulenberg}{Johnson et~al\mbox{.}}{2006}]%
        {johnson2006monitoring}
\bibfield{author}{\bibinfo{person}{Lloyd~D Johnson}, \bibinfo{person}{Patrick~M
  O'Malley}, \bibinfo{person}{Jerald~G Bachman}, {and} \bibinfo{person}{John~E
  Schulenberg}.} \bibinfo{year}{2006}\natexlab{}.
\newblock \showarticletitle{Monitoring the Future: National Results on
  Adolescent Drug Use. Overview of Key Findings 2005. NIH Publication No.
  06-5882.}
\newblock \bibinfo{journal}{\emph{National Institute on Drug Abuse (NIDA)}}
  (\bibinfo{year}{2006}).
\newblock


\bibitem[\protect\citeauthoryear{Joiner~Jr and Katz}{Joiner~Jr and
  Katz}{1999}]%
        {joiner1999contagion}
\bibfield{author}{\bibinfo{person}{Thomas~E Joiner~Jr} {and}
  \bibinfo{person}{Jennifer Katz}.} \bibinfo{year}{1999}\natexlab{}.
\newblock \showarticletitle{Contagion of depressive symptoms and mood:
  Meta-analytic review and explanations from cognitive, behavioral, and
  interpersonal viewpoints}.
\newblock \bibinfo{journal}{\emph{Clinical Psychology: Science and Practice}}
  \bibinfo{volume}{6}, \bibinfo{number}{2} (\bibinfo{year}{1999}),
  \bibinfo{pages}{149--164}.
\newblock


\bibitem[\protect\citeauthoryear{Jones, Remmerswaal, Verveer, Robinson,
  Franken, Wen, and Field}{Jones et~al\mbox{.}}{2019}]%
        {jones2019compliance}
\bibfield{author}{\bibinfo{person}{Andrew Jones}, \bibinfo{person}{Danielle
  Remmerswaal}, \bibinfo{person}{Ilse Verveer}, \bibinfo{person}{Eric
  Robinson}, \bibinfo{person}{Ingmar~HA Franken}, \bibinfo{person}{Cheng K~Fred
  Wen}, {and} \bibinfo{person}{Matt Field}.} \bibinfo{year}{2019}\natexlab{}.
\newblock \showarticletitle{Compliance with ecological momentary assessment
  protocols in substance users: a meta-analysis}.
\newblock \bibinfo{journal}{\emph{Addiction}} \bibinfo{volume}{114},
  \bibinfo{number}{4} (\bibinfo{year}{2019}), \bibinfo{pages}{609--619}.
\newblock


\bibitem[\protect\citeauthoryear{Julian}{Julian}{1997}]%
        {julian1997utilization}
\bibfield{author}{\bibinfo{person}{David~A Julian}.}
  \bibinfo{year}{1997}\natexlab{}.
\newblock \showarticletitle{The utilization of the logic model as a system
  level planning and evaluation device}.
\newblock \bibinfo{journal}{\emph{Evaluation and Program Planning}}
  \bibinfo{volume}{20}, \bibinfo{number}{3} (\bibinfo{year}{1997}),
  \bibinfo{pages}{251--257}.
\newblock


\bibitem[\protect\citeauthoryear{Klasnja, Hekler, Korinek, Harlow, and
  Mishra}{Klasnja et~al\mbox{.}}{2017}]%
        {klasnja2017toward}
\bibfield{author}{\bibinfo{person}{Predrag Klasnja}, \bibinfo{person}{Eric~B
  Hekler}, \bibinfo{person}{Elizabeth~V Korinek}, \bibinfo{person}{John
  Harlow}, {and} \bibinfo{person}{Sonali~R Mishra}.}
  \bibinfo{year}{2017}\natexlab{}.
\newblock \showarticletitle{Toward usable evidence: optimizing knowledge
  accumulation in HCI research on health behavior change}. In
  \bibinfo{booktitle}{\emph{Proceedings of the 2017 CHI conference on human
  factors in computing systems}}. ACM, \bibinfo{pages}{3071--3082}.
\newblock


\bibitem[\protect\citeauthoryear{Klasnja, Hekler, Shiffman, Boruvka, Almirall,
  Tewari, and Murphy}{Klasnja et~al\mbox{.}}{2015}]%
        {klasnja2015microrandomized}
\bibfield{author}{\bibinfo{person}{Predrag Klasnja}, \bibinfo{person}{Eric~B
  Hekler}, \bibinfo{person}{Saul Shiffman}, \bibinfo{person}{Audrey Boruvka},
  \bibinfo{person}{Daniel Almirall}, \bibinfo{person}{Ambuj Tewari}, {and}
  \bibinfo{person}{Susan~A Murphy}.} \bibinfo{year}{2015}\natexlab{}.
\newblock \showarticletitle{Microrandomized trials: An experimental design for
  developing just-in-time adaptive interventions.}
\newblock \bibinfo{journal}{\emph{Health Psychology}} \bibinfo{volume}{34},
  \bibinfo{number}{S} (\bibinfo{year}{2015}), \bibinfo{pages}{1220}.
\newblock


\bibitem[\protect\citeauthoryear{Kok, Schaalma, Ruiter, Van~Empelen, and
  Brug}{Kok et~al\mbox{.}}{2004}]%
        {kok2004intervention}
\bibfield{author}{\bibinfo{person}{Gerjo Kok}, \bibinfo{person}{Herman
  Schaalma}, \bibinfo{person}{Robert~AC Ruiter}, \bibinfo{person}{Pepijn
  Van~Empelen}, {and} \bibinfo{person}{Johannes Brug}.}
  \bibinfo{year}{2004}\natexlab{}.
\newblock \showarticletitle{Intervention mapping: protocol for applying health
  psychology theory to prevention programmes}.
\newblock \bibinfo{journal}{\emph{Journal of health psychology}}
  \bibinfo{volume}{9}, \bibinfo{number}{1} (\bibinfo{year}{2004}),
  \bibinfo{pages}{85--98}.
\newblock


\bibitem[\protect\citeauthoryear{Lane, Lin, Mohammod, Yang, Lu, Cardone, Ali,
  Doryab, Berke, Campbell, et~al\mbox{.}}{Lane et~al\mbox{.}}{2014}]%
        {lane2014bewell}
\bibfield{author}{\bibinfo{person}{Nicholas~D Lane}, \bibinfo{person}{Mu Lin},
  \bibinfo{person}{Mashfiqui Mohammod}, \bibinfo{person}{Xiaochao Yang},
  \bibinfo{person}{Hong Lu}, \bibinfo{person}{Giuseppe Cardone},
  \bibinfo{person}{Shahid Ali}, \bibinfo{person}{Afsaneh Doryab},
  \bibinfo{person}{Ethan Berke}, \bibinfo{person}{Andrew~T Campbell},
  {et~al\mbox{.}}} \bibinfo{year}{2014}\natexlab{}.
\newblock \showarticletitle{Bewell: Sensing sleep, physical activities and
  social interactions to promote wellbeing}.
\newblock \bibinfo{journal}{\emph{Mobile Networks and Applications}}
  \bibinfo{volume}{19}, \bibinfo{number}{3} (\bibinfo{year}{2014}),
  \bibinfo{pages}{345--359}.
\newblock


\bibitem[\protect\citeauthoryear{Lane, Mohammod, Lin, Yang, Lu, Ali, Doryab,
  Berke, Choudhury, and Campbell}{Lane et~al\mbox{.}}{2011}]%
        {lane2011bewell}
\bibfield{author}{\bibinfo{person}{Nicholas~D Lane}, \bibinfo{person}{Mashfiqui
  Mohammod}, \bibinfo{person}{Mu Lin}, \bibinfo{person}{Xiaochao Yang},
  \bibinfo{person}{Hong Lu}, \bibinfo{person}{Shahid Ali},
  \bibinfo{person}{Afsaneh Doryab}, \bibinfo{person}{Ethan Berke},
  \bibinfo{person}{Tanzeem Choudhury}, {and} \bibinfo{person}{Andrew~T
  Campbell}.} \bibinfo{year}{2011}\natexlab{}.
\newblock \showarticletitle{BeWell: A smartphone application to monitor, model
  and promote wellbeing}. In \bibinfo{booktitle}{\emph{5th International
  Conference on Pervasive Computing Technologies for Healthcare
  (PervasiveHealth2011)}}.
\newblock


\bibitem[\protect\citeauthoryear{Lazar, Koehler, Tanenbaum, and Nguyen}{Lazar
  et~al\mbox{.}}{2015}]%
        {lazar2015we}
\bibfield{author}{\bibinfo{person}{Amanda Lazar}, \bibinfo{person}{Christian
  Koehler}, \bibinfo{person}{Joshua Tanenbaum}, {and} \bibinfo{person}{David~H
  Nguyen}.} \bibinfo{year}{2015}\natexlab{}.
\newblock \showarticletitle{Why we use and abandon smart devices}. In
  \bibinfo{booktitle}{\emph{Proceedings of the 2015 ACM International Joint
  Conference on Pervasive and Ubiquitous Computing}}. ACM,
  \bibinfo{pages}{635--646}.
\newblock


\bibitem[\protect\citeauthoryear{Lee and Dey}{Lee and Dey}{2014}]%
        {lee2014real}
\bibfield{author}{\bibinfo{person}{Matthew~L Lee} {and}
  \bibinfo{person}{Anind~K Dey}.} \bibinfo{year}{2014}\natexlab{}.
\newblock \showarticletitle{Real-time feedback for improving medication
  taking}. In \bibinfo{booktitle}{\emph{Proceedings of the SIGCHI Conference on
  Human Factors in Computing Systems}}. ACM, \bibinfo{pages}{2259--2268}.
\newblock


\bibitem[\protect\citeauthoryear{Leonardelli and Lakin}{Leonardelli and
  Lakin}{2010}]%
        {leonardelli2010new}
\bibfield{author}{\bibinfo{person}{GEOFFREY~J Leonardelli} {and}
  \bibinfo{person}{Jessica~L Lakin}.} \bibinfo{year}{2010}\natexlab{}.
\newblock \showarticletitle{The new adventures of regulatory focus:
  Self-uncertainty and the quest for a diagnostic self-evaluation}.
\newblock \bibinfo{journal}{\emph{Handbook of the uncertain self}}
  (\bibinfo{year}{2010}), \bibinfo{pages}{249--263}.
\newblock


\bibitem[\protect\citeauthoryear{Lin, Mamykina, Lindtner, Delajoux, and
  Strub}{Lin et~al\mbox{.}}{2006}]%
        {lin2006fish}
\bibfield{author}{\bibinfo{person}{James~J Lin}, \bibinfo{person}{Lena
  Mamykina}, \bibinfo{person}{Silvia Lindtner}, \bibinfo{person}{Gregory
  Delajoux}, {and} \bibinfo{person}{Henry~B Strub}.}
  \bibinfo{year}{2006}\natexlab{}.
\newblock \showarticletitle{Fish`n'Steps: Encouraging physical activity with an
  interactive computer game}. In \bibinfo{booktitle}{\emph{International
  conference on ubiquitous computing}}. Springer, \bibinfo{pages}{261--278}.
\newblock


\bibitem[\protect\citeauthoryear{Lippman, Anderson~Moore, Guzman, Ryberg,
  McIntosh, Ramos, Caal, Carle, and Kuhfeld}{Lippman et~al\mbox{.}}{2014}]%
        {lippman2014flourishing}
\bibfield{author}{\bibinfo{person}{Laura~H Lippman}, \bibinfo{person}{Kristin
  Anderson~Moore}, \bibinfo{person}{Lina Guzman}, \bibinfo{person}{Renee
  Ryberg}, \bibinfo{person}{Hugh McIntosh}, \bibinfo{person}{Manica~F Ramos},
  \bibinfo{person}{Salma Caal}, \bibinfo{person}{Adam Carle}, {and}
  \bibinfo{person}{Megan Kuhfeld}.} \bibinfo{year}{2014}\natexlab{}.
\newblock \bibinfo{booktitle}{\emph{Flourishing children}}.
\newblock \bibinfo{publisher}{Springer}.
\newblock


\bibitem[\protect\citeauthoryear{Locke and Latham}{Locke and Latham}{1994}]%
        {locke1994goal}
\bibfield{author}{\bibinfo{person}{Edwin Locke} {and} \bibinfo{person}{Gary
  Latham}.} \bibinfo{year}{1994}\natexlab{}.
\newblock \showarticletitle{Goal-setting theory}.
\newblock \bibinfo{journal}{\emph{Organizational Behavior 1: Essential Theories
  of Motivation and Leadership}} (\bibinfo{year}{1994}),
  \bibinfo{pages}{159--183}.
\newblock


\bibitem[\protect\citeauthoryear{Locke and Latham}{Locke and Latham}{2006}]%
        {locke2006new}
\bibfield{author}{\bibinfo{person}{Edwin~A Locke} {and} \bibinfo{person}{Gary~P
  Latham}.} \bibinfo{year}{2006}\natexlab{}.
\newblock \showarticletitle{New directions in goal-setting theory}.
\newblock \bibinfo{journal}{\emph{Current directions in psychological science}}
  \bibinfo{volume}{15}, \bibinfo{number}{5} (\bibinfo{year}{2006}),
  \bibinfo{pages}{265--268}.
\newblock


\bibitem[\protect\citeauthoryear{Lopez-Larson, Bogorodzki, Rogowska, McGlade,
  King, Terry, and Yurgelun-Todd}{Lopez-Larson et~al\mbox{.}}{2011}]%
        {lopez2011altered}
\bibfield{author}{\bibinfo{person}{Melissa~P Lopez-Larson},
  \bibinfo{person}{Piotr Bogorodzki}, \bibinfo{person}{Jadwiga Rogowska},
  \bibinfo{person}{Erin McGlade}, \bibinfo{person}{Jace~B King},
  \bibinfo{person}{Janine Terry}, {and} \bibinfo{person}{Deborah
  Yurgelun-Todd}.} \bibinfo{year}{2011}\natexlab{}.
\newblock \showarticletitle{Altered prefrontal and insular cortical thickness
  in adolescent marijuana users}.
\newblock \bibinfo{journal}{\emph{Behavioural brain research}}
  \bibinfo{volume}{220}, \bibinfo{number}{1} (\bibinfo{year}{2011}),
  \bibinfo{pages}{164--172}.
\newblock


\bibitem[\protect\citeauthoryear{Lynn}{Lynn}{2001}]%
        {lynn2001impact}
\bibfield{author}{\bibinfo{person}{Peter Lynn}.}
  \bibinfo{year}{2001}\natexlab{}.
\newblock \showarticletitle{The impact of incentives on response rates to
  personal interview surveys: Role and perceptions of interviewers.}
\newblock \bibinfo{journal}{\emph{International Journal of Public Opinion
  Research}} (\bibinfo{year}{2001}).
\newblock


\bibitem[\protect\citeauthoryear{Majeed-Ariss, Baildam, Campbell, Chieng,
  Fallon, Hall, McDonagh, Stones, Thomson, and Swallow}{Majeed-Ariss
  et~al\mbox{.}}{2015}]%
        {majeed2015apps}
\bibfield{author}{\bibinfo{person}{Rabiya Majeed-Ariss},
  \bibinfo{person}{Eileen Baildam}, \bibinfo{person}{Malcolm Campbell},
  \bibinfo{person}{Alice Chieng}, \bibinfo{person}{Debbie Fallon},
  \bibinfo{person}{Andrew Hall}, \bibinfo{person}{Janet~E McDonagh},
  \bibinfo{person}{Simon~R Stones}, \bibinfo{person}{Wendy Thomson}, {and}
  \bibinfo{person}{Veronica Swallow}.} \bibinfo{year}{2015}\natexlab{}.
\newblock \showarticletitle{Apps and adolescents: a systematic review of
  adolescents’ use of mobile phone and tablet apps that support personal
  management of their chronic or long-term physical conditions}.
\newblock \bibinfo{journal}{\emph{Journal of medical Internet research}}
  \bibinfo{volume}{17}, \bibinfo{number}{12} (\bibinfo{year}{2015}).
\newblock


\bibitem[\protect\citeauthoryear{Mariakakis, Parsi, Patel, and
  Wobbrock}{Mariakakis et~al\mbox{.}}{2018}]%
        {mariakakis2018drunk}
\bibfield{author}{\bibinfo{person}{Alex Mariakakis}, \bibinfo{person}{Sayna
  Parsi}, \bibinfo{person}{Shwetak~N Patel}, {and} \bibinfo{person}{Jacob~O
  Wobbrock}.} \bibinfo{year}{2018}\natexlab{}.
\newblock \showarticletitle{Drunk User Interfaces: Determining Blood Alcohol
  Level through Everyday Smartphone Tasks}. In
  \bibinfo{booktitle}{\emph{Proceedings of the 2018 CHI Conference on Human
  Factors in Computing Systems}}. ACM, \bibinfo{pages}{234}.
\newblock


\bibitem[\protect\citeauthoryear{Marques and McKnight}{Marques and
  McKnight}{2009}]%
        {marques2009field}
\bibfield{author}{\bibinfo{person}{Paul~R Marques} {and}
  \bibinfo{person}{A~Scott McKnight}.} \bibinfo{year}{2009}\natexlab{}.
\newblock \showarticletitle{Field and laboratory alcohol detection with 2 types
  of transdermal devices}.
\newblock \bibinfo{journal}{\emph{Alcoholism: Clinical and Experimental
  Research}} \bibinfo{volume}{33}, \bibinfo{number}{4} (\bibinfo{year}{2009}),
  \bibinfo{pages}{703--711}.
\newblock


\bibitem[\protect\citeauthoryear{Meng, Hussain, Mohr, Czerwinski, and
  Zhang}{Meng et~al\mbox{.}}{2018}]%
        {meng2018exploring}
\bibfield{author}{\bibinfo{person}{Jingbo Meng}, \bibinfo{person}{Syed~Ali
  Hussain}, \bibinfo{person}{David~C Mohr}, \bibinfo{person}{Mary Czerwinski},
  {and} \bibinfo{person}{Mi Zhang}.} \bibinfo{year}{2018}\natexlab{}.
\newblock \showarticletitle{Exploring User Needs for a Mobile
  Behavioral-Sensing Technology for Depression Management: Qualitative Study}.
\newblock \bibinfo{journal}{\emph{Journal of medical Internet research}}
  \bibinfo{volume}{20}, \bibinfo{number}{7} (\bibinfo{year}{2018}),
  \bibinfo{pages}{e10139}.
\newblock


\bibitem[\protect\citeauthoryear{Metting, Schrage, Kocks, Sanderman, and
  van~der Molen}{Metting et~al\mbox{.}}{2018}]%
        {metting2018assessing}
\bibfield{author}{\bibinfo{person}{Esther Metting},
  \bibinfo{person}{Aaltje~Jantine Schrage}, \bibinfo{person}{Janwillem~WH
  Kocks}, \bibinfo{person}{Robbert Sanderman}, {and} \bibinfo{person}{Thys
  van~der Molen}.} \bibinfo{year}{2018}\natexlab{}.
\newblock \showarticletitle{Assessing the needs and perspectives of patients
  with asthma and chronic obstructive pulmonary disease on patient web portals:
  focus group study}.
\newblock \bibinfo{journal}{\emph{JMIR formative research}}
  \bibinfo{volume}{2}, \bibinfo{number}{2} (\bibinfo{year}{2018}),
  \bibinfo{pages}{e22}.
\newblock


\bibitem[\protect\citeauthoryear{Michie, Atkins, and West}{Michie
  et~al\mbox{.}}{2014a}]%
        {michie2014behavior}
\bibfield{author}{\bibinfo{person}{S Michie}, \bibinfo{person}{L Atkins}, {and}
  \bibinfo{person}{R West}.} \bibinfo{year}{2014}\natexlab{a}.
\newblock \showarticletitle{The behavior change wheel: a guide to designing
  interventions}.
\newblock \bibinfo{journal}{\emph{Great Britain: Silverback Publishing}}
  (\bibinfo{year}{2014}).
\newblock


\bibitem[\protect\citeauthoryear{Michie, Richardson, Johnston, Abraham,
  Francis, Hardeman, Eccles, Cane, and Wood}{Michie et~al\mbox{.}}{2013}]%
        {michie2013behavior}
\bibfield{author}{\bibinfo{person}{Susan Michie}, \bibinfo{person}{Michelle
  Richardson}, \bibinfo{person}{Marie Johnston}, \bibinfo{person}{Charles
  Abraham}, \bibinfo{person}{Jill Francis}, \bibinfo{person}{Wendy Hardeman},
  \bibinfo{person}{Martin~P Eccles}, \bibinfo{person}{James Cane}, {and}
  \bibinfo{person}{Caroline~E Wood}.} \bibinfo{year}{2013}\natexlab{}.
\newblock \showarticletitle{The behavior change technique taxonomy (v1) of 93
  hierarchically clustered techniques: building an international consensus for
  the reporting of behavior change interventions}.
\newblock \bibinfo{journal}{\emph{Annals of behavioral medicine}}
  \bibinfo{volume}{46}, \bibinfo{number}{1} (\bibinfo{year}{2013}),
  \bibinfo{pages}{81--95}.
\newblock


\bibitem[\protect\citeauthoryear{Michie, Van~Stralen, and West}{Michie
  et~al\mbox{.}}{2011}]%
        {michie2011behaviour}
\bibfield{author}{\bibinfo{person}{Susan Michie}, \bibinfo{person}{Maartje~M
  Van~Stralen}, {and} \bibinfo{person}{Robert West}.}
  \bibinfo{year}{2011}\natexlab{}.
\newblock \showarticletitle{The behaviour change wheel: a new method for
  characterising and designing behaviour change interventions}.
\newblock \bibinfo{journal}{\emph{Implementation science}} \bibinfo{volume}{6},
  \bibinfo{number}{1} (\bibinfo{year}{2011}), \bibinfo{pages}{42}.
\newblock


\bibitem[\protect\citeauthoryear{Michie, West, Campbell, Brown, and
  Gainforth}{Michie et~al\mbox{.}}{2014b}]%
        {michie2014abc}
\bibfield{author}{\bibinfo{person}{SF Michie}, \bibinfo{person}{Robert West},
  \bibinfo{person}{Rona Campbell}, \bibinfo{person}{Jamie Brown}, {and}
  \bibinfo{person}{Heather Gainforth}.} \bibinfo{year}{2014}\natexlab{b}.
\newblock \bibinfo{booktitle}{\emph{ABC of behaviour change theories}}.
\newblock \bibinfo{publisher}{Silverback Publishing}.
\newblock


\bibitem[\protect\citeauthoryear{Miller and Rose}{Miller and Rose}{2009}]%
        {miller2009toward}
\bibfield{author}{\bibinfo{person}{William~R Miller} {and}
  \bibinfo{person}{Gary~S Rose}.} \bibinfo{year}{2009}\natexlab{}.
\newblock \showarticletitle{Toward a theory of motivational interviewing.}
\newblock \bibinfo{journal}{\emph{American psychologist}} \bibinfo{volume}{64},
  \bibinfo{number}{6} (\bibinfo{year}{2009}), \bibinfo{pages}{527}.
\newblock


\bibitem[\protect\citeauthoryear{Miltenberger}{Miltenberger}{2011}]%
        {miltenberger2011behavior}
\bibfield{author}{\bibinfo{person}{Raymond~G Miltenberger}.}
  \bibinfo{year}{2011}\natexlab{}.
\newblock \bibinfo{booktitle}{\emph{Behavior modification: Principles and
  procedures}}.
\newblock \bibinfo{publisher}{Cengage Learning}.
\newblock


\bibitem[\protect\citeauthoryear{mobile}{mobile}{2018}]%
        {flurry2015}
\bibfield{author}{\bibinfo{person}{Flurry mobile}.}
  \bibinfo{year}{2018}\natexlab{}.
\newblock \bibinfo{title}{App Engagement: The Matrix Reloaded}.
\newblock
\newblock
\urldef\tempurl%
\url{http://flurrymobile.tumblr.com/post/113379517625/app-engagement-the-matrix-reloaded}
\showURL{%
\tempurl}


\bibitem[\protect\citeauthoryear{Mohr, Cuijpers, and Lehman}{Mohr
  et~al\mbox{.}}{2011}]%
        {mohr2011supportive}
\bibfield{author}{\bibinfo{person}{David~C Mohr}, \bibinfo{person}{Pim
  Cuijpers}, {and} \bibinfo{person}{Kenneth Lehman}.}
  \bibinfo{year}{2011}\natexlab{}.
\newblock \showarticletitle{Supportive accountability: a model for providing
  human support to enhance adherence to eHealth interventions}.
\newblock \bibinfo{journal}{\emph{Journal of medical Internet research}}
  \bibinfo{volume}{13}, \bibinfo{number}{1} (\bibinfo{year}{2011}).
\newblock


\bibitem[\protect\citeauthoryear{Nahum-Shani, Henderson, Lim, and
  Vinokur}{Nahum-Shani et~al\mbox{.}}{2014a}]%
        {nahum2014supervisor}
\bibfield{author}{\bibinfo{person}{Inbal Nahum-Shani},
  \bibinfo{person}{Melanie~M Henderson}, \bibinfo{person}{Sandy Lim}, {and}
  \bibinfo{person}{Amiram~D Vinokur}.} \bibinfo{year}{2014}\natexlab{a}.
\newblock \showarticletitle{Supervisor support: Does supervisor support buffer
  or exacerbate the adverse effects of supervisor undermining?}
\newblock \bibinfo{journal}{\emph{Journal of Applied Psychology}}
  \bibinfo{volume}{99}, \bibinfo{number}{3} (\bibinfo{year}{2014}),
  \bibinfo{pages}{484}.
\newblock


\bibitem[\protect\citeauthoryear{Nahum-Shani, Smith, Tewari, Witkiewitz,
  Collins, Spring, and Murphy}{Nahum-Shani et~al\mbox{.}}{2014b}]%
        {nahum2014just}
\bibfield{author}{\bibinfo{person}{Inbal Nahum-Shani},
  \bibinfo{person}{Shawna~N Smith}, \bibinfo{person}{Ambuj Tewari},
  \bibinfo{person}{Katie Witkiewitz}, \bibinfo{person}{Linda~M Collins},
  \bibinfo{person}{Bonnie Spring}, {and} \bibinfo{person}{S Murphy}.}
  \bibinfo{year}{2014}\natexlab{b}.
\newblock \showarticletitle{Just in time adaptive interventions (jitais): An
  organizing framework for ongoing health behavior support}.
\newblock \bibinfo{journal}{\emph{Methodology Center technical report}}
  \bibinfo{number}{14-126} (\bibinfo{year}{2014}).
\newblock


\bibitem[\protect\citeauthoryear{Nestor, Roberts, Garavan, and Hester}{Nestor
  et~al\mbox{.}}{2008}]%
        {nestor2008deficits}
\bibfield{author}{\bibinfo{person}{Liam Nestor}, \bibinfo{person}{Gloria
  Roberts}, \bibinfo{person}{Hugh Garavan}, {and} \bibinfo{person}{Robert
  Hester}.} \bibinfo{year}{2008}\natexlab{}.
\newblock \showarticletitle{Deficits in learning and memory: parahippocampal
  hyperactivity and frontocortical hypoactivity in cannabis users}.
\newblock \bibinfo{journal}{\emph{Neuroimage}} \bibinfo{volume}{40},
  \bibinfo{number}{3} (\bibinfo{year}{2008}), \bibinfo{pages}{1328--1339}.
\newblock


\bibitem[\protect\citeauthoryear{Nicholson, Wang, Airhihenbuwa, Mahoney, and
  Maney}{Nicholson et~al\mbox{.}}{1992}]%
        {nicholson1992predicting}
\bibfield{author}{\bibinfo{person}{Mary~E Nicholson}, \bibinfo{person}{MinQi
  Wang}, \bibinfo{person}{Collins~O Airhihenbuwa}, \bibinfo{person}{Beverly~S
  Mahoney}, {and} \bibinfo{person}{Dolores~W Maney}.}
  \bibinfo{year}{1992}\natexlab{}.
\newblock \showarticletitle{Predicting alcohol impairment: Perceived
  intoxication versus BAC}.
\newblock \bibinfo{journal}{\emph{Alcoholism: Clinical and Experimental
  Research}} \bibinfo{volume}{16}, \bibinfo{number}{4} (\bibinfo{year}{1992}),
  \bibinfo{pages}{747--750}.
\newblock


\bibitem[\protect\citeauthoryear{Patton, Stanford, and Barratt}{Patton
  et~al\mbox{.}}{1995}]%
        {patton1995factor}
\bibfield{author}{\bibinfo{person}{Jim~H Patton}, \bibinfo{person}{Matthew~S
  Stanford}, {and} \bibinfo{person}{Ernest~S Barratt}.}
  \bibinfo{year}{1995}\natexlab{}.
\newblock \showarticletitle{Factor structure of the Barratt impulsiveness
  scale}.
\newblock \bibinfo{journal}{\emph{Journal of clinical psychology}}
  \bibinfo{volume}{51}, \bibinfo{number}{6} (\bibinfo{year}{1995}),
  \bibinfo{pages}{768--774}.
\newblock


\bibitem[\protect\citeauthoryear{Peeke, Jones, and Stone}{Peeke
  et~al\mbox{.}}{1976}]%
        {peeke1976effects}
\bibfield{author}{\bibinfo{person}{Shirley~C Peeke}, \bibinfo{person}{Reese~T
  Jones}, {and} \bibinfo{person}{George~C Stone}.}
  \bibinfo{year}{1976}\natexlab{}.
\newblock \showarticletitle{Effects of practice on marijuana-induced changes in
  reaction time}.
\newblock \bibinfo{journal}{\emph{Psychopharmacology}} \bibinfo{volume}{48},
  \bibinfo{number}{2} (\bibinfo{year}{1976}), \bibinfo{pages}{159--163}.
\newblock


\bibitem[\protect\citeauthoryear{Perski, Blandford, West, and Michie}{Perski
  et~al\mbox{.}}{2016}]%
        {perski2016conceptualising}
\bibfield{author}{\bibinfo{person}{Olga Perski}, \bibinfo{person}{Ann
  Blandford}, \bibinfo{person}{Robert West}, {and} \bibinfo{person}{Susan
  Michie}.} \bibinfo{year}{2016}\natexlab{}.
\newblock \showarticletitle{Conceptualising engagement with digital behaviour
  change interventions: a systematic review using principles from critical
  interpretive synthesis}.
\newblock \bibinfo{journal}{\emph{Translational behavioral medicine}}
  \bibinfo{volume}{7}, \bibinfo{number}{2} (\bibinfo{year}{2016}),
  \bibinfo{pages}{254--267}.
\newblock


\bibitem[\protect\citeauthoryear{Petty and Cacioppo}{Petty and
  Cacioppo}{1986}]%
        {petty1986elaboration}
\bibfield{author}{\bibinfo{person}{Richard~E Petty} {and}
  \bibinfo{person}{John~T Cacioppo}.} \bibinfo{year}{1986}\natexlab{}.
\newblock \showarticletitle{The elaboration likelihood model of persuasion}.
\newblock In \bibinfo{booktitle}{\emph{Communication and persuasion}}.
  \bibinfo{publisher}{Springer}, \bibinfo{pages}{1--24}.
\newblock


\bibitem[\protect\citeauthoryear{Phillips}{Phillips}{2012}]%
        {phillips2012behaviorism}
\bibfield{author}{\bibinfo{person}{Denis~C Phillips}.}
  \bibinfo{year}{2012}\natexlab{}.
\newblock \showarticletitle{Behaviorism and behaviorist learning theories}.
\newblock In \bibinfo{booktitle}{\emph{Encyclopedia of the Sciences of
  Learning}}. \bibinfo{publisher}{Springer}, \bibinfo{pages}{438--442}.
\newblock


\bibitem[\protect\citeauthoryear{Prinstein}{Prinstein}{2007}]%
        {prinstein2007moderators}
\bibfield{author}{\bibinfo{person}{Mitchell~J Prinstein}.}
  \bibinfo{year}{2007}\natexlab{}.
\newblock \showarticletitle{Moderators of peer contagion: A longitudinal
  examination of depression socialization between adolescents and their best
  friends}.
\newblock \bibinfo{journal}{\emph{Journal of Clinical Child and Adolescent
  Psychology}} \bibinfo{volume}{36}, \bibinfo{number}{2}
  (\bibinfo{year}{2007}), \bibinfo{pages}{159--170}.
\newblock


\bibitem[\protect\citeauthoryear{Rabbi, Aung, Zhang, and Choudhury}{Rabbi
  et~al\mbox{.}}{2015}]%
        {Rabbi:2015:MAP:2750858.2805840}
\bibfield{author}{\bibinfo{person}{Mashfiqui Rabbi}, \bibinfo{person}{Min~Hane
  Aung}, \bibinfo{person}{Mi Zhang}, {and} \bibinfo{person}{Tanzeem
  Choudhury}.} \bibinfo{year}{2015}\natexlab{}.
\newblock \showarticletitle{MyBehavior: Automatic Personalized Health Feedback
  from User Behaviors and Preferences Using Smartphones}. In
  \bibinfo{booktitle}{\emph{Proceedings of the 2015 ACM International Joint
  Conference on Pervasive and Ubiquitous Computing}}
  \emph{(\bibinfo{series}{UbiComp '15})}. \bibinfo{publisher}{ACM},
  \bibinfo{address}{New York, NY, USA}, \bibinfo{pages}{707--718}.
\newblock
\showISBNx{978-1-4503-3574-4}
\urldef\tempurl%
\url{https://doi.org/10.1145/2750858.2805840}
\showDOI{\tempurl}


\bibitem[\protect\citeauthoryear{Rabbi, Aung, Gay, Reid, and Choudhury}{Rabbi
  et~al\mbox{.}}{2018}]%
        {rabbi2018feasibility}
\bibfield{author}{\bibinfo{person}{Mashfiqui Rabbi}, \bibinfo{person}{Min~SH
  Aung}, \bibinfo{person}{Geri Gay}, \bibinfo{person}{M~Cary Reid}, {and}
  \bibinfo{person}{Tanzeem Choudhury}.} \bibinfo{year}{2018}\natexlab{}.
\newblock \showarticletitle{Feasibility and Acceptability of Mobile
  Phone--Based Auto-Personalized Physical Activity Recommendations for Chronic
  Pain Self-Management: Pilot Study on Adults}.
\newblock \bibinfo{journal}{\emph{J Med Internet Res}} \bibinfo{volume}{20},
  \bibinfo{number}{10} (\bibinfo{year}{2018}), \bibinfo{pages}{e10147}.
\newblock


\bibitem[\protect\citeauthoryear{Raedeke and Dlugonski}{Raedeke and
  Dlugonski}{2017}]%
        {raedeke2017high}
\bibfield{author}{\bibinfo{person}{Thomas~D Raedeke} {and}
  \bibinfo{person}{Deirdre Dlugonski}.} \bibinfo{year}{2017}\natexlab{}.
\newblock \showarticletitle{High Versus Low Theoretical Fidelity Pedometer
  Intervention Using Social-Cognitive Theory on Steps and Self-Efficacy}.
\newblock \bibinfo{journal}{\emph{Research quarterly for exercise and sport}}
  \bibinfo{volume}{88}, \bibinfo{number}{4} (\bibinfo{year}{2017}),
  \bibinfo{pages}{436--446}.
\newblock


\bibitem[\protect\citeauthoryear{Ramirez-Valles, Zimmerman, and
  Newcomb}{Ramirez-Valles et~al\mbox{.}}{1998}]%
        {ramirez1998sexual}
\bibfield{author}{\bibinfo{person}{Jesus Ramirez-Valles},
  \bibinfo{person}{Marc~A Zimmerman}, {and} \bibinfo{person}{Michael~D
  Newcomb}.} \bibinfo{year}{1998}\natexlab{}.
\newblock \showarticletitle{Sexual risk behavior among youth: Modeling the
  influence of prosocial activities and socioeconomic factors}.
\newblock \bibinfo{journal}{\emph{Journal of health and social behavior}}
  (\bibinfo{year}{1998}), \bibinfo{pages}{237--253}.
\newblock


\bibitem[\protect\citeauthoryear{Rankin, Abrams, Barry, Bhatnagar, Clayton,
  Colombo, Coppola, Geyer, Glanzman, Marsland, et~al\mbox{.}}{Rankin
  et~al\mbox{.}}{2009}]%
        {rankin2009habituation}
\bibfield{author}{\bibinfo{person}{Catharine~H Rankin}, \bibinfo{person}{Thomas
  Abrams}, \bibinfo{person}{Robert~J Barry}, \bibinfo{person}{Seema Bhatnagar},
  \bibinfo{person}{David~F Clayton}, \bibinfo{person}{John Colombo},
  \bibinfo{person}{Gianluca Coppola}, \bibinfo{person}{Mark~A Geyer},
  \bibinfo{person}{David~L Glanzman}, \bibinfo{person}{Stephen Marsland},
  {et~al\mbox{.}}} \bibinfo{year}{2009}\natexlab{}.
\newblock \showarticletitle{Habituation revisited: an updated and revised
  description of the behavioral characteristics of habituation}.
\newblock \bibinfo{journal}{\emph{Neurobiology of learning and memory}}
  \bibinfo{volume}{92}, \bibinfo{number}{2} (\bibinfo{year}{2009}),
  \bibinfo{pages}{135--138}.
\newblock


\bibitem[\protect\citeauthoryear{Redding, Rossi, Rossi, Velicer, and
  Prochaska}{Redding et~al\mbox{.}}{2000}]%
        {redding2000health}
\bibfield{author}{\bibinfo{person}{Colleen~A Redding},
  \bibinfo{person}{Joseph~S Rossi}, \bibinfo{person}{Susan~R Rossi},
  \bibinfo{person}{Wayne~F Velicer}, {and} \bibinfo{person}{James~O
  Prochaska}.} \bibinfo{year}{2000}\natexlab{}.
\newblock \showarticletitle{Health behavior models}. In
  \bibinfo{booktitle}{\emph{International Electronic Journal of Health
  Education}}. Citeseer.
\newblock


\bibitem[\protect\citeauthoryear{Reynolds}{Reynolds}{1975}]%
        {reynolds1975primer}
\bibfield{author}{\bibinfo{person}{George~Stanley Reynolds}.}
  \bibinfo{year}{1975}\natexlab{}.
\newblock \showarticletitle{A primer of operant conditioning, Rev}.
\newblock  (\bibinfo{year}{1975}).
\newblock


\bibitem[\protect\citeauthoryear{Roth, Felsher, Reed, Goldshear, Truong,
  Garfein, and Simmons}{Roth et~al\mbox{.}}{2017}]%
        {roth2017potential}
\bibfield{author}{\bibinfo{person}{Alexis~M Roth}, \bibinfo{person}{Marisa
  Felsher}, \bibinfo{person}{Megan Reed}, \bibinfo{person}{Jesse~L Goldshear},
  \bibinfo{person}{Quan Truong}, \bibinfo{person}{Richard~S Garfein}, {and}
  \bibinfo{person}{Janie Simmons}.} \bibinfo{year}{2017}\natexlab{}.
\newblock \showarticletitle{Potential benefits of using ecological momentary
  assessment to study high-risk polydrug use}.
\newblock \bibinfo{journal}{\emph{mHealth}}  \bibinfo{volume}{3}
  (\bibinfo{year}{2017}).
\newblock


\bibitem[\protect\citeauthoryear{Rothman}{Rothman}{2000}]%
        {rothman2000toward}
\bibfield{author}{\bibinfo{person}{Alexander~J Rothman}.}
  \bibinfo{year}{2000}\natexlab{}.
\newblock \showarticletitle{Toward a theory-based analysis of behavioral
  maintenance.}
\newblock \bibinfo{journal}{\emph{Health Psychology}} \bibinfo{volume}{19},
  \bibinfo{number}{1S} (\bibinfo{year}{2000}), \bibinfo{pages}{64}.
\newblock


\bibitem[\protect\citeauthoryear{Rovniak, Melbourne, Hovell, Wojcik, Winett,
  and Martinez-Donate}{Rovniak et~al\mbox{.}}{[n.d.]}]%
        {Rovniak}
\bibfield{author}{\bibinfo{person}{Liza~S Rovniak}, \bibinfo{person}{;
  Melbourne}, \bibinfo{person}{F Hovell}, \bibinfo{person}{Janet~R Wojcik},
  \bibinfo{person}{Richard~A Winett}, {and} \bibinfo{person}{Ana~P
  Martinez-Donate}.} \bibinfo{year}{[n.d.]}\natexlab{}.
\newblock \bibinfo{booktitle}{\emph{{Enhancing Theoretical Fidelity: An
  E-mail-based Walking Program Demonstration}}}.
\newblock \bibinfo{type}{{T}echnical {R}eport}~2.
\newblock
\urldef\tempurl%
\url{https://journals-sagepub-com.ezp-prod1.hul.harvard.edu/doi/pdf/10.4278/0890-1171-20.2.85}
\showURL{%
\tempurl}


\bibitem[\protect\citeauthoryear{Schepis, Desai, Cavallo, Smith, McFetridge,
  Liss, Potenza, and Krishnan-Sarin}{Schepis et~al\mbox{.}}{2011}]%
        {schepis2011gender}
\bibfield{author}{\bibinfo{person}{Ty~S Schepis}, \bibinfo{person}{Rani~A
  Desai}, \bibinfo{person}{Dana~A Cavallo}, \bibinfo{person}{Anne~E Smith},
  \bibinfo{person}{Amanda McFetridge}, \bibinfo{person}{Thomas~B Liss},
  \bibinfo{person}{Marc~N Potenza}, {and} \bibinfo{person}{Suchitra
  Krishnan-Sarin}.} \bibinfo{year}{2011}\natexlab{}.
\newblock \showarticletitle{Gender differences in adolescent marijuana use and
  associated psychosocial characteristics}.
\newblock \bibinfo{journal}{\emph{Journal of addiction medicine}}
  \bibinfo{volume}{5}, \bibinfo{number}{1} (\bibinfo{year}{2011}),
  \bibinfo{pages}{65}.
\newblock


\bibitem[\protect\citeauthoryear{Schueller, Neary, O'Loughlin, and
  Adkins}{Schueller et~al\mbox{.}}{2018}]%
        {schueller2018discovery}
\bibfield{author}{\bibinfo{person}{Stephen~M Schueller},
  \bibinfo{person}{Martha Neary}, \bibinfo{person}{Kristen O'Loughlin}, {and}
  \bibinfo{person}{Elizabeth~C Adkins}.} \bibinfo{year}{2018}\natexlab{}.
\newblock \showarticletitle{Discovery of and interest in health apps among
  those with mental health needs: survey and focus group study}.
\newblock \bibinfo{journal}{\emph{Journal of medical Internet research}}
  \bibinfo{volume}{20}, \bibinfo{number}{6} (\bibinfo{year}{2018}),
  \bibinfo{pages}{e10141}.
\newblock


\bibitem[\protect\citeauthoryear{SCRAM}{SCRAM}{2018}]%
        {scram2017}
\bibfield{author}{\bibinfo{person}{SCRAM}.} \bibinfo{year}{2018}\natexlab{}.
\newblock \bibinfo{title}{SCRAM CAM (2017)}.
\newblock
\newblock
\urldef\tempurl%
\url{https://www.scramsystems.com/products/scram-continuous-alcohol-monitoring/}
\showURL{%
\tempurl}


\bibitem[\protect\citeauthoryear{Serre, Fatseas, Swendsen, and
  Auriacombe}{Serre et~al\mbox{.}}{2015}]%
        {serre2015ecological}
\bibfield{author}{\bibinfo{person}{Fuschia Serre}, \bibinfo{person}{Melina
  Fatseas}, \bibinfo{person}{Joel Swendsen}, {and} \bibinfo{person}{Marc
  Auriacombe}.} \bibinfo{year}{2015}\natexlab{}.
\newblock \showarticletitle{Ecological momentary assessment in the
  investigation of craving and substance use in daily life: a systematic
  review}.
\newblock \bibinfo{journal}{\emph{Drug and Alcohol Dependence}}
  \bibinfo{volume}{148} (\bibinfo{year}{2015}), \bibinfo{pages}{1--20}.
\newblock


\bibitem[\protect\citeauthoryear{Shield}{Shield}{2000}]%
        {shield2000critical}
\bibfield{author}{\bibinfo{person}{George Shield}.}
  \bibinfo{year}{2000}\natexlab{}.
\newblock \showarticletitle{A Critical Appraisal of Learning Technology Using
  Information and Communication Technologies.}
\newblock \bibinfo{journal}{\emph{Journal of technology Studies}}
  \bibinfo{volume}{26}, \bibinfo{number}{1} (\bibinfo{year}{2000}),
  \bibinfo{pages}{71--79}.
\newblock


\bibitem[\protect\citeauthoryear{Shiffman}{Shiffman}{2009}]%
        {shiffman2009ecological}
\bibfield{author}{\bibinfo{person}{Saul Shiffman}.}
  \bibinfo{year}{2009}\natexlab{}.
\newblock \showarticletitle{Ecological momentary assessment (EMA) in studies of
  substance use.}
\newblock \bibinfo{journal}{\emph{Psychological assessment}}
  \bibinfo{volume}{21}, \bibinfo{number}{4} (\bibinfo{year}{2009}),
  \bibinfo{pages}{486}.
\newblock


\bibitem[\protect\citeauthoryear{Shneidermana}{Shneidermana}{2016}]%
        {shneidermana2016dangers}
\bibfield{author}{\bibinfo{person}{Ben Shneidermana}.}
  \bibinfo{year}{2016}\natexlab{}.
\newblock \showarticletitle{The dangers of faulty, biased, or malicious
  algorithms requires independent oversight}.
\newblock \bibinfo{journal}{\emph{PNAS}} \bibinfo{volume}{113},
  \bibinfo{number}{48} (\bibinfo{year}{2016}), \bibinfo{pages}{13539}.
\newblock


\bibitem[\protect\citeauthoryear{Shrier, Burke, Kells, Scherer, Sarda,
  Jonestrask, Xuan, and Harris}{Shrier et~al\mbox{.}}{2018}]%
        {shrier2018pilot}
\bibfield{author}{\bibinfo{person}{Lydia~A Shrier}, \bibinfo{person}{Pamela~J
  Burke}, \bibinfo{person}{Meredith Kells}, \bibinfo{person}{Emily~A Scherer},
  \bibinfo{person}{Vishnudas Sarda}, \bibinfo{person}{Cassandra Jonestrask},
  \bibinfo{person}{Ziming Xuan}, {and} \bibinfo{person}{Sion~Kim Harris}.}
  \bibinfo{year}{2018}\natexlab{}.
\newblock \showarticletitle{Pilot randomized trial of MOMENT, a motivational
  counseling-plus-ecological momentary intervention to reduce marijuana use in
  youth}.
\newblock \bibinfo{journal}{\emph{mHealth}}  \bibinfo{volume}{4}
  (\bibinfo{year}{2018}).
\newblock


\bibitem[\protect\citeauthoryear{Shrier, Rhoads, Burke, Walls, and
  Blood}{Shrier et~al\mbox{.}}{2014}]%
        {shrier2014real}
\bibfield{author}{\bibinfo{person}{Lydia~A Shrier}, \bibinfo{person}{Amanda
  Rhoads}, \bibinfo{person}{Pamela Burke}, \bibinfo{person}{Courtney Walls},
  {and} \bibinfo{person}{Emily~A Blood}.} \bibinfo{year}{2014}\natexlab{}.
\newblock \showarticletitle{Real-time, contextual intervention using mobile
  technology to reduce marijuana use among youth: a pilot study}.
\newblock \bibinfo{journal}{\emph{Addictive behaviors}} \bibinfo{volume}{39},
  \bibinfo{number}{1} (\bibinfo{year}{2014}), \bibinfo{pages}{173--180}.
\newblock


\bibitem[\protect\citeauthoryear{Shrier, Walls, Rhoads, and Blood}{Shrier
  et~al\mbox{.}}{2013}]%
        {shrier2013individual}
\bibfield{author}{\bibinfo{person}{Lydia~A Shrier}, \bibinfo{person}{Courtney
  Walls}, \bibinfo{person}{Amanda Rhoads}, {and} \bibinfo{person}{Emily~A
  Blood}.} \bibinfo{year}{2013}\natexlab{}.
\newblock \showarticletitle{Individual and contextual predictors of severity of
  marijuana use events among young frequent users}.
\newblock \bibinfo{journal}{\emph{Addictive behaviors}} \bibinfo{volume}{38},
  \bibinfo{number}{1} (\bibinfo{year}{2013}), \bibinfo{pages}{1448--1456}.
\newblock


\bibitem[\protect\citeauthoryear{Simons, Correia, Carey, and Borsari}{Simons
  et~al\mbox{.}}{1998}]%
        {simons1998validating}
\bibfield{author}{\bibinfo{person}{Jeffrey Simons},
  \bibinfo{person}{Christopher~J Correia}, \bibinfo{person}{Kate~B Carey},
  {and} \bibinfo{person}{Brian~E Borsari}.} \bibinfo{year}{1998}\natexlab{}.
\newblock \showarticletitle{Validating a five-factor marijuana motives measure:
  Relations with use, problems, and alcohol motives.}
\newblock \bibinfo{journal}{\emph{Journal of Counseling Psychology}}
  \bibinfo{volume}{45}, \bibinfo{number}{3} (\bibinfo{year}{1998}),
  \bibinfo{pages}{265}.
\newblock


\bibitem[\protect\citeauthoryear{Simpson and Vuchinich}{Simpson and
  Vuchinich}{2000}]%
        {simpson2000temporal}
\bibfield{author}{\bibinfo{person}{Cathy~A Simpson} {and}
  \bibinfo{person}{Rudy~E Vuchinich}.} \bibinfo{year}{2000}\natexlab{}.
\newblock \showarticletitle{Temporal changes in the value of objects of choice:
  Discounting, behavior patterns, and health behavior}.
\newblock \bibinfo{journal}{\emph{Reframing health behavior change with
  behavioral economics}} (\bibinfo{year}{2000}), \bibinfo{pages}{193--215}.
\newblock


\bibitem[\protect\citeauthoryear{Sinha}{Sinha}{2008}]%
        {sinha2008chronic}
\bibfield{author}{\bibinfo{person}{Rajita Sinha}.}
  \bibinfo{year}{2008}\natexlab{}.
\newblock \showarticletitle{Chronic stress, drug use, and vulnerability to
  addiction}.
\newblock \bibinfo{journal}{\emph{Annals of the new York Academy of Sciences}}
  \bibinfo{volume}{1141}, \bibinfo{number}{1} (\bibinfo{year}{2008}),
  \bibinfo{pages}{105--130}.
\newblock


\bibitem[\protect\citeauthoryear{Skinner}{Skinner}{2011}]%
        {skinner2011behaviorism}
\bibfield{author}{\bibinfo{person}{Burrhus~Frederic Skinner}.}
  \bibinfo{year}{2011}\natexlab{}.
\newblock \bibinfo{booktitle}{\emph{About behaviorism}}.
\newblock \bibinfo{publisher}{Vintage}.
\newblock


\bibitem[\protect\citeauthoryear{Staddon and Cerutti}{Staddon and
  Cerutti}{2003}]%
        {staddon2003operant}
\bibfield{author}{\bibinfo{person}{John~ER Staddon} {and}
  \bibinfo{person}{Daniel~T Cerutti}.} \bibinfo{year}{2003}\natexlab{}.
\newblock \showarticletitle{Operant conditioning}.
\newblock \bibinfo{journal}{\emph{Annual review of psychology}}
  \bibinfo{volume}{54}, \bibinfo{number}{1} (\bibinfo{year}{2003}),
  \bibinfo{pages}{115--144}.
\newblock


\bibitem[\protect\citeauthoryear{Stephens, Babor, Kadden, Miller, and
  Group}{Stephens et~al\mbox{.}}{2002}]%
        {stephens2002marijuana}
\bibfield{author}{\bibinfo{person}{Robert~S Stephens},
  \bibinfo{person}{Thomas~F Babor}, \bibinfo{person}{Ronald Kadden},
  \bibinfo{person}{Michael Miller}, {and} \bibinfo{person}{Marijuana Treatment
  Project~Research Group}.} \bibinfo{year}{2002}\natexlab{}.
\newblock \showarticletitle{The Marijuana Treatment Project: rationale, design
  and participant characteristics}.
\newblock \bibinfo{journal}{\emph{Addiction}}  \bibinfo{volume}{97}
  (\bibinfo{year}{2002}), \bibinfo{pages}{109--124}.
\newblock


\bibitem[\protect\citeauthoryear{Stoyanov, Hides, Kavanagh, Zelenko,
  Tjondronegoro, and Mani}{Stoyanov et~al\mbox{.}}{2015}]%
        {stoyanov2015mobile}
\bibfield{author}{\bibinfo{person}{Stoyan~R Stoyanov}, \bibinfo{person}{Leanne
  Hides}, \bibinfo{person}{David~J Kavanagh}, \bibinfo{person}{Oksana Zelenko},
  \bibinfo{person}{Dian Tjondronegoro}, {and} \bibinfo{person}{Madhavan Mani}.}
  \bibinfo{year}{2015}\natexlab{}.
\newblock \showarticletitle{Mobile app rating scale: a new tool for assessing
  the quality of health mobile apps}.
\newblock \bibinfo{journal}{\emph{JMIR mHealth and uHealth}}
  \bibinfo{volume}{3}, \bibinfo{number}{1} (\bibinfo{year}{2015}).
\newblock


\bibitem[\protect\citeauthoryear{Suffoletto, Callaway, Kristan, Kraemer, and
  Clark}{Suffoletto et~al\mbox{.}}{2012}]%
        {suffoletto2012text}
\bibfield{author}{\bibinfo{person}{Brian Suffoletto}, \bibinfo{person}{Clifton
  Callaway}, \bibinfo{person}{Jeff Kristan}, \bibinfo{person}{Kevin Kraemer},
  {and} \bibinfo{person}{Duncan~B Clark}.} \bibinfo{year}{2012}\natexlab{}.
\newblock \showarticletitle{Text-message-based drinking assessments and brief
  interventions for young adults discharged from the emergency department}.
\newblock \bibinfo{journal}{\emph{Alcoholism: Clinical and Experimental
  Research}} \bibinfo{volume}{36}, \bibinfo{number}{3} (\bibinfo{year}{2012}),
  \bibinfo{pages}{552--560}.
\newblock


\bibitem[\protect\citeauthoryear{Sundar}{Sundar}{2007}]%
        {sundar2007social}
\bibfield{author}{\bibinfo{person}{S~Shyam Sundar}.}
  \bibinfo{year}{2007}\natexlab{}.
\newblock \showarticletitle{Social psychology of interactivity in human-website
  interaction}.
\newblock In \bibinfo{booktitle}{\emph{Oxford handbook of internet
  psychology}}.
\newblock


\bibitem[\protect\citeauthoryear{Theofanopoulou, Isbister, Edbrooke-Childs, and
  Slov{\'a}k}{Theofanopoulou et~al\mbox{.}}{2019}]%
        {theofanopoulou2019smart}
\bibfield{author}{\bibinfo{person}{Nikki Theofanopoulou},
  \bibinfo{person}{Katherine Isbister}, \bibinfo{person}{Julian
  Edbrooke-Childs}, {and} \bibinfo{person}{Petr Slov{\'a}k}.}
  \bibinfo{year}{2019}\natexlab{}.
\newblock \showarticletitle{A Smart Toy Intervention to Promote Emotion
  Regulation in Middle Childhood: Feasibility Study}.
\newblock \bibinfo{journal}{\emph{JMIR mental health}} \bibinfo{volume}{6},
  \bibinfo{number}{8} (\bibinfo{year}{2019}), \bibinfo{pages}{e14029}.
\newblock


\bibitem[\protect\citeauthoryear{Thompson}{Thompson}{2015}]%
        {thompson2015habituation}
\bibfield{author}{\bibinfo{person}{RF Thompson}.}
  \bibinfo{year}{2015}\natexlab{}.
\newblock \showarticletitle{Habituation}.
\newblock  (\bibinfo{year}{2015}).
\newblock


\bibitem[\protect\citeauthoryear{Trosclair-Lasserre, Lerman, Call, Addison, and
  Kodak}{Trosclair-Lasserre et~al\mbox{.}}{2008}]%
        {Trosclair-Lasserre}
\bibfield{author}{\bibinfo{person}{Nicole~M Trosclair-Lasserre},
  \bibinfo{person}{Dorothea~C Lerman}, \bibinfo{person}{Nathan~A Call},
  \bibinfo{person}{Laura~R Addison}, {and} \bibinfo{person}{Tiffany Kodak}.}
  \bibinfo{year}{2008}\natexlab{}.
\newblock \showarticletitle{Reinforcement magnitude: An evaluation of
  preference and reinforcer efficacy}.
\newblock \bibinfo{journal}{\emph{Journal of Applied Behavior Analysis}}
  \bibinfo{volume}{41}, \bibinfo{number}{2} (\bibinfo{year}{2008}),
  \bibinfo{pages}{203--220}.
\newblock


\bibitem[\protect\citeauthoryear{Tuten, DeFulio, Jones, and Stitzer}{Tuten
  et~al\mbox{.}}{2012}]%
        {tuten2012abstinence}
\bibfield{author}{\bibinfo{person}{Michelle Tuten}, \bibinfo{person}{Anthony
  DeFulio}, \bibinfo{person}{Hendr{\'e}e~E Jones}, {and}
  \bibinfo{person}{Maxine Stitzer}.} \bibinfo{year}{2012}\natexlab{}.
\newblock \showarticletitle{Abstinence-contingent recovery housing and
  reinforcement-based treatment following opioid detoxification}.
\newblock \bibinfo{journal}{\emph{Addiction}} \bibinfo{volume}{107},
  \bibinfo{number}{5} (\bibinfo{year}{2012}), \bibinfo{pages}{973--982}.
\newblock


\bibitem[\protect\citeauthoryear{Van~Berkel, Goncalves, Hosio, and
  Kostakos}{Van~Berkel et~al\mbox{.}}{2017}]%
        {van2017gamification}
\bibfield{author}{\bibinfo{person}{Niels Van~Berkel}, \bibinfo{person}{Jorge
  Goncalves}, \bibinfo{person}{Simo Hosio}, {and} \bibinfo{person}{Vassilis
  Kostakos}.} \bibinfo{year}{2017}\natexlab{}.
\newblock \showarticletitle{Gamification of mobile experience sampling improves
  data quality and quantity}.
\newblock \bibinfo{journal}{\emph{Proceedings of the ACM on Interactive,
  Mobile, Wearable and Ubiquitous Technologies}} \bibinfo{volume}{1},
  \bibinfo{number}{3} (\bibinfo{year}{2017}), \bibinfo{pages}{107}.
\newblock


\bibitem[\protect\citeauthoryear{Van~den Bos}{Van~den Bos}{2009}]%
        {van2009making}
\bibfield{author}{\bibinfo{person}{Kees Van~den Bos}.}
  \bibinfo{year}{2009}\natexlab{}.
\newblock \showarticletitle{Making sense of life: The existential self trying
  to deal with personal uncertainty}.
\newblock \bibinfo{journal}{\emph{Psychological Inquiry}} \bibinfo{volume}{20},
  \bibinfo{number}{4} (\bibinfo{year}{2009}), \bibinfo{pages}{197--217}.
\newblock


\bibitem[\protect\citeauthoryear{Vilardaga, Rizo, Zeng, Kientz, Ries, Otis, and
  Hernandez}{Vilardaga et~al\mbox{.}}{2018}]%
        {vilardaga2018user}
\bibfield{author}{\bibinfo{person}{Roger Vilardaga}, \bibinfo{person}{Javier
  Rizo}, \bibinfo{person}{Emily Zeng}, \bibinfo{person}{Julie~A Kientz},
  \bibinfo{person}{Richard Ries}, \bibinfo{person}{Chad Otis}, {and}
  \bibinfo{person}{Kayla Hernandez}.} \bibinfo{year}{2018}\natexlab{}.
\newblock \showarticletitle{User-centered design of learn to quit, a smoking
  cessation smartphone app for people with serious mental illness}.
\newblock \bibinfo{journal}{\emph{JMIR serious games}} \bibinfo{volume}{6},
  \bibinfo{number}{1} (\bibinfo{year}{2018}), \bibinfo{pages}{e2}.
\newblock


\bibitem[\protect\citeauthoryear{Villamar{\'\i}n-Salom{\'o}n and
  Brustoloni}{Villamar{\'\i}n-Salom{\'o}n and Brustoloni}{2010}]%
        {villamarin2010using}
\bibfield{author}{\bibinfo{person}{Ricardo~Mark Villamar{\'\i}n-Salom{\'o}n}
  {and} \bibinfo{person}{Jos{\'e}~Carlos Brustoloni}.}
  \bibinfo{year}{2010}\natexlab{}.
\newblock \showarticletitle{Using reinforcement to strengthen users' secure
  behaviors}. In \bibinfo{booktitle}{\emph{Proceedings of the SIGCHI Conference
  on Human Factors in Computing Systems}}. ACM, \bibinfo{pages}{363--372}.
\newblock


\bibitem[\protect\citeauthoryear{Vorauer}{Vorauer}{2006}]%
        {vorauer2006information}
\bibfield{author}{\bibinfo{person}{Jacquie~D Vorauer}.}
  \bibinfo{year}{2006}\natexlab{}.
\newblock \showarticletitle{An information search model of evaluative concerns
  in intergroup interaction.}
\newblock \bibinfo{journal}{\emph{Psychological Review}} \bibinfo{volume}{113},
  \bibinfo{number}{4} (\bibinfo{year}{2006}), \bibinfo{pages}{862}.
\newblock


\bibitem[\protect\citeauthoryear{Weary and Jacobson}{Weary and
  Jacobson}{1997}]%
        {weary1997causal}
\bibfield{author}{\bibinfo{person}{Gifford Weary} {and} \bibinfo{person}{Jill~A
  Jacobson}.} \bibinfo{year}{1997}\natexlab{}.
\newblock \showarticletitle{Causal uncertainty beliefs and diagnostic
  information seeking.}
\newblock \bibinfo{journal}{\emph{Journal of Personality and Social
  Psychology}} \bibinfo{volume}{73}, \bibinfo{number}{4}
  (\bibinfo{year}{1997}), \bibinfo{pages}{839}.
\newblock


\bibitem[\protect\citeauthoryear{Weatherly, McSweeney, and Swindell}{Weatherly
  et~al\mbox{.}}{1996}]%
        {weatherly1996within}
\bibfield{author}{\bibinfo{person}{Jeffrey~N Weatherly},
  \bibinfo{person}{Frances~K McSweeney}, {and} \bibinfo{person}{Samantha
  Swindell}.} \bibinfo{year}{1996}\natexlab{}.
\newblock \showarticletitle{WITHIN-SESSION RESPONSE PATTERNS ON CONJOINT
  VARIABLE-INTERVAL VARIABLE-TIME SCHEDULES}.
\newblock \bibinfo{journal}{\emph{Journal of the experimental analysis of
  behavior}} \bibinfo{volume}{66}, \bibinfo{number}{2} (\bibinfo{year}{1996}),
  \bibinfo{pages}{205--218}.
\newblock


\bibitem[\protect\citeauthoryear{Weegar and Pacis}{Weegar and Pacis}{2012}]%
        {weegar2012comparison}
\bibfield{author}{\bibinfo{person}{Mary~Anne Weegar} {and}
  \bibinfo{person}{Dina Pacis}.} \bibinfo{year}{2012}\natexlab{}.
\newblock \showarticletitle{A Comparison of two theories of
  learning-behaviorism and constructivism as applied to face-to-face and online
  learning}. In \bibinfo{booktitle}{\emph{Proceedings e-leader conference,
  Manila}}.
\newblock


\bibitem[\protect\citeauthoryear{Wen, Schneider, Stone, and Spruijt-Metz}{Wen
  et~al\mbox{.}}{2017}]%
        {wen2017compliance}
\bibfield{author}{\bibinfo{person}{Cheng K~Fred Wen}, \bibinfo{person}{Stefan
  Schneider}, \bibinfo{person}{Arthur~A Stone}, {and} \bibinfo{person}{Donna
  Spruijt-Metz}.} \bibinfo{year}{2017}\natexlab{}.
\newblock \showarticletitle{Compliance with mobile ecological momentary
  assessment protocols in children and adolescents: a systematic review and
  meta-analysis}.
\newblock \bibinfo{journal}{\emph{Journal of medical Internet research}}
  \bibinfo{volume}{19}, \bibinfo{number}{4} (\bibinfo{year}{2017}).
\newblock


\bibitem[\protect\citeauthoryear{Werbach and Hunter}{Werbach and
  Hunter}{2012}]%
        {werbach2012win}
\bibfield{author}{\bibinfo{person}{Kevin Werbach} {and} \bibinfo{person}{Dan
  Hunter}.} \bibinfo{year}{2012}\natexlab{}.
\newblock \bibinfo{booktitle}{\emph{For the win: How game thinking can
  revolutionize your business}}.
\newblock \bibinfo{publisher}{Wharton Digital Press}.
\newblock


\bibitem[\protect\citeauthoryear{Wolf, Risley, and Mees}{Wolf
  et~al\mbox{.}}{1963}]%
        {wolf1963application}
\bibfield{author}{\bibinfo{person}{Montrose Wolf}, \bibinfo{person}{Todd
  Risley}, {and} \bibinfo{person}{Hayden Mees}.}
  \bibinfo{year}{1963}\natexlab{}.
\newblock \showarticletitle{Application of operant conditioning procedures to
  the behaviour problems of an autistic child}.
\newblock \bibinfo{journal}{\emph{Behaviour Research and Therapy}}
  \bibinfo{volume}{1}, \bibinfo{number}{2-4} (\bibinfo{year}{1963}),
  \bibinfo{pages}{305--312}.
\newblock


\bibitem[\protect\citeauthoryear{Wolpe}{Wolpe}{1968}]%
        {wolpe1968psychotherapy}
\bibfield{author}{\bibinfo{person}{Joseph Wolpe}.}
  \bibinfo{year}{1968}\natexlab{}.
\newblock \showarticletitle{Psychotherapy by reciprocal inhibition}.
\newblock \bibinfo{journal}{\emph{Conditional reflex: a Pavlovian journal of
  research \& therapy}} \bibinfo{volume}{3}, \bibinfo{number}{4}
  (\bibinfo{year}{1968}), \bibinfo{pages}{234--240}.
\newblock


\bibitem[\protect\citeauthoryear{Wray, Merrill, and Monti}{Wray
  et~al\mbox{.}}{2014}]%
        {wray2014using}
\bibfield{author}{\bibinfo{person}{Tyler~B Wray}, \bibinfo{person}{Jennifer~E
  Merrill}, {and} \bibinfo{person}{Peter~M Monti}.}
  \bibinfo{year}{2014}\natexlab{}.
\newblock \showarticletitle{Using ecological momentary assessment (EMA) to
  assess situation-level predictors of alcohol use and alcohol-related
  consequences}.
\newblock \bibinfo{journal}{\emph{Alcohol research: current reviews}}
  \bibinfo{volume}{36}, \bibinfo{number}{1} (\bibinfo{year}{2014}),
  \bibinfo{pages}{19}.
\newblock


\bibitem[\protect\citeauthoryear{You, Lin, Li, Tsai, Huang, Lee, Wang, and
  Chu}{You et~al\mbox{.}}{2016}]%
        {you2016kediary}
\bibfield{author}{\bibinfo{person}{Chuang-Wen You}, \bibinfo{person}{Ya-Fang
  Lin}, \bibinfo{person}{Cheng-Yuan Li}, \bibinfo{person}{Yu-Lun Tsai},
  \bibinfo{person}{Ming-Chyi Huang}, \bibinfo{person}{Chao-Hui Lee},
  \bibinfo{person}{Hao-Chuan Wang}, {and} \bibinfo{person}{Hao-Hua Chu}.}
  \bibinfo{year}{2016}\natexlab{}.
\newblock \showarticletitle{KeDiary: Using Mobile Phones to Assist Patients in
  Recovering from Drug Addiction}. In \bibinfo{booktitle}{\emph{Proceedings of
  the 2016 CHI Conference on Human Factors in Computing Systems}}. ACM,
  \bibinfo{pages}{5704--5709}.
\newblock


\bibitem[\protect\citeauthoryear{You, Wang, Huang, Chen, Lin, Ho, Wang, Huang,
  and Chu}{You et~al\mbox{.}}{2015}]%
        {you2015soberdiary}
\bibfield{author}{\bibinfo{person}{Chuang-wen You}, \bibinfo{person}{Kuo-Cheng
  Wang}, \bibinfo{person}{Ming-Chyi Huang}, \bibinfo{person}{Yen-Chang Chen},
  \bibinfo{person}{Cheng-Lin Lin}, \bibinfo{person}{Po-Shiun Ho},
  \bibinfo{person}{Hao-Chuan Wang}, \bibinfo{person}{Polly Huang}, {and}
  \bibinfo{person}{Hao-Hua Chu}.} \bibinfo{year}{2015}\natexlab{}.
\newblock \showarticletitle{Soberdiary: A phone-based support system for
  assisting recovery from alcohol dependence}. In
  \bibinfo{booktitle}{\emph{Proceedings of the 33rd Annual ACM Conference on
  Human Factors in Computing Systems}}. ACM, \bibinfo{pages}{3839--3848}.
\newblock


\bibitem[\protect\citeauthoryear{Zang}{Zang}{2007}]%
        {Zang2007}
\bibfield{author}{\bibinfo{person}{A Zang}.} \bibinfo{year}{2007}\natexlab{}.
\newblock \showarticletitle{{Satiation, Habituation, and Elasticity: An
  Economic Analysis}}.
\newblock  (\bibinfo{year}{2007}).
\newblock
\urldef\tempurl%
\url{https://digitalcommons.iwu.edu/cgi/viewcontent.cgi?article=2093{\&}context=jwprc}
\showURL{%
\tempurl}


\bibitem[\protect\citeauthoryear{Zichermann and Cunningham}{Zichermann and
  Cunningham}{2011}]%
        {zichermann2011gamification}
\bibfield{author}{\bibinfo{person}{Gabe Zichermann} {and}
  \bibinfo{person}{Christopher Cunningham}.} \bibinfo{year}{2011}\natexlab{}.
\newblock \bibinfo{booktitle}{\emph{Gamification by design: Implementing game
  mechanics in web and mobile apps}}.
\newblock \bibinfo{publisher}{" O'Reilly Media, Inc."}.
\newblock


\bibitem[\protect\citeauthoryear{Zuckerman and Gal-Oz}{Zuckerman and
  Gal-Oz}{2014}]%
        {zuckerman2014deconstructing}
\bibfield{author}{\bibinfo{person}{Oren Zuckerman} {and}
  \bibinfo{person}{Ayelet Gal-Oz}.} \bibinfo{year}{2014}\natexlab{}.
\newblock \showarticletitle{Deconstructing gamification: evaluating the
  effectiveness of continuous measurement, virtual rewards, and social
  comparison for promoting physical activity}.
\newblock \bibinfo{journal}{\emph{Personal and ubiquitous computing}}
  \bibinfo{volume}{18}, \bibinfo{number}{7} (\bibinfo{year}{2014}),
  \bibinfo{pages}{1705--1719}.
\newblock


\end{thebibliography}
\bibliographystyle{ACM-Reference-Format}
\end{document}